\documentclass[twocolumn]{aa}
\usepackage{amsmath}
\usepackage{newtxtext,newtxmath}
\usepackage{xcolor,hyperref}
\usepackage[T1]{fontenc}

\usepackage{etoolbox}
\BeforeBeginEnvironment{tabular}{\begin{center}\footnotesize}
\AfterEndEnvironment{tabular}{\end{center}}

\newcommand{\qso}{J1120+0641}

\newcommand\ions[2]{#1$\;${\textsc{#2}}}
\newcommand{\hb}{H$\beta$}
\newcommand{\oiii}{[\ions{O}{iii}] $\lambda\lambda4959,5007$}
\newcommand{\oiiib}{[\ions{O}{iii}] $\lambda4959$}
\newcommand{\oiiia}{[\ions{O}{iii}] $\lambda5007$}
\newcommand{\cii}{[\ions{C}{ii}] $158\mu$m}

\newcommand{\ha}{H$\alpha$}

\newcommand{\niia}{[\ions{N}{ii}] ${\lambda6583}$}

\newcommand{\DELS}{DELS J0411$-$0907}
\newcommand{\VDES}{VDES J0020$-$3653}

\newcommand{\0}{\phantom{0}}
\newcommand\micron{\mbox{$\mu$m}}
\usepackage{microtype}

\begin{document}

\title{GA-NIFS and EIGER: A merging quasar host at $z=7$ \\with an overmassive black hole }
\titlerunning{A merging quasar host at $z=7$ with an overmassive black hole}
\authorrunning{M.A. Marshall et al.}

\author{Madeline A. Marshall\inst{\ref{LANL}}\thanks{Email: \href{mailto:mmarshall@lanl.gov}{mmarshall@lanl.gov}} \and
Minghao Yue\inst{\ref{MIT}} \and 
Anna-Christina Eilers\inst{\ref{MIT}} \and
Jan Scholtz\inst{\ref{RM1},\ref{RM2}} \and 
Michele Perna\inst{\ref{SA}} \and \\
Chris J. Willott \inst{\ref{CW}} \and 
Roberto Maiolino \inst{\ref{RM1},\ref{RM2},\ref{RM3}} \and 
Hannah \"Ubler\inst{\ref{RM1},\ref{RM2},\ref{HU}} \and 
Santiago Arribas\inst{\ref{SA}} \and
Andrew J. Bunker\inst{\ref{oxford}}\and\\
Stephane Charlot \inst{\ref{SC}} \and 
Bruno Rodr\'iguez Del Pino\inst{\ref{SA}} \and 
Torsten B\"{o}ker\inst{\ref{TB}} \and 
Stefano Carniani\inst{\ref{pisa}} \and 
Chiara Circosta\inst{\ref{IRAM}} \and\\
Giovanni Cresci\inst{\ref{INAF}}\and
Francesco D'Eugenio\inst{\ref{RM1},\ref{RM2}} \and
Gareth C. Jones\inst{\ref{oxford}}\and 
Giacomo Venturi\inst{\ref{pisa}} \and 
Rongmon Bordoloi\inst{\ref{Rongmon}} \and \\
Daichi Kashino\inst{\ref{Daichi}}\and 
Ruari Mackenzie\inst{\ref{Raurie}} \and 
Jorryt Matthee\inst{\ref{Jorryt}} \and 
Rohan Naidu\inst{\ref{MIT}} \and 
Robert A. Simcoe\inst{\ref{MIT}}
}

\institute{Los Alamos National Laboratory, Los Alamos, NM 87545, USA\label{LANL} \and
MIT Kavli Institute for Astrophysics and Space Research, 77 Massachusetts Ave., Cambridge, MA 02139, USA\label{MIT}
\and
Kavli Institute for Cosmology, University of Cambridge, Madingley Road, Cambridge, CB3 0HA, UK\label{RM1}\and
Cavendish Laboratory - Astrophysics Group, University of Cambridge, 19 JJ Thomson Avenue, Cambridge, CB3 0HE, UK\label{RM2}\and
Centro de Astrobiolog\'{\i}a (CAB), CSIC-INTA, Ctra. de Ajalvir km 4, Torrej\'on de Ardoz, E-28850, Madrid, Spain \label{SA}\and
National Research Council of Canada, Herzberg Astronomy \& Astrophysics Research Centre, 5071 West Saanich Road, Victoria, BC V9E 2E7, Canada\label{CW} \and
Department of Physics and Astronomy, University College London, Gower Street, London WC1E 6BT, UK\label{RM3}\and
Max-Planck-Institut für extraterrestrische Physik, Gießenbachstraße 1, 85748 Garching, Germany\label{HU}\and
University of Oxford, Department of Physics, Denys Wilkinson Building, Keble Road, Oxford OX13RH, United Kingdom\label{oxford}\and
Sorbonne Universit\'e, CNRS, UMR 7095, Institut d’Astrophysique de Paris, 98 bis bd Arago, 75014 Paris, France\label{SC}\and
European Space Agency, c/o STScI, 3700 San Martin Drive, Baltimore, MD 21218, USA\label{TB}\and
Scuola Normale Superiore, Piazza dei Cavalieri 7, I-56126 Pisa, Italy\label{pisa}\and
Institut de Radioastronomie Millimétrique (IRAM), 300 rue de la Piscine, 38400 Saint-Martin-d’Hères, France\label{IRAM}\and
INAF - Osservatorio Astrofisico di Arcetri, largo E. Fermi 5, 50127 Firenze, Italy\label{INAF}\and
Department of Physics, North Carolina State University, Raleigh, NC 27695-8202, USA\label{Rongmon}\and
National Astronomical Observatory of Japan, 2-21-1 Osawa, Mitaka, Tokyo 181-8588, Japan\label{Daichi}\and
Laboratoire d’astrophysique, Ecole Polytechnique Fédérale de Lausanne (EPFL), Observatoire, CH-1290 Versoix, Switzerland\label{Raurie}\and
Institute of Science and Technology Austria (ISTA), Am Campus 1, 3400 Klosterneuburg, Austria\label{Jorryt}
}

\abstract
{
{The James Webb Space Telescope is revolutionising our ability to understand the host galaxies and local environments of high-$z$ quasars. Here we obtain a comprehensive understanding of the host galaxy of the $z=7.08$ quasar \qso\ by combining NIRSpec integral field spectroscopy with NIRCam photometry of the host continuum emission.
Our emission-line maps reveal that this quasar host is undergoing a merger with a bright companion galaxy. The quasar host and the companion have similar dynamical masses of $\sim10^{10}M_\odot$, suggesting that this is a major galaxy interaction.
Through detailed quasar subtraction and SED fitting using the NIRCam data, we obtained an estimate of the host stellar mass of $M_{\ast}=(3.0^{+2.5}_{-1.4})\times10^9M_\odot$, with $M_{*}=(2.7^{+0.5}_{-0.5})\times10^9M_\odot$ for the companion galaxy.
Using the \hb\ Balmer line, we estimated a virial black hole mass of  $M_{\rm{BH}}=(1.9^{+2.9}_{-1.1})\times10^9 M_\odot$.
Thus, \qso\ has an extreme black hole--stellar mass ratio of $M_{\rm{BH}}/M_\ast=0.63^{+0.54}_{-0.31}$, which is $\sim3$ dex larger than expected by the local scaling relations between black hole and stellar mass.
\qso\ is powered by an overmassive black hole with the highest reported black hole--stellar mass ratio in a quasar host that is currently undergoing a major merger. These new insights highlight the power of JWST for measuring and understanding these extreme first quasars.
}}


\maketitle

\section{Introduction} \label{sec:intro}
High-redshift quasars at $z\geq6$ powered by highly accreting supermassive black holes (BHs) with masses up to $\sim10^9M_\odot$ \citep[e.g.][]{fan_2000,fan_2001,Fan2003,willott_2009,Willott2010,Kashikawa2015,Banados2016,Banados2017,Banados2022,Matsuoka2018,Wang2019,Yang2023} are some of the most extreme objects in the Universe, and they raise serious questions about the nature of the first BH seeds and BH growth mechanisms. 
However, much is yet to be fully understood about these objects due to limitations in the ability to study their host galaxies, BH properties, and environments.

Prior to the launch of the James Webb Space Telescope \citep[JWST;][]{Gardner2006,Gardner2023,McElwain2023}, the host galaxies of these high-$z$ quasars could only be seen in the rest-frame far-infrared (FIR), detected in the sub-millimetre with the Plateau de Bure Interferometer \citep[PdBI; e.g.][]{Bertoldi2003,Walter2003}, later upgraded to the Northern Extended Millimetre Array \citep[NOEMA; e.g.][]{Banados2015,Mazzucchelli2017}, and the Atacama Large Millimeter Array \citep[ALMA; e.g.][]{Wang2013,willott_2013b}.  This emission traces the cold gas and dust, allowing the host 
dynamical masses ($10^{10}$--$10^{11}M_\odot$),
dust masses ($10^7$--$10^9M_\odot$),
radii ($\sim$1--5 kpc), obscured star-formation rates (SFRs; $10$--$2700 M_\odot/$yr), and their various morphologies to be measured
\citep{Walter2009,Wang2013,Venemans2015,Venemans2017a,Willott2017,Izumi2018,Pensabene2020,Neeleman2021, Salvestrini2025,Mazzucchelli2025}. These observations have shown that there is a variety in high-$z$ quasar host galaxy properties.
For a more comprehensive understanding of the full picture of these host galaxies, however, observations of their stellar emission as well as other gas phase (i.e. ionised) components are required.

In the rest-frame ultraviolet (UV)--optical, where the majority of the stellar emission occurs, the bright quasar significantly outshines the host galaxy \citep[e.g.][]{schmidt_1963,mcleod_1994,dunlop_2003,hutchings_2003,floyd_2013}.
At high-$z$, the size of the galaxies becomes small relative to the point spread function (PSF) of current telescopes, so the faint host galaxy emission is hidden underneath the bright quasar point source \citep[e.g.][]{Mechtley2012}. Thus, to detect the host, deep imaging with sufficiently high spatial resolution to decompose the galaxy and quasar emission is required. However, even with the resolution of the Hubble Space Telescope (HST), this quasar subtraction process had been unsuccessful in obtaining any high-$z$ host detections \citep{Mechtley2012,Decarli2012,McGreer2014,Marshall2019c}.

With a spatial resolution of $\sim0\farcs04$ at 1$\mu$m, which is almost three times better than that of HST, JWST has provided the first rest-frame UV/optical high-$z$ quasar host galaxy detections.
\citet{Ding2023} detected the hosts of two low-luminosity ($M_{\rm{UV}}\simeq-23.9$ mag and $-23.7$ mag) quasars at $z\simeq6.4$ by performing detailed PSF modelling and quasar removal from images from the Near-Infrared Camera \citep[NIRCam;][]{Rieke2023}.
Using a similar approach, \citet{Yue2023} detected the host galaxies of three of their sample of six luminous ($M_{\rm{UV}}\simeq-29$ to $-26$ mag) $z\gtrsim6$ quasars. 
\citet{Stone2023} have also claimed a successful host detection of a quasar at $z=6.25$.
Thus, with JWST these host detections are now possible, and measurement of their stellar properties can begin.

Notably, by measuring the stellar continuum of these hosts, their stellar masses can be calculated. One key question is whether early BHs grew more efficiently than their hosts, in which case they should be comparatively more massive than their present day equivalents, and therefore they would fall above the local BH--stellar mass relation \citep[e.g.][]{Magorrian1998,Marconi2003,Haring2004,Kormendy2013,Reines2015,Greene2020}. \citet{Yue2023} measured their three quasars, finding them to have BH--stellar mass ratios of $\sim0.15$, 1--2 dex larger than expected from the local \citet{Reines2015} and \citet{Kormendy2013} BH--stellar mass relations; however, these BHs are overly massive, even when considering selection effects.
The lower luminosity quasars presented in \citet{Ding2023} are consistent with the local BH--stellar mass relations of \citet{Haring2004} and \citet{Bennert2011} when accounting for selection effects.
Some observations of confirmed and candidate $z>3$ active galactic nuclei (AGNs; not in the quasar regime) support the picture of overmassive BHs relative to the local relation \citep{Uebler2023,Harikane2023,Maiolino2023,Pacucci2023}.
However, \citet{Sun2024} found no evidence of the evolution of the BH--stellar mass relation up to $z\simeq4$.
Selection effects, measurement uncertainties, and low sample sizes make a redshift evolution of the BH--stellar mass relation difficult to confirm \citep{Li2024}. 

Another key piece of this puzzle is obtaining accurate measurements of BH masses. Prior to JWST, all high-$z$ quasar BH mass estimates relied on the \ions{Mg}{ii} and \ions{C}{iv} emission lines. However, \ions{Mg}{ii}- and \ions{C}{iv}-based reverberation mapping calibrations are more uncertain than those based on the broad hydrogen lines \citep[e.g.][]{Trevese2014,Kaspi2021} and therefore give less reliable single-epoch BH mass estimates \citep[e.g.][]{shen2008,Shen2012,Peterson2009,Coatman2016,Farina2022}.
The hydrogen lines from the quasar's broad-line region (BLR) have previously been unobservable beyond $z\gtrsim4$. With JWST, \hb\ is now observable up to $z=9.8$ and \ha\ up to $z=7.0$ with the Near-Infrared Spectrograph \citep[NIRSpec;][]{Jakobsen2022,Boeker2023} and to even higher redshifts with the Mid-Infrared Instrument \citep[MIRI;][]{Rieke2015}. This allows us to determine the BH masses of these quasars more accurately \citep[e.g.][]{Eilers2023,Yang2023a,Marshall2023} and to obtain a clearer understanding of the BH growth history and ongoing accretion.

Alongside measuring the quasar host properties, for a full picture of this rapid early BH growth, we must understand their environments. Some theories have suggested that intense BH growth can be caused via galaxy--galaxy mergers and local kiloparsec-scale interactions \citep[e.g.][]{Sanders1988,Hopkins2006}, which may be more prevalent in the early Universe. Mergers are also suggested to be a potential cause of the observed BH--host scaling relations \citep[e.g.][]{Matteo2005,Croton2006b}.
ALMA has uncovered major galaxy interactions around high-$z$ quasars at rates of up to 50\% \citep{Trakhtenbrot2017}, with companions found at $\simeq8$--60 kpc separations \citep[e.g.][]{Wagg2012,Decarli2017,Venemans2020}.  These companions have been less frequently detected in the rest-frame UV \citep{willott_2005}, likely due to their dusty nature \citep{Trakhtenbrot2017}.
With spatially resolved spectroscopic capabilities and coverage in the rest-frame optical where dust is less prevalent, JWST has the power to uncover the local environments of high-$z$ quasars in greater detail than ever reported before.

The NIRSpec Integral Field Unit \citep[IFU;][]{Boeker2022} can reveal detailed kinematics around quasars on the kiloparsec-scale \citep[e.g.][]{Perna2023,Marshall2023}. \citet{Marshall2023} studied two $z=6.8$ quasars and discovered three kiloparsec-scale companion galaxies; one companion is merging with one of the quasars, while two are merging with the other.
The quasar--galaxy merger PJ308–21 at $z\simeq6.2$ discovered with ALMA \citep{Decarli2019} has been observed with the NIRSpec IFU to study the ionised gas and stellar emission \citep{Loiacono2024} and the interplay of gas, dust, star formation, and nuclear activity \citep{Decarli2024}.
\citet{Perna2023} used the IFU to investigate a quasar at $z=3.3$ that lives in an extreme environment with multiple companion galaxies and found that one of these companions is hosting an obscured AGN.
At $z = 7.15$, \citet{Uebler2024} discovered an AGN within a galaxy--galaxy merger, with two BHs expected to merge in the future.
After detailed quasar subtraction, NIRCam images from \citet{Yue2023} also revealed extended emission structures, companion galaxies, and/or potential merger signatures around the majority of their six $z\gtrsim6$ quasars. 
NIRSpec IFU and NIRCam imaging of an AGN at $z=5.55$ showed that the AGN has two faint nearby companions \citep{Uebler2023,Ji2024}.
The $z=8.7$ AGN of \citet{Larson2023} appears asymmetric in their NIRCam images, with
companion galaxies that imply a major merger.
\citet{Matthee2024} found that around the faint ($M_{\rm{UV}}\simeq-20$ to $-17$ mag) AGNs at $z\simeq4$--6 discovered with the NIRCam, the majority show at least one spatially separated companion, while similar AGNs in \citet{Maiolino2023} show evidence of two broad components of H$\alpha$, which they suggest could be signatures of merging BHs. 
\citet{Perna2023b} found that companions are routinely observed around AGNs at $3<z<6$ within a radius of $\sim 10$~kpc and that a significant fraction of them could host accreting BHs.
Thus, while companion galaxies are not observed around all high-$z$ quasars, these observations support the picture of major mergers being potential drivers of rapid early BH growth. 

At $z=7.08$, ULAS J112001.48+064124.3 (hereafter \qso) was the first quasar discovered at $z>7$ \citep{Mortlock2011}. This was the highest-$z$ quasar known until the 2018 discovery of ULAS J134208.10+092838.61, at $z=7.54$ \citep{Banados2017}, and it remains the fourth highest-$z$ quasar known to date \citep[after][]{Yang2020,Wang2021b}, thus making it one of the most targeted high-$z$ quasars with JWST so far.
As part of the Emission-line galaxies and Intergalactic Gas in the Epoch of Reionization (EIGER) project, \qso\ was observed with the NIRCam with deep F115W, F200W, and F356W imaging and slitless F356W grism spectroscopy \citep{Yue2023}.
\qso\ was also imaged with the NIRCam F210M, F360M, and F480M filters \citep{Stone2024}.
\citet{Bosman2024} observed \qso\ with the MIRI Medium-Resolution Spectrometer (MRS), obtaining IFU spectra from 4.9--27.9$\mu$m.
These observations have led to two key results.
Firstly, \citet{Yue2023} detected the stellar emission from the host of \qso, providing the first stellar mass measurement of $M_\ast=6.5\substack{+4.5 \\ -3.3}\times10^{9}M_\odot$.
Secondly, with measurements of the broad hydrogen lines, we can be more confident in the BH mass estimates.
\citet{Yue2023} measured $M_{\rm{BH}}=(1.19\pm0.08)\times10^9 M_\odot$ from \hb, and \citet{Bosman2024} measured  $M_{\rm{BH}}=(1.55\pm 0.22)\times10^9 M_\odot$ from \ha. Accounting for the estimated 0.43 dex scatter in the hydrogen-based BH mass scaling relations \citep{Vestergaard2006}, these estimates become $M_{\rm{BH}}=\left(1.2\substack{+2.2\\-0.8}\right)\times10^9 M_\odot$ from \hb\ and $M_{\rm{BH}}=\left(1.6\substack{+3.2\\-1.1}\right)\times10^9 M_\odot$ from \ha.
With a BH mass of $\sim10$--$25\%$ of the stellar mass estimate, these measurements imply that \qso\ lies significantly above the local BH--host mass relation, which predicts ratios of only $\sim0.1\%$ \citep{Greene2020}.

Within the Galaxy Assembly with NIRSpec Integral Field Spectroscopy (GA-NIFS) programme, \qso\ was observed with the NIRSpec IFU, obtaining high-resolution (R$\sim$2700) spectra from 2.9--5.3$\mu$m across the $3''\times3$'' field of view (FOV).
In this paper, we present the GA-NIFS IFU data for \qso, showing the detailed emission-line structures around the quasar. By combining this spatially resolved emission-line data with the detailed photometry from the EIGER NIRCam imaging, we are able to gain a much clearer picture of this system, discovering that it is undergoing a major merger with a companion galaxy and estimating the quasar host stellar mass more accurately.
This paper is organised as follows.
We discuss the observations, our data reduction, and analysis techniques in Section \ref{sec:data}.
In Section \ref{sec:IFUresults} we show the results from the NIRSpec IFU data, including the emission-line structure and kinematics, dynamical mass estimates, and BH mass measurements.
In Section \ref{sec:NIRCamResults} we show updated results from the NIRCam data, including the resulting spectral energy distribution (SED) fitting when combining both the NIRSpec IFU and NIRCam imaging. We include a discussion in Section \ref{sec:Discussion} and conclude in Section \ref{sec:Conclusions}.

Throughout this work we adopt the WMAP9 cosmology \citep{Hinshaw2013} as included in \textsc{AstroPy} \citep{Astropy2013}, with $H_0=69.32$ km / (Mpc s), $\Omega_m=0.2865$, and $\Omega_\Lambda=0.7134$. All magnitudes are on the AB system.

\section{Observations and data analysis}
\label{sec:data}
\subsection{NIRSpec IFU data reduction and analysis}
For our NIRSpec IFU data, we follow the same reduction and analysis procedure as in \citet{Marshall2023}, with several updates. Here we give an overview, but refer the reader to \citet{Marshall2023} for full details. To summarise our data analysis, once we have reduced the data, we first perform background and continuum subtraction. We then remove the point-source quasar line emission from the data cube via detailed PSF modelling. Using this quasar-subtracted cube we make \hb\ and \oiii\ emission-line maps to study the extended host kinematics.
We separately integrate the original continuum-subtracted cube to obtain a quasar spectrum, which we fit with a detailed model in order to measure the BH mass.

\subsubsection{NIRSpec IFU observations}
The quasar \qso\ was observed as part of the GA-NIFS  Guaranteed Time Observations (GTO) programme\footnote{\url{https://ga-nifs.github.io/}}.
In order to save overhead time, the NIRSpec observations were combined with those using MIRI in programme \#1263 (PI L. Colina).
It was observed with the NIRSpec IFU, which provides spectroscopy over a 3\arcsec$\times$3\arcsec\ FOV in each of the $0\farcs1\times0\farcs1$ spatial elements, with a grating/filter pair of G395H/F290LP, giving a spectral resolution of $R\sim2700$ over the wavelength range 2.87--5.27$\mu$m.
The observations were taken with a NRSIRS2 readout pattern \citep{Rauscher2017} with 25 groups per integration and one integration per exposure, using a 6-point medium cycling dither pattern, with a total on-source exposure time of 11029 seconds.

\qso\ was observed on December 11, 2022, at a position angle (PA) of 70.97 degrees. 
We constrained the allowable PA window to minimise the leakage of light through the micro-shutter assembly (MSA) from bright sources. 
We also offset the quasar within the detector by $-0\farcs7$ from the centre to `mind the gap'. The physical gap between the two NIRSpec detectors results in a range of unobservable wavelengths from $\sim4.0$--4.2$\mu$m which varies with IFU slice \citep{Boeker2022}. At this redshift, the quasar's \oiii\ fall around this detector gap, and so offsetting the quasar within the detector allowed these lines to be observed.

\subsubsection{IFU data reduction}

The IFU data was reduced with the JWST calibration pipeline version 1.8.2 \citep{Bushouse2022}, using the context file {\sc jwst\_1068.pmap}. We applied additional corrections to improve the data quality: correcting the count-rate frames such that the background base-level is consistent with zero counts per second; subtracting the $1/f$ noise \citep{Moseley2010}, by modelling and then subtracting it from each column (i.e. along the spatial axis); and implementing a modified version of the {\sc lacosmic} algorithm \citep{Dokkum2001} in the JWST pipeline for outlier detection \citep[details in][]{D'Eugenio2024}; full details of the data reduction process can be found in \citet{Perna2023}.

\begin{figure*}
\begin{center}
\includegraphics[scale=0.8]{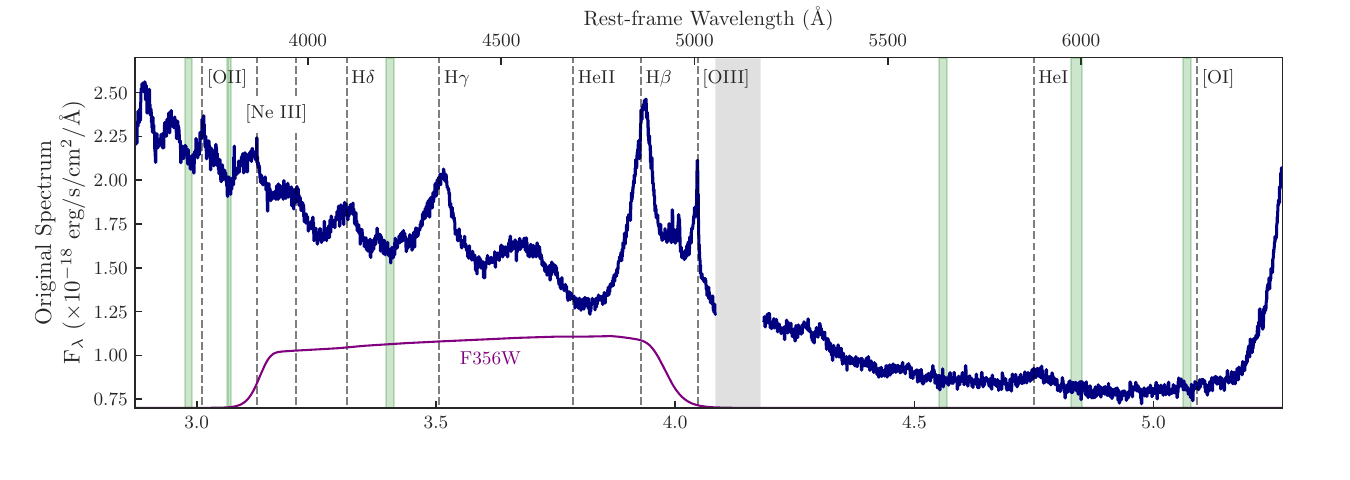}
\vspace{-0.7cm}

\includegraphics[scale=0.8]{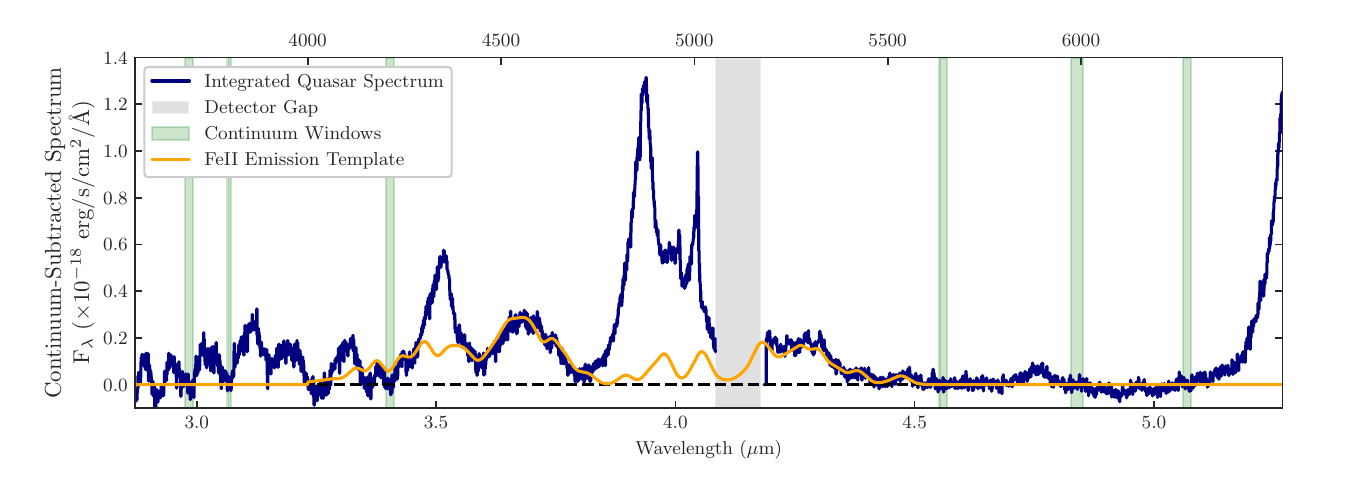}
\vspace{-0.3cm}
\caption{Integrated spectrum of \qso\ from the IFU data (blue). 
The spectrum is integrated over an aperture of radius $0\farcs35$ centred on the peak of the quasar emission. 
An aperture flux correction of 1.23$\times$ has been applied. 
The top panel shows the spectrum integrated from the original background-subtracted data cube.
The bottom panel shows the spectrum integrated from the continuum-subtracted data cube.
The green shaded regions show the continuum windows used to model and subtract the continuum emission (Section \ref{sec:Continuum}).
At this redshift, $z=7.08$, the \oiii\ doublet falls just blueward of the detector gap (grey shaded region), and the broad \ha\ line falls just off the red edge of detector.
The purple curve shows the wavelength coverage of the NIRCam F356W filter used in the EIGER images. 
The yellow curve shows the \citet{Park2022} iron emission template, with redshift and normalisation taken from our best spectral model fit (Section \ref{sec:BHfitting}).
}
\label{fig:FullSpectrum}
\end{center}
\end{figure*}

In this work we used the {\sc cube\_build} step in the pipeline to produce combined data from the various dithers.
We created 0.05\arcsec/pixel cubes using the `drizzle' weighting---this optimises the spatial resolution of the data. However, these cubes are affected by oscillations in the extracted spectra due to re-sampling effects \citep[see][]{Law2023,Perna2023}. 
These oscillations (or `wiggles') impact the accurate determination of the continuum and emission-line profiles at the spaxel level, which is crucial for both PSF modelling and subtraction, and consequently, for studying the quasar host galaxy. To address this, we applied the same method detailed in \citet{Perna2023}, and subtracted the wiggles in each individual spaxel in the close surroundings of the quasar.
The resulting integrated (background-subtracted) spectrum for \qso\ is shown in Figure \ref{fig:FullSpectrum}, showing the full wavelength coverage from 2.87--5.27$\mu$m, corresponding to rest-frame 3550--6520\AA\ at $z=7.085$.

The noise estimation from the ‘ERR’ cube produced by the pipeline underestimates the actual noise in the data. We therefore rescaled the ERR cube
to match the noise estimated from the continuum emission \citep[as in e.g.][]{Uebler2023,DelPino2024,Jones2024,Lamperti2024}.
For each spaxel we measured the root mean square (RMS) noise across 3.78--3.83$\mu$m, just blueward of \hb, and compare this to the mean ERR value across the same region. In spaxels where the ratio of RMS to the mean ERR is greater than 1, we multiplied the ERR spectrum by this fraction. This results in a final error spectrum that is a median of $50\%$ larger than the original ERR array.
However, we note that because of the uncertainty introduced via the wiggle subtraction, continuum subtraction, and various interpolations/re-samplings (i.e. smoothing the data), this will also be an under-estimate of the true noise.

\begin{figure}
\begin{center}
\vspace{-0.4cm}
\includegraphics[scale=0.8]{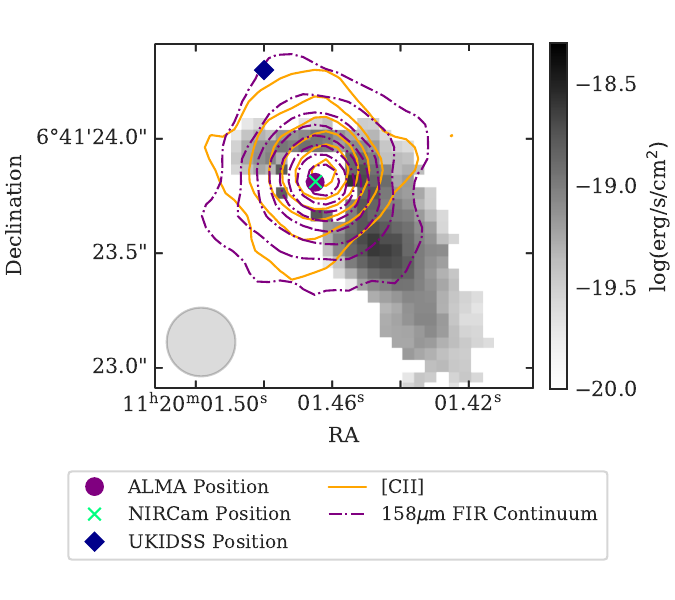}
\vspace{-0.7cm}
\caption{Quasar-subtracted \oiiia\ emission map from the NIRSpec IFU (grey image) compared to the \cii\ emission map (orange contours) and the rest-frame 158\micron\ FIR continuum emission map (purple contours) from ALMA (see Section \ref{sec:ALMA}). 
The \cii\ contours are linearly spaced from 5$\sigma$ to 26$\sigma$, where $\sigma=16\mu$Jy.
The FIR continuum contours are linearly spaced from 5$\sigma$ to 42$\sigma$, where $\sigma=7\mu$Jy/beam. 
The grey circle depicts the approximate PSF of the ALMA observations, with beam diameter $0\farcs3$. The green cross marks the peak of the quasar emission from the NIRSpec IFU data that has been astrometrically aligned to the NIRCam imaging, which is aligned to Gaia DR2 \citep{Yue2023}. The purple circle marks the peak of the 158\micron\ FIR continuum emission reported by \citet{Venemans2017a}. The blue diamond marks the original quasar position from UKIDSS quoted by \citet{Mortlock2011}.
}
\label{fig:ALMAoverlay}
\end{center}
\end{figure}

Due to an uncertainty in the specified astrometry of our IFU observations, we astrometrically aligned our IFU data to the EIGER data. We integrated our cube across the wavelength range covered by the F356W filter, and matched the quasar location in the resulting image to that in the EIGER F356W image, which has been aligned to the Gaia DR2 catalogue \citep{Gaia2018} with the quasar located at R.A. 11:20:1.464, Dec +6.41.23.783.
In Figure \ref{fig:ALMAoverlay} we show this quasar peak from the NIRCam data compared to the peak of the 158\micron\ FIR continuum emission from ALMA and the quasar position from its discovery with the UK Infrared Telescope (UKIRT) Infrared Deep Sky Survey (UKIDSS). 
In the ALMA study of \citet{Venemans2017a}, the FIR continuum emission was found to peak at R.A. 11:20:01.465; Dec +06:41:23.810, offset $\simeq0\farcs5$ to the south-west of the quoted quasar position from UKIDSS \citep{Mortlock2011}, R.A. 11:20:01.480; Dec +06:41:24.300.
From the NIRCam imaging, the quasar location is consistent with the ALMA location, also offset from the UKIDSS location.
UKIDSS is astrometrically aligned to 2MASS, with an astrometric accuracy of 0\farcs1 \citep{Lawrence2007}.
The UKIDSS images have a pixel scale of 0\farcs4 \citep{Lawrence2007}, and so the coarse spatial resolution may also contribute to the large astrometric offset.
ALMA is astrometrically aligned with the International Celestial Reference Frame to an accuracy of $<0\farcs001$; the astrometry of individual sources depends on frequency, baseline, and signal-to-noise ratio $\rm{S/N}$, which for this target results in an expected position uncertainty of $0\farcs02$ (ALMA Cycle 7 Technical Handbook\footnote{\url{https://almascience.eso.org/documents-and-tools/cycle7/alma-technical-handbook}}).
The NIRCam images are aligned to Gaia DR2, with accuracy of 0\farcs0015 \citep{Anderson2021}.
We conclude that the higher spatial resolution and more accurately aligned NIRCam and ALMA images give a more accurate position of the quasar, with the original UKIDSS quasar position offset 0\farcs5 to the north-east from its true location.

\subsubsection{Background and continuum subtraction}
\label{sec:Continuum}
For our analysis, we first subtracted the background and continuum emission from our quasar cube in each spaxel. 
We selected a background region with circular aperture of radius 0\farcs25 near a corner of the IFU FOV chosen to avoid the PSF of the quasar and the extended emission as well as the higher-noise pixels around the edge of the FOV. We measured the median spectrum across spaxels within this aperture and fit this spectrum with a polynomial function of fourth order. We then subtracted this polynomial curve from each spaxel to subtract the background.

For the continuum subtraction, we considered the flux in the emission-line- and iron-continuum-free windows at rest-frame 3790--3800\AA, 4200--4220\AA,
5630--5650\AA, and 5970--6000\AA, following \citet{Kuraszkiewicz2002} and \citet{Kovacevic2010}, as well as additional windows near the blue and red edges of the spectral range at 3680--3700\AA\ and 6260--6280\AA\ to better constrain the continuum across the full wavelength range. 
For each spaxel we measured the $\rm{S/N}$ across these continuum windows and modelled and subtracted the continuum if the median $\rm{S/N}\geq1.5$. We find that spaxels with median $\rm{S/N}<1.5$ across their continuum windows have negligible contribution to the continuum in the final integrated spectra.
For spaxels with median $\rm{S/N}\geq1.5$ in the continuum windows, we measured the mean flux and wavelength in each continuum window and interpolated between these means with \textsc{scipy}'s \textsc{interp1d} function using a spline interpolation of second order. We evaluated the interpolation function across the full wavelength range, producing a full continuum model in each spaxel that was then subtracted from the spectrum. Each spaxel was modelled independently, and this created a continuum-subtracted data cube. 
Figure \ref{fig:FullSpectrum} shows the quasar spectrum as integrated from the original cube as well as from the continuum-subtracted cube, for comparison. Throughout the remainder of this work, we only use the continuum-subtracted cube.

\subsubsection{IFU quasar subtraction}
\label{sec:IFUsubtraction}
To study the extended narrow-line emission from the host galaxy system, we must first subtract the quasar's large contribution to the emission. The narrow-line emission can be emitted both by the quasar, photoionised by the AGN and shocks, and gas throughout the system that has been photoionised by star formation. We remove the bright unresolved narrow-line emission from the nucleus to study the spatial distribution and kinematics of the extended host system.
The quasar BLR is spatially unresolved, and so its emission traces the PSF of the instrument; we use this fact to perform our quasar subtraction.

\textit{Step 1, Measuring the spatial PSF:}  We used \textrm{QDeblend3D} \citep{Husemann2013,Husemann2014}, which uses the relative strength of quasar broad lines in each spaxel to determine the spatial PSF shape.  After continuum subtraction (Section \ref{sec:Continuum}), in each spaxel we measured the mean of the broad-line flux between rest-frame 4800--4830\AA\ and 4880--4910\AA, on both sides of the peak of \hb.  These spectral windows cover the broad-line wings, and are free from any narrow lines that would bias the measurement. These broad-line wing fluxes were normalised to that of the central brightest quasar spaxel, giving the 2D PSF. 

For our final PSF shape, we selected spaxels in which the $\rm{S/N}>10$ for at least three consecutive sampled wavelengths within the broad-line region, and then expanded this region spatially by selecting any adjacent spaxels with $\rm{S/N}>2$ at two consecutive wavelengths, following the \textsc{find\_signal\_in\_cube} algorithm of \citet{SunGithub} \citep[see also][]{Sun2018}.
To reduce the effect of any artefacts and companion galaxies on the measured PSF shape, we then excluded any spaxels that are located more than 10$\sigma$ from the quasar peak, where $\sigma=\rm{FWHM}/(2\sqrt{2\ln{2}})$ and the full width at half maximum (FWHM) is the diffraction limit of the telescope at the observed wavelength of \hb, $0\farcs152$. 
Our measured spatial PSF is shown in Figure \ref{fig:PSF}.

\textit{Step 2, Creating a 3D quasar model cube:} 
For our 1D quasar spectral model, we used the spectrum of the brightest spaxel, the centre of the quasar emission, under the assumption that the host line emission in that spaxel is negligible relative to the dominant quasar emission.
This spectrum includes both the narrow and broad-line quasar emission, and so both the unresolved quasar narrow and broad emission is subtracted.
We created a 3D quasar cube by scaling the quasar spectrum by the 2D PSF, and subtracted this from our original continuum-subtracted cube to create our host galaxy cube.
We are able to ignore the variation of the PSF with wavelength, because we only analyse the \hb\ and nearby \oiii\ lines.

To account for the additional uncertainty associated with the quasar subtraction, we created a new error cube \citep[see][for more details]{Marshall2023}. We generated 200 realisations of the continuum-subtracted cube with normally distributed noise, taken from the original noise cube. We then performed the quasar subtraction on each of the realisations.
In each spaxel, we measured the standard deviation across the 200 realisations at each wavelength both before and after quasar subtraction.
We then measured the mean standard deviation across the rest-frame 50~\AA\ surrounding \hb\ before and after quasar subtraction, $\sigma_{\rm{original}}$ and $\sigma_{\rm{quasar\ subtracted}}$.
We multiplied the original noise cube by $\sigma_{\rm{quasar\ subtracted}}/\sigma_{\rm{original}}$, in spaxels where $\sigma_{\rm{quasar\ subtracted}}>\sigma_{\rm{original}}$; this results in a maximum increase of a factor of 2$\times$.

\subsubsection{Emission-line fitting for galaxy line maps}
\label{sec:MapFitting}
To create maps of the galaxy's spatially resolved \hb\ and \oiii\ emission from the quasar-subtracted cube, we fit each spectrum as a series of Gaussians, with \textsc{AstroPy's} Levenberg-Marquardt algorithm.
We fit the three narrow lines each as a single Gaussian,
with their mean velocity and line widths constrained to be equal in velocity space, as we assume they arise from the same physical region with the same kinematics. 
We tied the \oiii\ amplitudes to have the standard ratio of 1:2.98 \citep{Storey2000}. The resulting line flux is taken to be the integrated Gaussian model.
Our velocity maps show the median velocity $v_{50}$, a non-parametric measure for the line centre, and the velocity dispersion $w_{80}$, the width containing 80\% of the line flux. 
The velocity dispersion maps are derived after removing the instrumental width in quadrature at each spaxel, where for NIRSpec in the G395H/F290LP configuration, the instrumental FWHM at \hb\ is approximately $115$ km/s at this redshift \citep{Jakobsen2022}.

Our \oiii\ line maps include only spaxels with $\rm{S/N}>10$, and any subsequently adjacent spaxels with $\rm{S/N}>1.5$, following the \textsc{find\_signal\_in\_cube} algorithm of \citet{SunGithub}. Given the much lower $\rm{S/N}$ for \hb, we included spaxels with $\rm{S/N}>2$, and any subsequently adjacent spaxels with $\rm{S/N}>1$; this \hb\ flux map is only shown for visual reference and not used to infer any properties about the quasar system.
For these measured line maps, we also need to exclude heavily corrupted spaxels caused by the quasar subtraction, with spaxels in which the quasar flux significantly dominated over the host galaxy having significant noise and artefacts. 
We therefore excluded the central $5\times5$ spaxels surrounding the quasar peak.
We find an additional region of corrupted spaxels to the south-east of the quasar, within $\simeq7$ spaxels of the peak. We find that in this region fits to the spurious spectra result in \oiiia\ velocity offset greater than $300$ km/s from the quasar redshift, which are not seen throughout the rest of the quasar-subtracted cube. Thus, we also excluded spaxels with an \oiiia\ velocity offset greater than $300$ km/s to excluded this region of significant contamination. The resulting quasar-subtracted emission-line maps are presented in Figure \ref{fig:RegionMaps}.
       
\subsubsection{Integrated emission-line fitting for quasar properties}
\label{sec:BHfitting}
To create an integrated quasar spectrum and fit a model, we used the Markov chain Monte Carlo (MCMC)-based technique of \textsc{QubeSpec}\footnote{\url{https://github.com/honzascholtz/Qubespec}}.
First, we chose to integrate the continuum-subtracted data cube spatially across an aperture with radius $0\farcs35$, centred on the peak of the quasar emission. This was chosen to maximise our $\rm{S/N}$ while containing the majority of the quasar flux, by containing the main core of the PSF, yet not including the ring of reduced flux that occurs prior to the second radial PSF peak (see Figure \ref{fig:PSF}). In \citet{Marshall2023} we found that for two quasars, the average fraction of the quasar flux contained within a $0\farcs35$ radius aperture was 81.4\% at similar wavelengths; we corrected our aperture flux by this same correction factor of 1.23$\times$ to maintain consistency. 

With QubeSpec, we then fit this integrated spectrum around the \hb\ line. 
The \oiii\ lines are fit with two Gaussian components: one narrow `galaxy' component, and one broader component from ionised galactic outflows. The \hb\ line was also allowed to have both a narrow and outflow component; in our model the narrow and outflow components were tied to have the same redshift and line width across all three \oiii\ and \hb\ emission lines, as we assumed they arise from the same physical region with the same kinematics. 
While outflows are typically blue-shifted, we did not constrain this to be the case in our fit and allowed the narrow and outflow component fits to be completely independent. We constrained the amplitude of \oiiib\ to be 2.98 times less than that of \oiiia\ \citep{Storey2000} for both components.

For the \hb\ BLR line component, we considered two separate models.
For our first model, we followed the method of \citet{Marshall2023} by fitting the broad \hb\ line as a broken power law (BPL) model:
\begin{equation}
f(x)=
\begin{cases}
(x/x_{\rm{break}})^{-\alpha_1}: x<x_{\rm{break}}\\(x/x_{\rm{break}})^{-\alpha_2}: x>x_{\rm{break}},
\end{cases}
\end{equation}
convolved with a Gaussian with standard deviation $\sigma$ \citep[as in e.g.][]{Nagao2006,Cresci2015}. 
\citet{Marshall2023} found that this model provided a better fit to the BLR signal of \VDES\ and \DELS\ than a single Gaussian, and so we used this BPL model for consistency.
However, \citet{Yue2023} chose to fit their NIRCam grism \hb\ spectroscopy of \qso\ with two Gaussian components, as did \citet{Bosman2024} for their MIRI spectrum. Thus, to provide a clearer comparison to these studies, we also included a model where the \hb\ BLR is fit as two independent Gaussian components. 

Finally, we also included a model for the iron emission blends by using templates from \citet{BG92}, \citet{VeronCetty2004}, and \citet{Park2022}. 
To choose a best-fitting iron model, we used QubeSpec to fit the integrated spectrum in the iron emission regions of rest-frame 4450--4650\AA\ and 5350--5600\AA\ with each of the different templates. We find that the \citet{Park2022} model provides the best fit to the spectrum in the main iron-continuum windows surrounding \hb, at rest-frame $\sim4400$--4700\AA\ and $\sim5200$--5600\AA, with the other two models having a template shape that provides a much poorer match to the observed spectrum \citep[see][for a comparison of the templates]{Park2022}. Therefore, we chose to use this \citet{Park2022} iron template in our final model. We used the values of the peak flux and FWHM parameters from this \citet{Park2022} iron-only MCMC fit as constraints in the full fit to reduce the range of the free parameters. The iron emission was constrained to have the same velocity offset as the \hb\ BLR in the full fit. 

We ran our QubeSpec MCMC for 20000 iterations.
We have one model with a double Gaussian (DB) \hb\ BLR and one model with a BPL \hb\ BLR. We include results from these two similarly performing models in this work to show how the choice of model impacts the measured BH properties.

In this fitting of the integrated quasar spectrum, we measure the quasar \oiiia\ line to have a redshift of $z=7.0790\pm0.0004$, and the \hb\ BLR component to have a redshift of $z=7.0949\pm0.0002$.
We find that the \oiiia\ line from the integrated host galaxy spectrum, with the quasar removed, peaks at $z=7.0804\pm0.0028$.
In the discovery spectrum, \citet{Mortlock2011} measured a redshift of $z=7.085\pm0.003$ from \ions{Si}{iii}], \ions{C}{iii}] and \ions{Mg}{ii}.
From the ALMA \cii\ measurements of \citet{Venemans2017a}, the redshift is measured to be $z=7.0851\pm0.0005$.
\citet{Bosman2024} measure a similar variety of redshifts depending on the specific emission line, 
with $z=7.092\pm0.002$ from \ha, $z=7.097\substack{+0.004\\-0.002}$ from Pa-$\alpha$, $z=7.098\pm0.003$ from Pa-$\beta$, $z=7.0735\substack{+0.0025\\-0.0015}$ from Pa-$\delta$, and $z=7.111\pm0.003$ from \ions{O}{i}, excluding their estimates from blended lines. Because of this variation in redshift estimates, throughout this work we use the host galaxy \oiiia\ redshift of $z=7.0804\pm0.0028$ when we quote velocities and create our velocity maps.

\subsection{NIRCam imaging}
The NIRCam imaging of \qso\ used in this work was taken as part of the Emission-line galaxies and Intergalactic Gas in the Epoch of Reionization (EIGER) project (Proposal ID: 1243, PI: Lilly). We use the reduced NIRCam images of the quasar provided by \citet{Yue2023}. The corresponding observations and data reduction have been described in \citet{Kashino2022}, \citet{Matthee2023}, \citet{Eilers2023} and \citet{Yue2023}, which we briefly summarise here.

The EIGER observations of \qso\ contain four individual visits,
which deliver NIRCam imaging in the F115W, F200W, and F356W bands, forming a $3'\times6'$ mosaic around the quasar.
The quasar is covered by all of the four visits.
Each visit consists of three dither positions following the INTRAMODULEX primary dither pattern.
The exposure time per visit is 4381s for the F115W imaging, 5959s for the F200W imaging, and 1578s for the F356W imaging.

The imaging data were reduced using the {\texttt{jwst}} pipeline version 1.8.4.
We first ran {\texttt{Detector1Pipeline}} to generate the rate files,
then ran {\texttt{Image2Pipeline}} to obtain calibrated images.
For astrometry, we first aligned the calibrated images to each other using {\texttt{tweakwcs}}, 
then combined all the images and calibrate the absolute astrometry to the {\em Gaia} DR2 catalogue \citep{Gaia2018}. 
We corrected for $1/f$ noise, masked snowballs, subtracted the wisp patterns, and removed cosmic rays from the images using custom codes.
We then ran {\texttt{Image3Pipeline}} to stack images within the same visit and the same module.
We used a pixel size of $0\farcs03$ for the F356W images and $0\farcs015$ for the F115W and the F200W images, which keep super-Nyquist sampling of the PSF \citep[e.g.][]{zhuangshen2023}.
The final product of the image reduction includes twelve images, corresponding to three filters and four visits.

In addition to the quasar images, we also used the NIRCam PSF models provided by \citet{Yue2023} (see their figure 1). The PSF models were constructed using images of bright stars in EIGER observations. \citet{Yue2023} also provided the error maps of the PSF models, which reflect the spatial and temporal variations of the PSFs.

The NIRCam quasar images, PSFs, and PSF errors are used to perform quasar subtraction and reveal the quasar host emission in broad bands via image fitting in Section \ref{sec:NIRCamResults}.
We do this by fitting the NIRCam images with a point source (for the quasar) and two S\'ersic profile models (one for the quasar host and one for the companion galaxy), using the MCMC-based technique of \textsc{psfMC} \citep{Mechtley2016}, a 2D surface brightness modelling software designed for quasar--host decomposition.
The detailed methods and results are described in Section \ref{subsec:nircamsub}.

\citet{Stone2023} reported NIRCam imaging for \qso\ in the F210M, F360M, and F480M bands, as part of the GTO program \#1205. However, \citet{Stone2023} did not detect the host galaxy component of \qso\ in these bands, likely due to the limited depth of their observations. For this reason, we do not include the NIRCam images from \citet{Stone2023} in this work.

\section{Spectral properties from the IFU}
\label{sec:IFUresults}
\subsection{Emission-line structure and kinematics}
\label{sec:Structure}

\begin{figure*}
\begin{center}
\vspace{-0.3cm}
\includegraphics[scale=0.8]{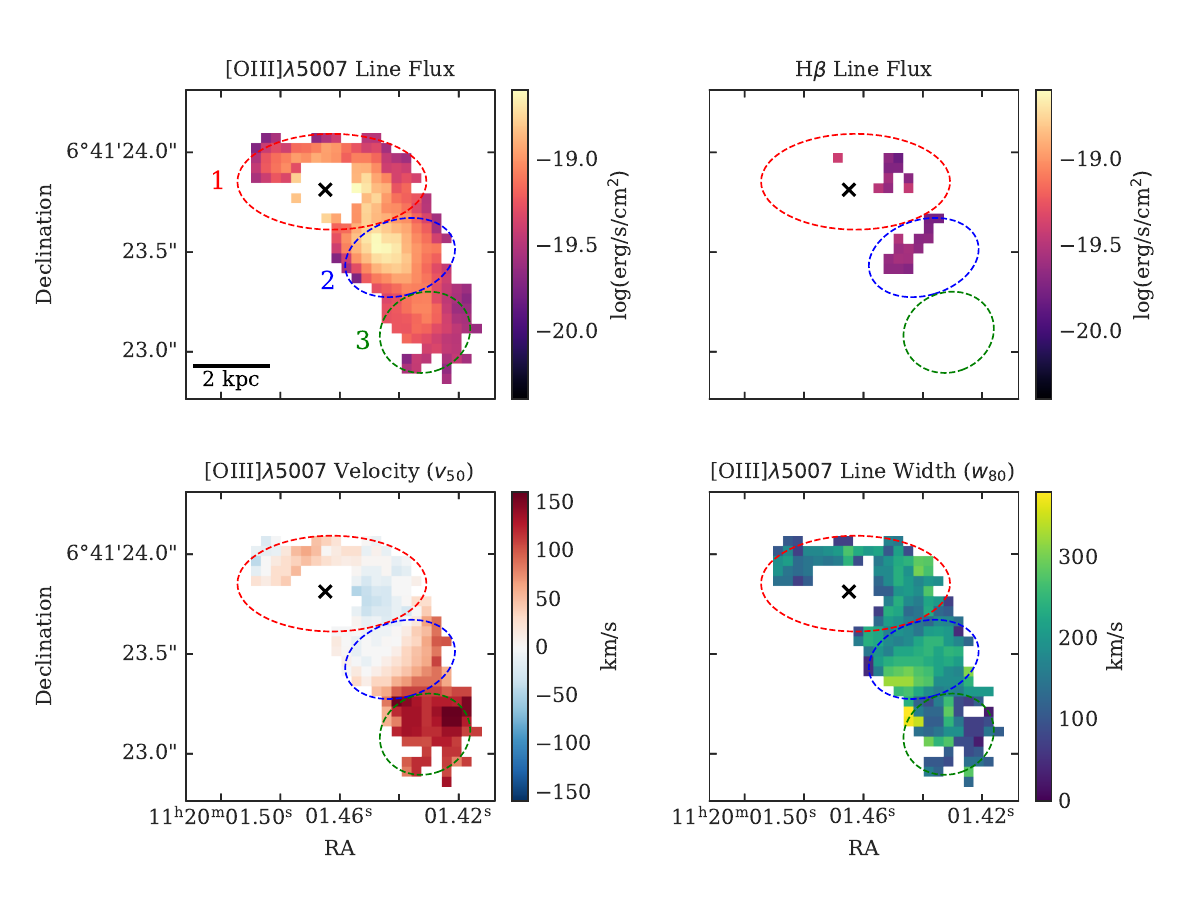}
\vspace{-0.6cm}
\caption{Emission-line regions surrounding \qso\ showing flux and kinematic maps after the subtraction of the quasar emission. The top-left and -right panels show the flux of the \oiiia\ and \hb\ lines, respectively, from the integrated flux of the fitted Gaussian in each spaxel.
The bottom-left and -right panels show our \oiiia\ kinematic maps, depicting the non-parametric central velocity of the line ($v_{50}$) relative to the quasar host redshift of $z=7.0804\pm0.0028$ and the line width ($w_{80}$), respectively. 
Three emission-line regions are highlighted by coloured ellipses, and the crosses show the location of the quasar.}
\label{fig:RegionMaps}
\end{center}
\end{figure*}

\begin{figure}
\begin{center}
\vspace{-0.4cm}
\includegraphics[scale=0.8]{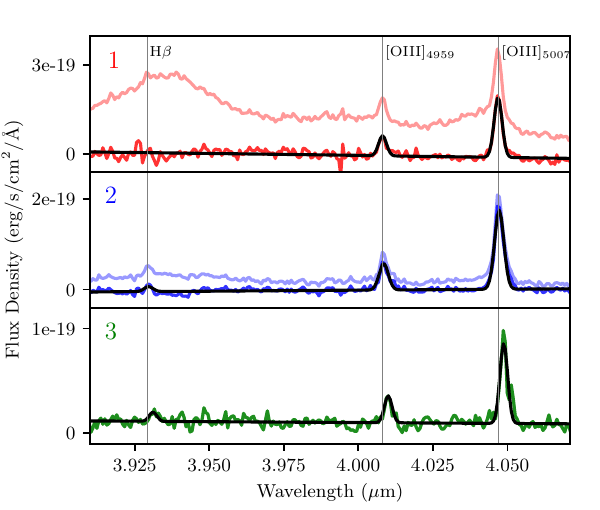}
\caption{Quasar-subtracted spectra integrated over the three spatial regions shown in Figure \ref{fig:RegionMaps} (opaque coloured lines) along with our best-fit Gaussian models for the \hb\ and \oiii\ emission lines (black).
The analogous spectra from the non-quasar-subtracted cube are also plotted for comparison (transparent coloured lines), showing the necessity of quasar subtraction to accurately measure the emission from Regions 1 and 2. All spectra have been continuum subtracted.
The vertical lines mark the location of the \hb\ and \oiii\ lines at the redshift of the quasar host galaxy, Region 1, as measured from the fit to this spectrum, $z=7.0804\pm0.0028$.
For the host Region 1, we exclude the central $5\times5$ pixels surrounding the quasar peak as well as nearby spaxels with \oiiia\ velocity offset $>300$ km/s, as these are highly corrupted by the quasar subtraction and introduce significant noise and artefacts. This means that we slightly underestimate the total flux in this region; however, the fluxes are significantly more reliable than if these most corrupted spaxels were included.}
\label{fig:RegionSpectra}
\end{center}
\end{figure}

In Figure \ref{fig:RegionMaps} we present the quasar-subtracted \oiiia\ flux map of the \qso\ field. This shows two distinct peaks in the flux distribution, corresponding to the host galaxy Region 1, and a second emission-line structure Region 2 to the south-west of the quasar. 
There is also a much smaller and fainter region of flux further to the south-west, Region 3. 
This appears to have an asymmetric morphology, with a velocity that extends to higher velocities beyond 
the southern edge of Region 2.
We define each of these three regions by elliptical apertures as shown in Figure \ref{fig:RegionMaps}.
These apertures, with radius 0\farcs2--0\farcs5, are selected based on visual inspection of the \oiiia\ flux map to approximately encompass the $\rm{S/N}>1.5$ \oiiia\ flux. The edge between the Region 1 and 2 apertures is chosen to lie along the line of decreased flux between the two peaks, while the edge between Regions 2 and 3 is chosen to be where the velocity gradient in Region 2 transitions to an area of roughly constant velocity.

Based on the \oiiia\ map, Regions 1 and 2 are most likely two distinct galaxies. The flux in Region 1 is centrally concentrated around the quasar location (albeit corrupted by the quasar in the core). The flux in Region 2 peaks $\sim$0\farcs42 to the south-west of the quasar, with a generally smooth flux profile decreasing radially from the peak.
There is a noticeable decrease in \oiiia\ flux between these two peaks---Regions 1 and 2 are two physically distinct emission-line regions, albeit with very small spatial and velocity offset.
Region 2 shows a smooth \oiiia\ velocity profile, from a negative velocity offset on the north-eastern edge to a positive velocity offset on the south-western edge. This indicates that Region 2 is a rotational structure.
Region 2 is also clearly detected in NIRCam photometry in the F115W, F200W, and F356W filters (see Section \ref{subsec:nircamsub}) and thus must have moderate stellar continuum emission in the rest-frame UV--optical.
This is not consistent with the properties of off-galaxy `\ha\ blobs', clumps of gas with significant emission-line fluxes but weak or undetected optical continuum \citep{Pan2020,Ji2021}.
The presence of stellar continuum, low offset velocities and line widths (Table \ref{tab:HostFlux}), and the smooth flux and velocity structure are generally inconsistent with Region 2 being a clump of gas ejected or ionised by the central quasar.
It also has significant line flux and physical size, comparable to that of the host galaxy, and so Region 2 is most likely to be a separate companion galaxy rather than a sub-clump of the host galaxy. 
We thus conclude that Region 2 is a companion galaxy near the quasar host.
Further to the south-west, Region 3 may be a tail of extended gas emanating from Region 2, or it could be a second companion galaxy. Its irregular morphology may be physical, caused by the galaxy--galaxy interaction, or due to a lack of S/N in our observations for this faint region---deeper observations would be required to determine its physical nature.

We spatially integrated the spectra within the ellipses covering these regions, which are shown in Figure \ref{fig:RegionSpectra}. 
These three regions are clearly at a very similar redshift to the quasar, as the doublet \oiii\ lines are detected in all of the regions.
We applied the same line fitting method to these integrated region spectra as we did for the individual spaxels in Section \ref{sec:MapFitting}. The measured line properties are given in Table \ref{tab:HostFlux}. In Figure \ref{fig:RegionMaps} we also present the \hb\ flux map, as well as a map of the \oiiia\ velocity $v_{50}$ and velocity dispersion $w_{80}$.

\begin{table*}
\caption{Fluxes and velocities for the \hb\ and \oiii\ galaxy line components 
from fits to the quasar-subtracted spectra presented in Figure \ref{fig:RegionSpectra}, which have been integrated over the three emission-line regions shown in Figure \ref{fig:RegionMaps}.}
\begin{tabular}{llllllllll}
\hline 
\hline
 & $\0\0F_{\rm{H\beta}} $ 
 & $F_{\rm{[O\textsc{iii}]{\lambda4959}}}$  
 & $F_{\rm{[O\textsc{iii}]{\lambda5007}}}$  
 & $F_{\rm{[O\textsc{iii}]{\lambda5007}}}/F_{\rm{H\beta}} $  
 & $\0\0V_r$  & $\0V_\sigma$ \\
  & \multicolumn{3}{c}{$(10^{-18}\rm{erg/s/cm}^2)$ [S/N]} & &\multicolumn{2}{c}{(km/s)}\\
  \hline
Region 1 (Host) & $<2.1$ & $2.3\pm0.6$ [4.1] & $6.8\pm0.6$ [12] & $>3.2$ &\0\00  ($z=7.0804$) &$\086\pm49$ \\
Region 2 (Companion) & $\0\00.4\pm0.1$ [3.6] & $2.3\pm0.2$ [12] & $6.8\pm0.1$ [55] & $\016.4\pm4.6$ & $\030\pm49$ &$104\pm49$ \\
Region 3 & $<0.4$ & $0.8\pm0.1$ [8.2] & $2.5\pm0.1$ [23] & $>7.0$ &$143\pm49$ &$\083\pm49$ \\\hline
\end{tabular} 
\tablefoot{The values in braces give the integrated S/N of each line.
Upper limits for undetected lines are 3$\sigma$ limits on the line flux. 
The velocity $V_r$ is relative to the redshift of the quasar host, obtained from the integrated spectrum of Region 1, $z=7.0804\pm0.0028$.
Uncertainties on the velocity and velocity dispersion ($V_\sigma$) assume an uncertainty of $\pm1$ wavelength element, $\pm6.65$\AA\ or 49 km/s.
The velocity dispersions have been corrected for an instrumental broadening of FWHM$_{\rm{inst}}=115$ km/s.
The fluxes of these extended sources have not been multiplied by the aperture correction.}
\label{tab:HostFlux}
\end{table*}

The host galaxy, Region 1, has the core region surrounding the quasar missing in the emission-line maps, with no narrow flux seen in this region of the quasar subtracted data cube. This is because the quasar subtraction technique is not capable of accurately recovering any host flux in this region, underneath the bright and dominant quasar. Realistically, the host galaxy will have narrow-line emission within this region, and so our integrated flux will underestimate the total flux due to this missing contaminated region. Inferring properties of the host is thus difficult due to this contamination.
We measure that the quasar host is elongated in the east--west direction, extending across $\sim1''$, where $1''$ corresponds to $5.33$ kpc at $z=7.085$, with the quasar located approximately in the middle along this axis.
The velocity dispersion for the host from the integrated spectrum is measured to be $92\pm49$ km/s.
To best estimate the properties of the host galaxy, we fit the \oiiia\ emission-line map with a S\'ersic profile model with psfMC, using our IFU PSF estimate from our quasar subtraction as the PSF model. 
The best-fit S\'ersic model for the host galaxy has effective radius $r_{\rm{eff}}= 0\farcs361\pm0\farcs001$, axis ratio $b/a=0.41\pm0.01$, and position angle $\rm{PA} = 77.5\pm0.6$ deg defined counter-clockwise from north. We fixed the S\'ersic index to be $n=1$, as this parameter can only be poorly constrained in this difficult PSF subtraction process \citep[see e.g.][]{Ding2023,Yue2023}. 

We also fit a S\'ersic model to the close companion galaxy Region 2. The best-fit S\'ersic profile to the \oiiia\ emission of the companion galaxy with fixed $n=1$ has $r_{\rm{eff}}= 0\farcs346\pm0\farcs005$, $b/a=0.42\pm0.01$, and $\rm{PA} = 35.4\pm0.5$ deg, with the peak offset 0\farcs2355 or 1.25 kpc to the west and 0\farcs415 or 2.2 kpc to the south of the quasar host peak. 
The velocity map in Figure \ref{fig:RegionMaps} clearly shows a velocity gradient across the companion, with the side closest to the quasar blueshifted relative to the far side with velocity difference $\sim100$ km/s.
From the fit to the integrated region spectrum (Figure \ref{fig:RegionSpectra}),
the companion has a negligible velocity offset of only $32\pm49$ km/s from the quasar host, and a similar velocity dispersion of $104\pm49$ km/s. 
With this similar velocity, and the 2D sky offset of only 0\farcs477 corresponding to 2.5 kpc, this companion very easily satisfies the common merger criterion of having a projected distance of $<20/h$ kpc, where $h$ is the dimensionless Hubble constant $h=H_0/(100$ km/s/Mpc), and a velocity difference $\Delta V < 500$ km/s \citep[e.g.][]{Patton2000,Conselice2009}.
The companion galaxy is measured to have very similar \oiii\ and \hb\ flux as the host galaxy (see Table \ref{tab:HostFlux})---quantifying this flux ratio more precisely is difficult due to the missing flux from the quasar core.

The third emission region directly to the south of Region 2 is four times fainter than the host and main companion, and it does not have a clear regular shape. 
The velocity is offset by $145\pm49$ km/s from the host galaxy, with a similarly narrow line width of $83\pm49$ km/s.
As seen from the \oiiia\ velocity map in Figure \ref{fig:RegionMaps}, in the 2D plane of the sky this region extends spatially beyond the redshifted edge of the companion galaxy, and it has a velocity that extends gradually larger than that of the edge of the companion. This third region may be a tail of gas extending from the main companion or a third companion galaxy.

To understand the excitation mechanisms, we considered the flux ratios of the narrow emission lines. From Table \ref{tab:HostFlux}, the \oiiia/\hb\ flux ratio is $16.4\pm4.6$ for the main companion galaxy, with a limit of $>3.4$ and $>7.0$ for Regions 1 and 3 where \hb\ was not detected.  These cannot be easily classified using the traditional BPT diagram \citep{Baldwin1981} with the existing data. Our spectra do not cover the \ha\ and \niia\ lines. With the spectral resolution of the MIRI data, the narrow \ha\ line could not be decomposed from the broad component, and \niia\ was not detected \citep{Bosman2024}. 
In Appendix \ref{sec:BPT} we plot the BPT diagram for the three regions, with horizontal ranges showing their \oiiia/\hb\ flux ratios. 
With such a high \oiiia/\hb\ ratio, the companion galaxy and Region 3 are likely photoionised by an AGN, lying above both the \citet{Kewley2001} and \citet{Kauffmann2003} demarcation curves. 
We saw no evidence of additional broad lines nor a separate point source out to $\lesssim5.3$\micron\ ($\lesssim6500$\AA\ rest-frame) that would clearly indicate a secondary type~1 AGN. It is instead most likely that the quasar is photoionising the nearby regions \citep[`cross-ionisation'; e.g.][]{Moran1992,daSilva2011,Merluzzi2018,Keel2019,Moiseev2023,Protusova2024}.
However, we cannot rule out the possibility of there being a faint reddened or obscured AGN in Region 2.
The host galaxy could lie in the confusion region in the upper-left low-metallicity area of the star-forming branch, where both star-forming galaxies and AGNs lie at high-$z$ \citep[e.g.][]{Cameron2023,Scholtz2023}. However, given the presence of the bright quasar, AGN photoionisation is most likely dominant.

\subsection{Dynamical mass}

\begin{table*}
\caption{Dynamical masses and values used in their calculation for the quasar host and companion galaxy.
}
\begin{tabular}{lllllllllllll}
\hline 
\hline 
& $R$
& $\sigma$
& $b/a$
& $i$
 & $M_{\rm{dyn, disp}}$ 
 & $M_{\rm{dyn, rot}}$
 & $M_{\rm{dyn, vir}}$ \\
 
 & (kpc) & (km/s) & & (deg) & \multicolumn{3}{c}{($10^{10}\rm{M}_\odot$) } \\

 \hline

Region 1 (Host) &$1.93\pm0.03$ & $88\pm{ 49 }$ & $0.40 \pm 0.01$ & $66\pm1$ &0.5$\substack{ +0.8\\-0.4 }$ & 1.3$\substack{ +1.9\\-1.1 }$ & 1.7$\substack{ +2.6\\-1.4 }$\\
Region 2 (Companion)& $1.83\pm0.02$ & $104\pm{ 49 }$ & $0.42 \pm 0.01$ & $65\pm1$ &0.7$\substack{ +0.8\\-0.5 }$ & 1.8$\substack{ +2.1\\-1.3 }$ & 2.3$\substack{ +2.8\\-1.7 }$\\
\hline
\end{tabular} 
\tablefoot{Three dynamical masses are calculated, assuming that the galaxies are dispersion dominated, $M_{\rm{dyn, disp}}$, rotation dominated, $M_{\rm{dyn, rot}}$, or virialised, ${M}_{\rm{dyn, vir}}$---our best estimate is the rotation dominated measurement.
Here, $R$ is the S\'ersic radius found in the best fit to the \oiiia\ emission from psfMC.
The axis ratio $b/a$ is also measured from this S\'ersic fit, with the inclination of the galaxy $i$ calculated as $\cos(i) = b/a$.
The intrinsic integrated line velocity dispersion $\sigma$ is taken from the \oiiia\ line properties given in Table \ref{tab:HostFlux}.}
\label{tab:DynamicalMasses}
\end{table*}

The \oiiia\ emission-line maps allowed us to estimate the dynamical mass of the host galaxy and the companion. Due to the contamination from the quasar resulting in a corrupted central host core, we cannot perform detailed dynamical modelling of the host. Instead, we follow the method from \citet{Decarli2018} used commonly in quasar host studies with ALMA. 
We considered three dynamical model assumptions for the galaxies:\\
dispersion dominated, 
\begin{equation}
M_{\rm{dyn, disp}}=\frac{3}{2} \frac{R\sigma^2}{G};
\end{equation}
rotation-dominated thin disc \citep[see e.g.][]{Willott2015},
\begin{equation}
M_{\rm{dyn, rot}}=\frac{R}{G} \left(\frac{0.75~ \rm{FWHM}}{\sin(i)}\right)^2;
\label{eq:rot}
\end{equation}
and the virial approximation \citep{Cappellari2006,VanDerWel2022},
\begin{equation}
{M}_{\rm{dyn, vir}}\approx \frac{5R{\sigma}^{2}}{G}.
\end{equation}
Here, $R$ is the radius of the galaxy, $\sigma$ the integrated velocity dispersion, which relates to the line FWHM as $\sigma=\rm{FWHM}/(2\sqrt{2\ln{2}})$, $i$ is the inclination angle of the assumed thin disc, and $G$ is the gravitational constant.
For the radius $R$, we used the S\'ersic half-light radius found in the best fit to the \oiiia\ emission from psfMC; $0\farcs361$ or 1.92 kpc for the host and $0\farcs345$ or 1.84 kpc for the companion. Given the irregular shape of the observed \oiiia\ emission from Region 3, we did not attempt to model its dynamical mass.
The inclination $i$ is calculated from the axis ratio, $\cos(i) = b/a$.
For the line width, we used the values measured from Table \ref{tab:HostFlux} for the \oiiia\ line. These quantities are all provided in Table \ref{tab:DynamicalMasses}, alongside the resulting dynamical masses.

For the quasar host galaxy, we estimate dynamical masses of $M_{\rm{dyn}}=0.5$--$1.7\times10^{10}M_\odot$ from the three assumptions.
Most high-$z$ quasar host studies assume a rotating disc geometry \citep[e.g.][]{Wang2013,Willott2015,Decarli2018}, and so we adopt the rotation-dominated thin disc measurement using Equation \ref{eq:rot} as our best estimate of the dynamical mass: $M_{\rm{dyn}}=\left(1.3\substack{ +1.9\\-1.1 }\right)\times10^{10}M_\odot$ for the quasar host.
The companion galaxy shows a clear velocity gradient in \oiiia\ (Figure \ref{fig:RegionMaps}), indicating a rotational velocity structure, and so we use the rotational estimate of $M_{\rm{dyn}}=\left(1.8\substack{ +2.1\\-1.3}\right)\times10^{10}M_\odot$ for the companion. 
Our host dynamical mass estimate is consistent with the dynamical mass upper limit estimated from ALMA \cii\ observations, $M_{\rm{dyn}}< (4.3\pm0.9)\times 10^{10}M_\odot$ \citep{Venemans2017a}.
These dynamical mass estimates indicate that this is a `major' galaxy merger, between galaxies of similar mass.

\subsection{Black hole properties}
\subsubsection{Black hole mass}
\label{sec:BHmass}

\begin{figure*}
\begin{center}
\includegraphics[scale=0.8]{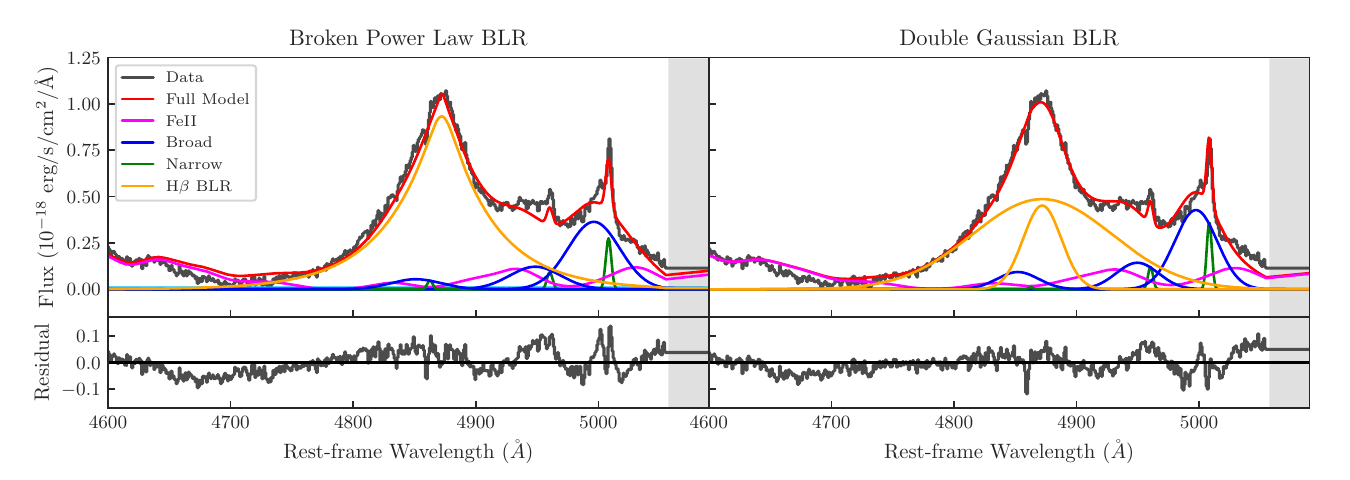}
\vspace{-0.3cm}
\caption{Integrated quasar spectrum for \qso\ (black) showing the region around \hb.  The spectrum is integrated over a radius of $0\farcs35$, and no correction for the loss of flux in this aperture has been applied (although this is accounted for in the related calculations). 
The best model fit (red) is shown alongside the narrow-line components (green), the broader outflow components (blue), the iron emission template (pink), and the model for the BLR (orange). The lower panels show the residual of the model fit. 
Both models assume the iron emission follows the \citet{Park2022} template.
The left panels show a model assuming that the \hb\ BLR is described by a BPL model.
The right panels show a model assuming that the \hb\ BLR is described by two Gaussian profiles (the DG model).
The shaded grey regions show where the detector gap falls, resulting in no coverage of those wavelengths.
}
\label{fig:BHSpectra}
\end{center}
\end{figure*}

In Figure \ref{fig:BHSpectra} we show the integrated quasar spectrum around \hb\ alongside the two best-fitting models: 
a BPL \hb\ BLR model and a DG \hb\ BLR model.
For these two models, the average redshift of the narrow-line components of \hb\ and \oiii\ (which are constrained to have the same velocity) is $z=7.079\pm0.001$.
Both models result in a similar fit, with the DG model performing slightly better with lower residuals in the left wing of the \hb\ line and the region around \oiiib. 
The poor residuals in the fits around rest-frame 4960\AA\ are likely due to the iron blend template imperfectly matching the true emission from this source; Figure \ref{fig:FullSpectrum} shows the iron emission model compared with the integrated quasar spectrum over the full wavelength range, which shows an imperfect match in some other regions of \ions{Fe}{ii} emission.

Both models find a clear outflow component in the \oiii\ and \hb\ lines, fit by a Gaussian with velocity offset of $\Delta v = -579\substack{+42\\-55}$ km/s ($\Delta v = -650\substack{+89\\-193}$ km/s) from the narrow-line component, and with $\rm{FWHM}=2490\substack{+10\\-400}$ km/s ($\rm{FWHM}=2450\substack{+40\\-360}$ km/s), for the DG (BPL) model.
This outflow is spatially unresolved at the location of the quasar and is removed in the quasar subtraction process---from the low velocity offsets and velocity dispersions in the quasar-subtracted line maps (Figure \ref{fig:RegionMaps}) there is no signature of extended outflows or any indication that Regions 2 and 3 are outflowing gas. The quasar-driven outflow will be studied in depth alongside those of the full GA-NIFS quasar sample (Venturi et al., in prep.).

To calculate the BH mass, we used the commonly employed scaling relation
\begin{equation}
\label{eq:MBH_5100}
M_{\rm{BH,5100}} = a ~\left( \frac{\lambda L_{\rm{5100}}}{10^{44} \rm{erg~ s}^{-1}} \right)^{b} \left( \frac{\rm{FWHM}_{\rm{H\beta}}}{10^{3} \rm{km~ s}^{-1}} \right)^{2} \rm{M}_\odot,
\end{equation}
which uses the continuum luminosity of the AGN at rest-frame $\lambda=5100$\AA,  $L_{5100}$, the line width of the \hb\ BLR component, $\rm{FWHM}_{\rm{H\beta}}$, and the constants $a$ and $b$ that are calibrated empirically from reverberation mapping studies.
We consider the calibration $a=(4.4\pm0.2) \times10^6$, $b=0.64\pm0.02$ from \citet{Greene2005}, and alternatively  $a=(8.1\pm0.4)\times10^6$, $b=0.50\pm0.06$ from \citet{Vestergaard2006}.
We could not measure $L_{5100}$, as at the quasar redshift $5100$\AA\ rest-frame falls into the gap between the two detectors. We instead used the value measured from the NIRCam grism spectra, of $\lambda L_{5100}=1.76\times10^{46}$ erg/s \citep{Yue2023}.

Alternatively, we considered the pure-\hb\ scaling relation, which uses the \hb\ BLR luminosity $L_{\rm{H\beta}}$ instead of the continuum luminosity:
\begin{equation}
\label{eq:MBH_HB}
M_{\rm{BH,H\beta}} = a ~\left( \frac{ L_{\rm{H\beta}}}{10^{42} \rm{erg~ s}^{-1}} \right)^{b} \left( \frac{\rm{FWHM}_{\rm{H\beta}}}{10^{3} \rm{km~ s}^{-1}} \right)^{2} \rm{M}_\odot.
\end{equation}
Again $a$ and $b$ are calibrated from low-$z$ reverberation mapping studies; we considered both
 $a=(3.6~\pm~0.2) ~\times~ 10^6$, $b=0.56~\pm~0.02$ from \citet{Greene2005}, and $a=(4.7~\pm~0.3)~\times~ 10^6$, $b=0.63~\pm~0.06$ from \citet{Vestergaard2006}. These scaling relations Equations \ref{eq:MBH_5100} and \ref{eq:MBH_HB} have an estimated scatter of 0.43 dex \citep{Vestergaard2006}.

\begin{table*}
\begin{center}
\caption{Estimated BH masses ($M_{\rm{BH}}$), Eddington ratios ($\lambda_{\rm{Edd}}$), and the FWHMs and luminosities used in their calculation for the quasar \qso.
}
\begin{tabular}{lllllllllll}

 \hline \hline
& FWHM$_{\rm{H\beta}}$  & $L_{\rm{H\beta}}$  & \multicolumn{4}{c}{$M_{\rm{BH,H\beta}}$} & \multicolumn{4}{c}{$\lambda_{\rm{Edd,H\beta}}$} \\

Model & & & 5100\AA, G & 5100\AA, V & \hb, G & \hb, V\rlap{\tablefootmark{*}}  & 5100\AA, G & 5100\AA, V & \hb, G & \hb, V\rlap{\tablefootmark{*}}  \\
& (km/s) & $(10^{44} \rm{erg/s})$ & \multicolumn{4}{c}{$(10^9 M_\odot)$}\\
\hline 
BPL& $4104\substack{ +405\\-355 }$ &$5.2\substack{ +0.6\\-0.4 }$&$2.0\substack{ +0.4\\-0.3 } $&$1.8\substack{ +0.4\\-0.3 }$&$2.0\substack{ +0.5\\-0.4 }$&$4.1\substack{ +1.0\\-0.7 }$&$0.6\substack{ +0.1\\-0.1 }$&$0.7\substack{ +0.1\\-0.1 }$ &$0.6\substack{ +0.1\\-0.1 }$&$0.3\substack{ +0.1\\-0.1 }$\\
DG& $3901\substack{ +354\\-253 }$ &$5.4\substack{ +0.3\\-0.5 }$&$1.8\substack{ +0.3\\-0.2 } $&$1.6\substack{ +0.3\\-0.2 }$&$1.8\substack{ +0.4\\-0.3 }$&$3.7\substack{ +0.8\\-0.6 }$&$0.7\substack{ +0.1\\-0.1 }$&$0.8\substack{ +0.1\\-0.1 }$ &$0.7\substack{ +0.1\\-0.1 }$&$0.3\substack{ +0.1\\-0.1 }$\\

\hline
\end{tabular}
\label{tab:BHproperties}
\end{center}
\tablefoot{Full width half maximum and luminosity $L_{\rm{H\beta}}$ are taken from the \hb\ BLR model fit to the integrated quasar spectrum, from each of the two models shown in Figure \ref{fig:BHSpectra}. 
The models use the \citealt{Park2022} iron template and have different assumptions in the assumed \hb\ BLR shape (BPL and DG). There are four calculated BH masses for each model; using the 5100\AA\ luminosity of $\lambda L_{5100}=1.76\times10^{46}$ erg/s (Equation \ref{eq:MBH_5100}), and using the \hb\ luminosity (Equation \ref{eq:MBH_HB}), assuming the \citet{Greene2005} (G) and \citet{Vestergaard2006} (V) empirical calibrations. The quoted measurement  uncertainties do not include the uncertainties in the parameters $a$ and $b$ used in these calibrations or the scatter in the relations of 0.43 dex. The Eddington ratio $\lambda_{\rm{Edd}}=L_{\rm{Bol}}/L_{\rm{Edd}}$ is calculated for each of these BH masses, assuming a bolometric luminosity of $L_{\rm{Bol}}=1.63\times10^{47} \rm{erg~s^{-1}}$ from \citet{Yue2023}.}
\tablefoottext{*}{\footnotesize The \citet{Vestergaard2006} \hb-only masses are significantly larger than the three other estimates, and so we exclude these as outliers in our analysis, including them here only for reference.}
\end{table*}

The resulting \hb\ BH mass estimates are given in Table \ref{tab:BHproperties}, alongside the luminosities and FWHMs used in the calculations.
We find that for the two BLR models (BPL and DG), the median 5100\AA-based measurement from Equation \ref{eq:MBH_5100} 
is $M_{\rm{BH,5100}}=\left(1.8\substack{+0.4\\-0.3}\right)\times10^9 M_\odot$.
For the \citet{Greene2005} \hb-only based measurement from Equation \ref{eq:MBH_HB}, 
the median for the two models is  $M_{\rm{BH,H\beta,G}}=\left(1.9\substack{+0.5\\-0.4}\right)\times10^9 M_\odot$.
However, the \citet{Vestergaard2006}  \hb-only based measurements from Equation \ref{eq:MBH_HB} are significantly larger, 
with median $M_{\rm{BH,H\beta,V}}=\left(3.9\substack{+1.0\\-0.7}\right)\times10^9 M_\odot$.
Both of these empirical relations are based on low-$z$ AGNs, and not high-$z$ luminous quasars; these sources require extrapolation beyond the luminosity range of the observed low-$z$ calibration sample. Thus, it seems likely that the disagreement of the \citet{Vestergaard2006}  \hb-only masses is due to this relation diverging from the \citet{Greene2005} relation at AGN luminosities beyond the observed regime. Because three of the four relations produce BH masses that are in close agreement, we decide to disregard the \citet{Vestergaard2006} \hb-only masses as outliers.

We now compare how the different modelling assumptions alter the measured BH mass.
Both the DG and BPL models provide very similar mass estimates, with the BPL model producing slightly larger estimates but being consistent within $1\sigma$. 
The DG model measures a slightly larger \hb\ luminosity than the BPL model, while the FWHM of the line is slightly lower for the DG model, although these estimates agree to within $1\sigma$.
The quoted uncertainties for each parameter are lower for the DG model, which has lower residuals around the \hb\ line.

Taking the median of the BPL and DG masses over each of the three estimates per model, we obtain a best estimate BH mass of $M_{\rm{BH}}=\left(1.9\substack{+0.4\\-0.3}\right)\times10^9 M_\odot$. 
These uncertainties include only observational errors and do not include the scatter in the scaling relations Equations \ref{eq:MBH_5100} and \ref{eq:MBH_HB}, estimated to be 0.43 dex from \citet{Vestergaard2006}. Including the scaling relation uncertainties, we estimate $M_{\rm{BH}}=\left(1.9\substack{+2.9\\-1.1}\right)\times10^9 M_\odot$.

Finally, we consider the effect of the host galaxy emission on the measured BH mass. We performed the same QubeSpec fitting process on the quasar model spectrum---the spectrum from the brightest spaxel in the cube, where the host contribution is negligible---with appropriate flux scaling correction.
We also performed the QubeSpec fitting on an approximate host-subtracted spectrum---we integrated the quasar-subtracted cube over the same 0\farcs35 aperture, and measured and removed its \oiii\ line profile from the full integrated spectrum.
We find that the BH masses measured from these spectra are 0.3 and 0.1 dex lower than our best estimate, respectively, although they are consistent within the $1\sigma$ measurement uncertainties; we conclude that the host galaxy emission does not significantly impact our BH mass measurement.

In \citet{Yue2023}, the \hb\ BH mass is measured from NIRCam grism spectra as $M_{\rm{BH}}=(1.19\pm0.08)\times10^9 M_\odot$. Since we assume the same $L_{5100}$, this lower estimate occurs due to their lower \hb\ FWHM, which is $3337\substack{+95\\-111}$ km/s, relative to our $3901\substack{ +354\\-253 }$ km/s from the DG model. While our BH mass and FWHM estimates are slightly larger, these are consistent within 2$\sigma$. 
Their NIRCam grism spectra has a spectral resolution of $\sim1600$ at 4$\mu$m. In comparison, our IFU observations have a spectral resolution of $\sim2700$, which allowed us to conduct more detailed modelling. Our observations also include the \oiii\ lines, while in the NIRCam grism these fall on the edge of the detector and cannot be well measured; these lines play a large role in constraining our MCMC fit, particularly with the iron template parameters.
\citet{Yue2023} assumed a \hb\ model composed of two Gaussians, which should be most consistent with our DG model. However, we also allowed a narrow and an outflow component to the \hb\ line, 
but they did not, and this choice will make a small difference.
Overall, with these differences in the measured spectra and modelling approaches, it is reasonable that we are measuring slightly different BH masses for the same \hb\ line, although we reiterate that our masses are consistent within 2$\sigma$, and well below the systematic uncertainties (i.e. 0.43 dex).

With MIRI's MRS, \citet{Bosman2024} obtained a spectrum of \qso\ that covers the \ha, Pa-$\alpha$,  Pa-$\beta$ and  Pa-$\gamma$ emission lines.
Using similar scaling relations to Equation \ref{eq:MBH_HB}, these lines were used to calculate BH masses that are:
${M}_{{{{\rm{BH}}}},{{{\rm{H}}}}\alpha }=(1.55\pm 0.22)\times10^9 M_\odot $,
${M}_{{{{\rm{BH}}}},{{{\rm{Pa}}}}\alpha }=1.0\substack{ +0.3\\-0.2}\times10^9 M_{\odot }$, and ${M}_{{{{\rm{BH}}}},{{{\rm{Pa}}}}\beta }=0.87\substack{ +0.21\\-0.17}\times10^9 M_{\odot }$.
They also measured the BH mass using the combined Paschen-series lines and the rest-frame infrared continuum to be $M_{\rm{BH,IR}}=(0.89\pm0.14)\times10^9M_\odot$.
These measurements have not been corrected for dust extinction, as they found by comparing the \ha\ and Paschen-series lines that minimal BLR extinction is present, of order $E(B-V)\lesssim0.1$ \citep{Bosman2024}.
The \ha\ BH mass is similar to our \hb\ BH mass estimate, which we have also not corrected for dust attenuation; this is consistent with the quasar emission being minimally affected by dust.
The Paschen-based BH masses are lower than our \hb\ mass, although they are all consistent within $1\sigma$ when considering the additional uncertainty from the scaling relations.
Overall, we conclude that the BH mass measurements measured by \citet{Bosman2024} with MIRI MRS are consistent with our \hb\ BH mass measurement.

Finally, we compare the JWST BH mass estimates to the previous estimates from ground-based telescopes.
From the \ions{Mg}{ii} emission line, the BH mass was measured to be ${M}_{{{{\rm{BH}}}},{{{\rm{Mg}}}}\,{{{\rm{II}}}}}=(1.35\pm 0.04)\times1{0}^{9} M_\odot$ \citep{Yang2021a}, which is consistent with our estimates. We note that the scaling relations for \ions{Mg}{ii} BH mass estimates have a scatter of $\sim0.55$ dex \citep{Vestergaard2009}, which is not included in the quoted uncertainties.
From the \ions{C}{iv} emission line, \citet{Farina2022} measured a BH mass of ${M}_{{{{\rm{BH}}}},{{{\rm{C}}}}\,{{{\rm{IV}}}}}=(2.40\pm0.05)\times1{0}^{9}\,{M}_{\odot }$. 
\citet{Bosman2024} found that this \ions{C}{iv} BH mass is not consistent with their Paschen-based BH masses, although it is consistent with their \ha\ mass.
However, when considering that there is a scatter in the \ions{C}{iv} scaling relations of $\sim0.36$--0.40 dex \citep{Vestergaard2006,Shen2012b}, this \ions{C}{iv} BH mass is consistent with all of the above quoted BH masses.

Combining all of these independent observations and methods to estimate BH mass, we can be confident that the BH mass of \qso\ is generally in the range of $\sim1$--$2\times10^9M_\odot$, with the best estimate from this work being $M_{\rm{BH}}=(1.9\substack{+0.4\\-0.3})\times10^9 M_\odot$. 

\subsubsection{Eddington ratio}
To estimate the Eddington ratio $\lambda_{\rm{Edd}}=L_{\rm{Bol}}/L_{\rm{Edd}}$, we use the bolometric luminosity from \citet{Yue2023}, $L_{\rm{Bol}}=1.63\times10^{47} \rm{erg~s^{-1}}$, as calculated
from the rest-frame 5100\AA\ luminosity using the conversion~ $L_{\rm{Bol}} = 9.26 \times \lambda L_{5100}$ \citep{Shen2011}.
We calculated the Eddington luminosity using
\begin{equation}
\begin{split}
L_{\rm{Edd}} &= \frac{4\pi G m_{p}c M_{\rm{BH}}}{\sigma_T}= 1.26\times10^{38} \left(\frac{M_{\rm{BH}}}{\rm{M}_\odot}\right) \rm{erg~s^{-1}},
\end{split}
\end{equation}
where $G$ is the gravitational constant, $m_{p}$ the proton mass, and $\sigma_T$ the Thomson scattering cross-section.
The Eddington ratios for each of our BH mass estimates are given in Table \ref{tab:BHproperties}.
For our 5100\AA\ \citet{Greene2005} and \citet{Vestergaard2006}, and \hb-only \citet{Greene2005} BH mass estimates, the Eddington ratio ranges from 0.6--0.8 for each of the models.
For our best BH mass estimate of 
 $M_{\rm{BH}}=\left(1.9\substack{+0.4\\-0.3}\right)\times10^9 M_\odot$, we estimate $\lambda_{\rm{Edd}}=0.7\pm0.1$, or $0.7\substack{ +1.1\\-0.4}$ when including the 0.43 dex scatter in the BH scaling relations.
\citet{Bosman2024} estimate $\lambda_{\rm{Edd}}=0.9\pm0.1$ from \ha, using a slightly larger bolometric luminosity of $L_{\rm{Bol}}=1.705\times10^{47} \rm{erg~s^{-1}}$ from \citet{Farina2022}, and \citet{Yue2023} estimate $\lambda_{\rm{Edd}}=1.08$ from \hb, using the same $L_{\rm{Bol}}$ and a lower $M_{\rm{BH}}$; both are consistent with our measurement.

\section{Host galaxy properties from NIRCam}
\label{sec:NIRCamResults}

\subsection{Quasar subtraction and host modelling} \label{subsec:nircamsub}
From the \oiii\ emission-line maps we discovered that the \qso\ system is comprised of both a quasar host galaxy and a merging companion galaxy.
When \citet{Yue2023} performed their detailed quasar subtraction on NIRCam images, they modelled this host system with only one S\'ersic profile.
They were able to reveal the continuum emission surrounding \qso, with clear detections in F356W and F200W, and a tentative detection in F115W.
They found $m_{\rm{F200W}}=24.43\pm0.10$ mag and $m_{\rm{F356W}}=24.45\pm0.20$ mag, which resulted in a stellar mass estimate of $M_\ast=6.5\substack{+4.5\\-3.3} \times10^{9}M_\odot$ from SED fitting. 
The peak of this F200W and F356W continuum emission was offset by $0\farcs5$ from the quasar---consistent with the location of the companion galaxy. These initial magnitude and mass estimates from \citet{Yue2023} included both of the host and companion components. In this work, we use the structure found from the IFU to improve the modelling of the system and the subsequent measured host stellar mass.

We used \textsc{psfMC} to fit the NIRCam images of {\qso}. We used a point source to describe the emission from the quasar, 
and used two exponential disks (S\'ersic profiles with indices $n=1$) to describe the extended emission from the quasar host galaxy (Region 1) and the close companion (Region 2). 
We note that \qso\ was visited four times in the EIGER program;
following the method in \citet{Yue2023} we fit the images from the four visits individually, and take the mean and standard deviation of all of the MCMC samples from the four visits as the best-values and uncertainties of the model parameters. 
This approach allowed us to take into account the systematic uncertainties caused by inter-visit differences (e.g. the changes of PSF between the different pointings, which each have the quasar in a different detector location).
If the magnitude error of the component is smaller than 0.3 mag, we consider the component to be detected in the image.

\begin{figure}
    \centering
    \includegraphics[width=0.75\linewidth, trim={0 0.5cm 0 0}, clip]{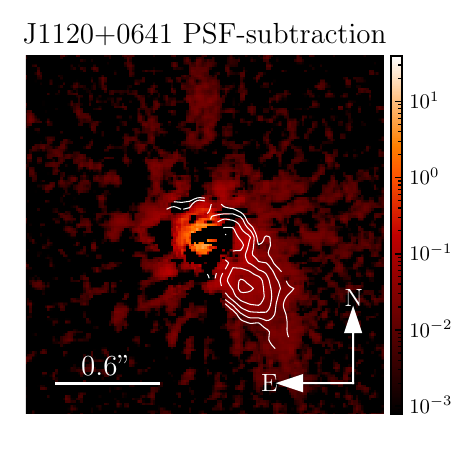}
    \caption{Overlay plot of the F200W PSF-subtracted image (background image with colour bar) and the {\oiiia} map (white contour). The emission in the PSF-subtracted F200W image largely overlaps with the {\oiiia} line map, indicating that the continuum and line emission originate from the same objects, i.e. the quasar host and the close companion. The marked scale of 0\farcs6 corresponds to a physical scale of 3.2 kpc at the redshift of the quasar.}
    \label{fig:F200W_O3_overlay}
\end{figure}

We first tried to fit the NIRCam images with all parameters allowed to vary (point source position and magnitude, S\'ersic positions, magnitudes, major axis radii $r_a$, minor axis radii $r_b$, and position angles, with fixed S\'ersic indices $n=1$). However, the parameters related to Regions 1 and 2 (e.g. their sizes and ellipticities) did not converge to physical values. We thus
fixed the spatial positions and morphological parameters of Regions 1 and 2 to the values measured in Section \ref{sec:Structure} from running psfMC on our \oiiia\ emission-line maps. 
In Figure \ref{fig:F200W_O3_overlay} we show the F200W PSF-subtracted image overlaid with the \oiiia\ line map.
The PSF-subtracted {\oiiia} and F200W maps largely overlap with each other, confirming that the extended continuum and line emission in \qso\ come from the same sources, the host galaxy and the close companion. 
The residual images also show faint positive flux in the north and south-west; the south-west structure corresponds to Region 3 seen in the IFU images, while the northern structure is not covered by the IFU.  Additional residuals are most likely due to the host and companion not being ideally modelled by these S\'ersic profiles.

\begin{figure*}
    \centering
    \includegraphics[width=0.99\linewidth, trim={0 2cm 0 0}, clip]{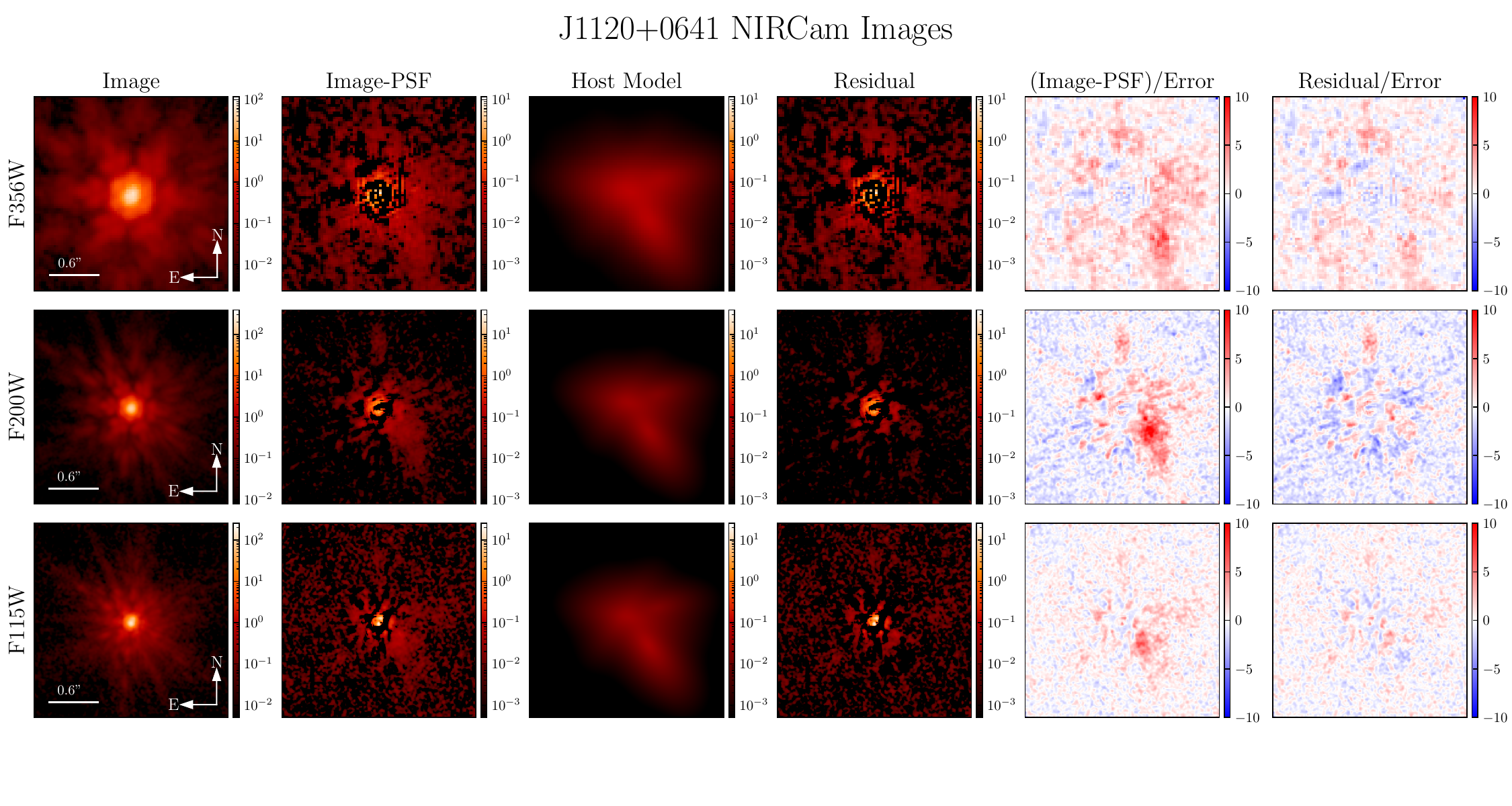}
    \caption{Results from the NIRCam PSF subtraction for \qso. From left to right, the original NIRCam images, the PSF-subtracted images, the model for the host and the companion, the image fitting residual (i.e. the subtraction of the full model from the original image), the PSF-subtracted images divided by the error images, and the residual images divided by the error images. The extended emission is modelled as two S\'ersic profiles using the positions and morphologies as measured from the \oiiia\ maps (Section \ref{sec:Structure}). The PSF-subtracted images clearly show the extended emission around the quasar. }
    \label{fig:nircamimage}
\end{figure*}

The full results of the image fitting are shown in Figure \ref{fig:nircamimage}, with the extracted magnitudes given in Table \ref{tab:SEDfitting}.
The PSF-subtracted images clearly show extended emission in all three bands, 
confirming the results from \citet{Yue2023}. 
For the host galaxy (Region 1), the image fitting gives
$m_\text{F115W}=25.94\pm0.20$ mag, $m_\text{F200W}=25.48\pm0.37$ mag and $m_\text{F356W}=25.72\pm0.47$ mag. 
Given the above-mentioned detection criteria, Region 1 is only detected in F115W and has non-detections in F200W and F356W.
For the companion (Region 2), we measure $m_\text{F115W}=25.32\pm0.12$ mag, $m_\text{F200W}=24.95\pm0.11$ mag and $m_\text{F356W}=25.11\pm0.18$ mag, and so the companion is detected in all three bands.
The quasar (i.e. the point source) has $m_\text{F115W}=20.366\pm0.003$ mag, $m_\text{F200W}=19.886\pm0.002$ mag and $m_\text{F356W}= 19.632\pm0.003$ mag.
The host and the companion are $\sim5$ magnitudes fainter than the quasar.
While their magnitudes are difficult to measure relative to the bright flux of the quasar, \citet{Yue2023} tested this methodology on mock quasar host observations and found that biases from this fitting process are low, smaller than the random error, providing confidence in these magnitude measurements.

\begin{table*}
    \centering
    \caption{Extracted NIRCam magnitudes for the quasar host and companion galaxy, and their stellar mass ($M_\ast$), fraction of stars formed in the recent starburst ($f_{\rm{starburst}}$), and metallicity ($Z$) measured from the subsequent SED fitting.}
    \begin{tabular}{lllllll}
    \hline 
    \hline
         & $m_\text{F115W}$ & $m_\text{F200W}$  & $m_\text{F356W}$  & $M_\ast$ & $f_{\rm{starburst}}$ & $\log(Z/Z_\odot)$ \\
         & \multicolumn{3}{c}{(mag)} & ($\times10^9 M_\odot$) &&\\
         \hline
        Region 1 (Host) & $25.94\pm0.20$ & $25.48\pm0.37$* & $25.72\pm0.47$* & $3.0\substack{+2.5\\-1.4}$ & $0.57\substack{+0.21\\-0.29}$ & $-0.50\substack{+0.24\\-0.24}$\\
        Region 2 (Companion) & $25.32\pm0.12$ & $24.95\pm0.11$  & $25.11\pm0.18$ & $2.7\substack{+0.5\\-0.5}$ & $0.99\substack{+0.01\\-0.01}$ & $0.09\substack{+0.08\\-0.69}$\\
\hline
\end{tabular} 
\tablefoot{Magnitudes marked with a `*' denote unsuccessful detections, with a successful detection classified as one where the magnitude error is $<0.3$ mag.}
\label{tab:SEDfitting}
\end{table*}

\subsection{Spectral energy distribution fitting: Stellar population properties}
\label{sec:CombinedResults}

We fit the SEDs of the quasar host and the companion using {\texttt{Prospector}} \citep{Johnson2021} to constrain their stellar masses and other properties.
{\texttt{Prospector}} is based on Flexible Stellar Population Synthesis (FSPS) and uses photoionisation predictions for nebular emission from {\texttt{Cloudy}} \citep[e.g.][]{Byler2017}. The input observations are the host and companion {\hb}, {\oiiib}, and {\oiiia} line fluxes from Table \ref{tab:HostFlux}, as well as their F115W, F200W, and F356W magnitudes from Table \ref{tab:SEDfitting}. We fit the SEDs of Region 1 (the host galaxy) and Region 2 (the companion) separately. Although we report non-detections for Region 1 in the F200W and the F356W bands, we still included these magnitudes and their errors when performing SED fitting for Region 1. This is because measurements with large uncertainties are preferred over upper limits in {\texttt{Prospector}}, as recommended in their documentation\footnote{https://prospect.readthedocs.io/en/latest/faq.html}.

We used a Chabrier initial mass function \citep{Chabrier2003} and assumed a dust attenuation following the \citet{Calzetti2000} law.
We considered two star-formation history (SFH) models when fitting the SED, namely,
a delayed-$\tau$ model $(\text{SFR}(t)\propto te^{-t/\tau})$, 
and a delayed-$\tau$ plus a starburst model.
We also included an empirical template for emission-line flux produced by AGNs as included in \texttt{Prospector}, which is based on the observations of \citet{Richardson2014}.
The free parameters and their priors of this SED model include:
(1) the stellar mass $M_*$ with a log-uniform prior at $[10^8M_\odot, 10^{12}M_\odot]$;
(2) the stellar metallicity $\log (Z/Z_\odot)$ with a uniform prior at $[-2, 0.2]$;
(3) the starting time of the star formation $t_\text{age}$ with a uniform prior at $[0, t(z)]$, where $t(z)$ is the age of the universe at the quasar's redshift;
(4) the exponential decay timescale $\tau$ with a uniform prior at $[0.01\text{Myr}, 20\text{Myr}]$;
(5) the dust attenuation (quantified as the optical depth at 5500{\AA}, $\tau_{5500}$) with a uniform prior at $[0, 2]$; 
(6) the gas-phase metallicity $\log (Z_g/Z_\odot)$ with a uniform prior at $[-2, 0.5]$; 
(7) the ionisation parameter $\log U$ with a uniform prior at $[-3, 1]$, and;
(8) the flux scaling factor applied to the AGN emission-line template, with a uniform prior at [$10^{-6},10^{-2}$].
For the delayed-$\tau$ plus starburst SFH, there are two additional parameters:
(9) the fraction of stellar mass produced by the starburst, $f_\text{burst}$, with a uniform prior at $[0,1]$,
and (10) the time of the starburst, $t_\text{burst}$, with a uniform prior at $[0, t(z)]$, which is also required to be later than $t_\text{age}$.

\begin{figure*}
\centering
    \includegraphics[width=0.48\linewidth]{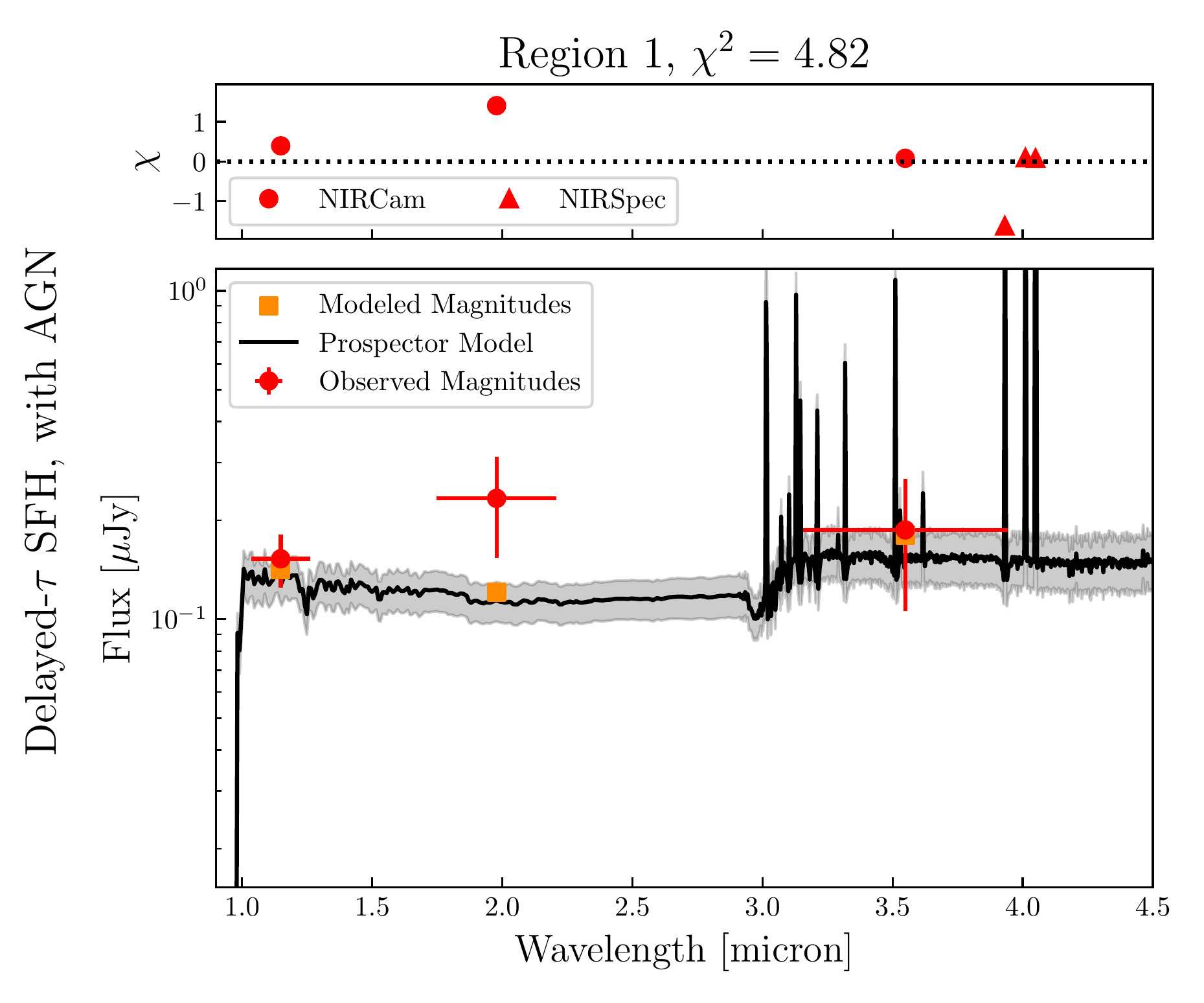}
    \includegraphics[width=0.48\linewidth]{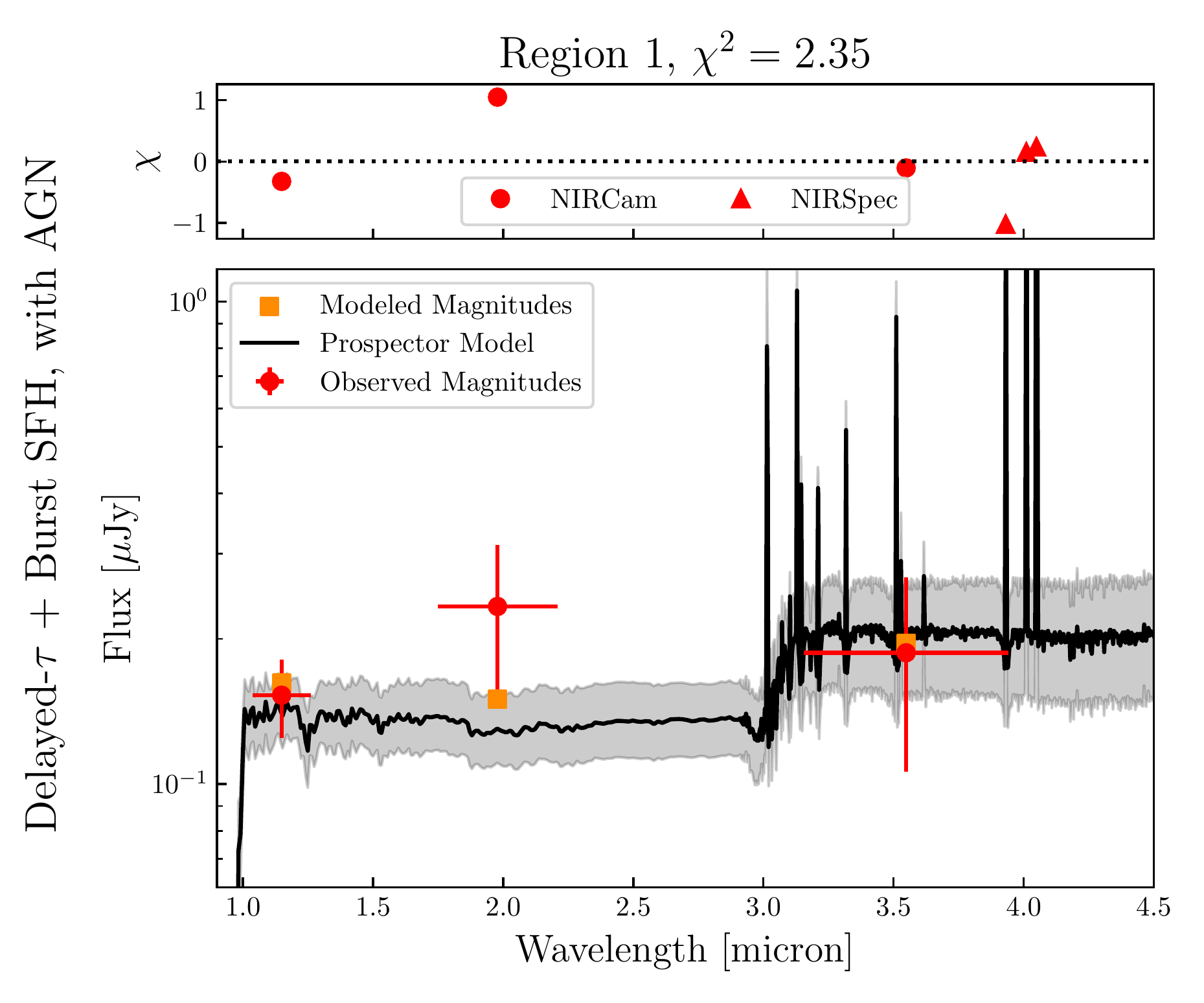}
    \includegraphics[width=0.48\linewidth]{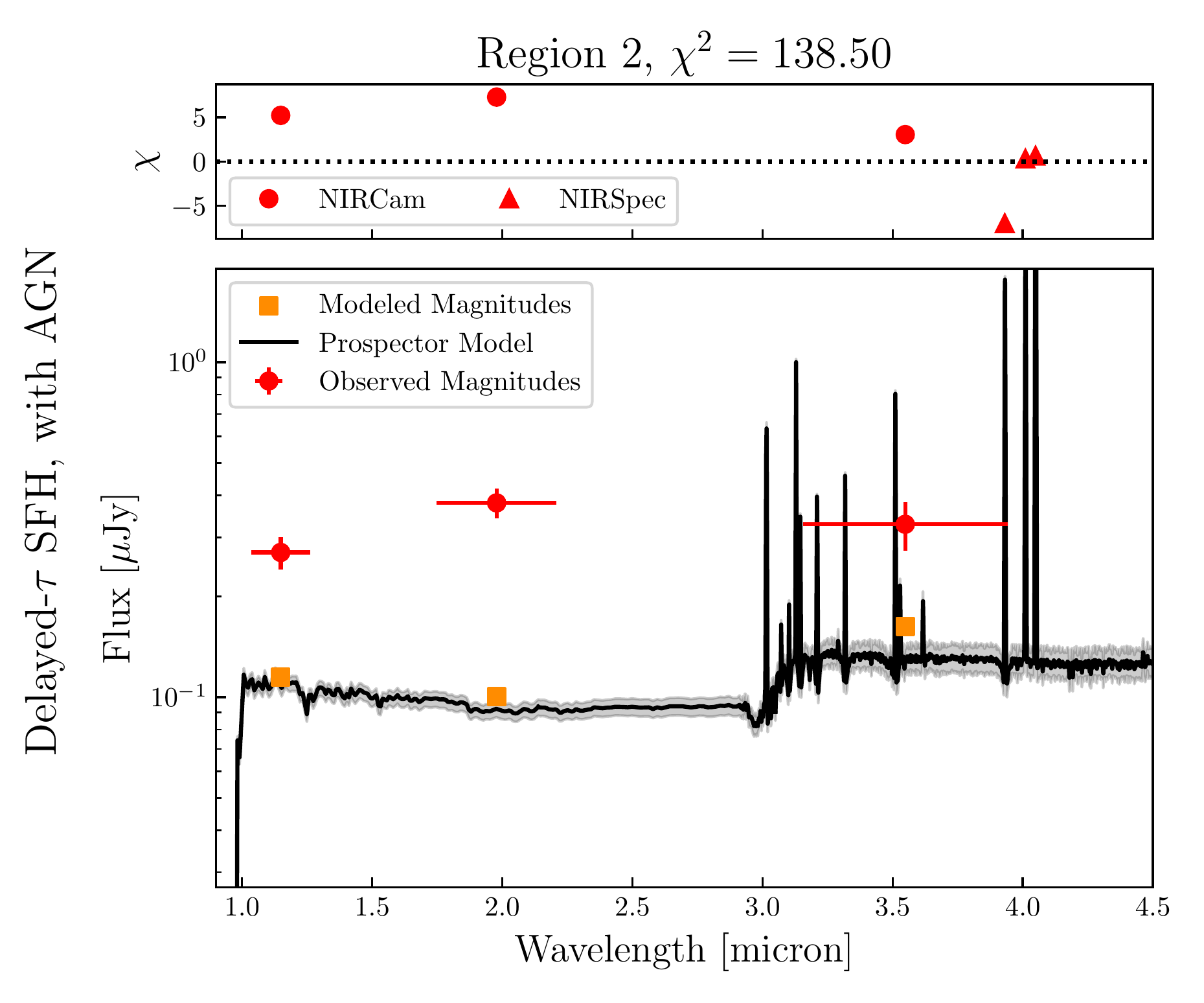}
    \includegraphics[width=0.48\linewidth]{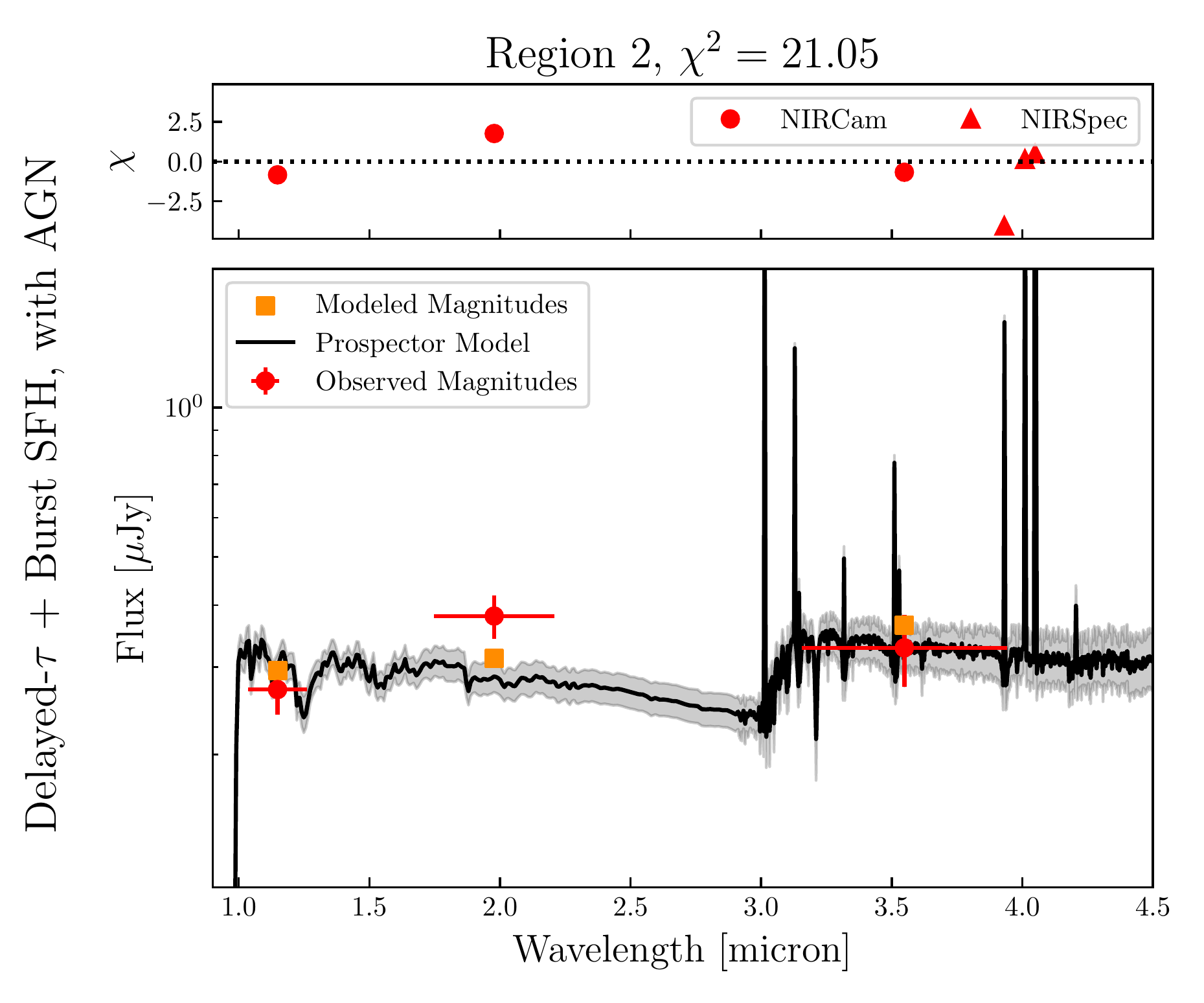}

    \caption{Spectral energy distribution fitting for the host galaxy Region 1 (upper panels) and for the companion galaxy Region 2 (lower panels). The left and right columns correspond to a single delayed-$\tau$ SFH and a delayed-$\tau$ plus a starburst SFH, respectively. The red circles represent the NIRCam broad-band fluxes from image fitting, and the red triangles represent the observed {\hb} and {\oiii} line fluxes. The solid black line and the gray shaded region mark the \texttt{Prospector} modelled spectrum and its $1\sigma$ error, while the orange squares show the resulting model photometry. }
    \label{fig:sed_allregion}
\end{figure*}

The result of the SED fitting is shown in Figure \ref{fig:sed_allregion}. For both Region 1 and Region 2, the delayed-$\tau$ plus a starburst SFH gives the best description to the observed fluxes. 
We therefore used the delayed-$\tau$ plus a starburst SFH to infer the properties of the quasar host and its companion. 
The estimated physical parameters are listed in Table \ref{tab:SEDfitting}.
The results of the SED fitting suggest that \qso\ might have a complex SFH that cannot be well-described by a single SFH component. 
For example, Figure \ref{fig:sed_allregion} shows that the \hb\ line is overestimated for both Regions 1 and 2 in both the delayed-$\tau$ and delayed-$\tau$ plus starburst models.
The F200W fluxes for both regions are underestimated in both SFH models. This is likely due to an overestimation of the measured F200W flux in the quasar subtraction process, likely caused by slight inaccuracies in the PSF models, and the galaxies not being perfectly modelled by S\'ersic profiles.

According to the SED models, the inferred stellar mass is $M_{*,\rm{host}}=\left(3.0\substack{+2.5\\-1.4}\right)\times10^9M_\odot$ for Region 1 and $M_{*,\rm{companion}}=\left(2.7\substack{+0.5\\-0.5}\right)\times10^9M_\odot$ for Region 2. 
The fraction of stellar mass formed by the starburst component is $f_{\rm{starburst}}=0.57\substack{+0.21\\-0.29}$ for Region 1 and is $0.99\pm0.01$ for Region 2. 
This suggests that the majority of the stars in the companion galaxy formed in a single past starburst, with little recent star formation.
The SED models also provide constraints on the stellar metallicities, which are based on FSPS stellar population models.
The metallicity of Region 1 is estimated to be $\log(Z/Z_\odot)=-0.50\pm0.24$, and the metallicity for Region 2 is $0.09\substack{+0.08\\-0.69}$. 
Our analysis demonstrates how we can constrain the SFH of quasar host galaxies by combining the broad-band fluxes from NIRCam imaging and emission-line fluxes from NIRSpec IFU.

As discussed in Section \ref{sec:Structure}, both Regions 1 and 2 are likely photoionised by the central quasar based on the BPT diagnostic (Appendix \ref{sec:BPT}). Including the AGN emission-line template in our SED fitting allowed us to account for this excess non-star-formation photoionisation, in order to more accurately model the stellar population.
If we exclude the AGN emission-line template and re-fit the SED models, assuming that the emission lines are entirely from star formation, we obtain slightly larger stellar masses of
$M_{*,\rm{host}}=\left(3.1\substack{+2.0\\-1.3}\right)\times10^9M_\odot$ for Region 1 and $M_{*,\rm{companion}}=\left(4.5\substack{+1.0\\-0.8}\right)\times10^9M_\odot$ for Region 2. 
These estimates are consistent within our uncertainties (within 1$\sigma$ for Region 1 and 2$\sigma$ for Region 2), and the slight differences do not affect our conclusions. We can be confident that the host galaxy stellar mass is $M_{*,\rm{host}}\simeq3\times10^9M_\odot$, with a broadly similar mass for the companion galaxy.

We end this section by comparing the results from this work and that from \citet{Yue2023}. \citet{Yue2023} performed SED fitting using only the F200W and F356W magnitudes of the entire extended emission, which included both the quasar host and the companion. They assumed a delayed-$\tau$ SFH and inferred a stellar mass of $M_*=\left(6.5\substack{+4.4\\-3.3}\right)\times10^9M_\odot$. 
Our host galaxy stellar mass estimate $M_{*,\rm{host}}=\left(3.0\substack{+2.5\\-1.4}\right)\times10^9M_\odot$, and the combined total system mass  $M_{*,\rm{host}}+M_{*,\rm{companion}}=\left(7.6\substack{+3.0\\-2.1}\right)\times10^9M_\odot$ 
are consistent with this original estimate, although our measurements have lower uncertainty due to the improved SED fitting from the addition of the NIRSpec IFU data.
In this work, we show that a single delayed-$\tau$ model does not adequately describe the observations when we combine broad-band photometry and emission-line fluxes, and demonstrate that the SFH likely contains a starburst component. We are also able to constrain the metallicities of the quasar host galaxy and its companion.
These comparisons emphasise the importance of emission-line measurements in understanding the co-evolution of high-redshift BHs and their host galaxies.

\section{Discussion}
\label{sec:Discussion}
\subsection{Mass distribution in the system: Presence of an overmassive black hole }

Using the PdBI, \citet{Venemans2012} detected the \cii\ emission-line and FIR-continuum luminosity of the host of \qso,
unresolved at the 2$''$ resolution. They estimated a  SFR between 160--440 $M_\odot$/yr, a total dust mass of 0.67--$5.7\times10^8M_\odot$, and a dynamical mass of $M_{\rm{dyn}}<3.6\times10^{10} \sin(i)^{-2}M_\odot$.
ALMA observations at higher $0\farcs23$ resolution improved on these measurements, although the emission is still only marginally resolved \citep{Venemans2017a}.
They estimated a SFR of 105--340 $M_\odot$/yr, a dust mass of $(0.8$--$4)\times10^8M_\odot$, and an upper limit of the dynamical mass of $M_{\rm{dyn}}<(4.3\pm0.9)\times10^{10}M_\odot$.
From upper limits on the CO(2--1) and CO(7--6) lines, \citet{Venemans2017a} estimate upper limits of the molecular gas mass of $M_{\rm{gas}}<4\times10^{10}M_\odot$.
With higher resolution observations allowing for more accurate dynamical modelling, these mass estimates could likely be improved. 
However, it is reasonable to conservatively assume that the host total gas mass and dynamical mass are $<4\times10^{10}M_\odot$, with a dust mass of $M_{\rm{dust}}<10^{9}M_\odot$.
As we show in Section \ref{sec:ALMA}, the majority of the \cii\ and FIR continuum emission is from the host galaxy and not the companion, and so these measurements are unlikely to be significantly biased by the presence of the previously unknown companion.
From our \oiii\ emission-line maps, we estimate the quasar host to have a dynamical mass of $M_{\rm{dyn}}=\left(1.3\substack{ +1.9\\-1.1 }\right)\times10^{10}M_\odot$, with the companion galaxy having $M_{\rm{dyn}}=\left(1.8\substack{ +2.1\\-1.3}\right)\times10^{10}M_\odot$. These are consistent with the dynamical mass and gas mass upper limits from ALMA.

From this work we estimate a stellar mass for the host galaxy of $M_*=\left(3.0\substack{+2.5\\-1.4}\right)\times10^9M_\odot$.
In the original \citet{Yue2023} work, they estimated a host stellar mass of $\left(6.5\substack{+4.4\\-3.3}\right) \times 10^9 M_\odot$. The primary differences in these estimations are due to an improved understanding of the system morphology (i.e. the presence of two separate galaxies instead of a single host), and an improved exponential decay plus starburst SFH model in the SED fitting. Overall, we can be confident that the host stellar mass is $10^9\lesssim M_\ast \lesssim 10^{10}M_\odot$.
Thus, the stellar mass is greater than the dust mass of $M_{\rm{dust}}<10^{9}M_\odot$, and less than our best dynamical mass estimate of $M_{\rm{dyn}}=\left(1.3\substack{ +1.9\\-1.1 }\right)\times10^{10}M_\odot$.
If we assume that the gas mass can be estimated as $M_{\rm{gas}}=M_{\rm{dyn}}-M_\ast$, we estimate $M_{\rm{gas}}\simeq\left(0.9\substack{ +1.9\\-0.9}\right)\times10^{10}M_\odot$,
consistent with the upper limit of the gas mass from ALMA of $M_{\rm{gas}}<4\times10^{10}M_\odot$.
This gives an estimate of the gas fraction of $M_{\rm{gas}}/M_{\rm{dyn}}\simeq0.76$, while the dust fraction is $M_{\rm{dust}}/M_{\rm{dyn}}<0.08$.

For \qso, we can now be confident in the accuracy of the measured BH mass, with many independent estimates including three studies based on the broad hydrogen lines from the JWST observations of \citet{Yue2023}, \citet{Bosman2024} and this work. As discussed in Section \ref{sec:BHmass}, we conclude that $M_{\rm{BH}}\simeq1$--$2\times10^9M_\odot$. The estimates based on hydrogen lines are $M_{\rm{BH}}=(0.87$--$1.9)\times10^{9}M_\odot$, including our best estimate $M_{\rm{BH}}=\left(1.9\substack{+0.4\\-0.3}\right)\times10^9 M_\odot$. The largest estimate is from \ions{C}{iv}, $M_{\rm{BH}}=2.4\times10^{9}\,{M}_{\odot }$ \citep{Farina2022}.

With a stellar mass estimate of $M_*=\left(3.0\substack{+2.5\\-1.4}\right)\times10^9M_\odot$, and hydrogen-based BH mass estimates of $M_{\rm{BH}}=(0.87$--$1.9)\times10^{9}M_\odot$, the BH--stellar mass ratio for \qso\ is very high. 
For the minimum BH mass of $M_{\rm{BH}}=\left(0.87\substack{+0.21\\ -0.17}\right)\times10^{9}M_\odot$ from Paschen-$\beta$ \citep{Bosman2024}, $M_{\rm{BH}}/M_\ast = 0.29\substack{+0.25\\ -0.15}$.
The estimate of $M_{\rm{BH}}=(1.55\pm 0.22)\times10^9 M_\odot$ from \citet{Bosman2024}, the largest hydrogen-based estimate excluding our measurement, corresponds to a ratio of $M_{\rm{BH}}/M_\ast = 0.52\substack{+0.44\\ -0.25}$. 
For our \hb\ BH mass estimate of $M_{\rm{BH}}=\left(1.9\substack{+0.4\\-0.3}\right)\times10^9 M_\odot$, we get a ratio of $M_{\rm{BH}}/M_\ast = 0.63\substack{+0.54\\ -0.31}$.  Including the uncertainty in the BH mass scaling relations of 0.43 dex, which is not included in the \citet{Bosman2024} measurements, we obtain a ratio of $M_{\rm{BH}}/M_\ast = 0.63\substack{+1.10\\ -0.47}$. 

\begin{figure}
\begin{center}
\includegraphics[scale=0.8]{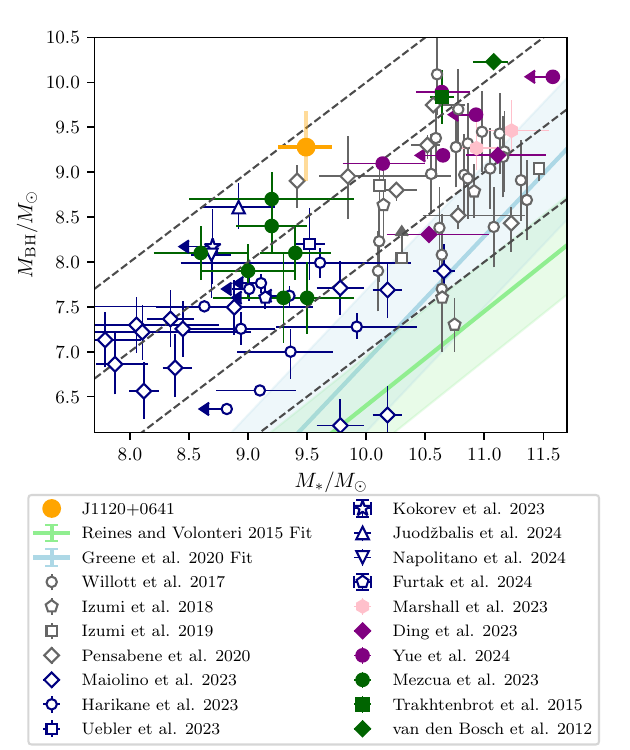}
\caption{Black hole--stellar mass relation showing the overmassive BH in \qso\ (gold). 
Purple points show $z\gtrsim6$ quasars with stellar mass estimates from JWST \citep{Ding2023,Yue2023}. Pink points show dynamical mass measurements for $z\simeq6$ quasars as measured with JWST \citep{Marshall2023}, while grey points show dynamical mass measurements as measured with ALMA \citep{Willott2017,Izumi2018,Izumi2019,Pensabene2020}. 
Blue points show measurements for high-$z$ AGNs discovered with JWST \citep{Maiolino2023,Harikane2023,Uebler2023,Kokorev2023,Juodzbalis2024,Furtak2024,Napolitano2024}.
Green points show samples of low-$z$ sources with large $M_{\rm{BH}}/M_\ast$ ratios \citep{Mezcua2023,Trakhtenbrot2015,VanDenBosch2012}.
The grey dashed lines depict $M_{\rm{BH}}/M_\ast$ ratios of 0.1\%, 1\%, 10\%, and 100\%.
The coloured curves show the local \citet{Reines2015} and \citet{Greene2020} relations, with the shaded regions showing the $\pm1\sigma$ scatter in the relations, for comparison.
}
\label{fig:BHstellarMass}
\end{center}
\end{figure}

We plot the BH--stellar mass relation in Figure \ref{fig:BHstellarMass}, showing this best estimate for \qso, alongside a range of data for comparison. The BH--stellar mass ratio is clearly much larger than  expected from the local BH-stellar mass relations of \citet{Reines2015} and \citet{Greene2020}. 
Applying the \citet{Greene2020} scaling relation of $M_{\rm{BH}}$--$M_\ast$ to a stellar mass of $M_*=\left(3.0\substack{+2.5\\-1.4}\right)\times10^9M_\odot$, the BH mass would be estimated as $M_{\rm{BH}}=\left(1.5\substack{+10.4 \\ -0.2}\right)\times10^6M_\odot$, with $M_{\rm{BH}}/M_\ast =0.0005\substack{+0.02 \\ -0.0005}$.
Instead applying the \citet{Reines2015} $M_{\rm{BH}}$--$M_{\ast}$ scaling relation, the BH mass would be estimated as $M_{\rm{BH}}=\left(0.71\substack{+2.75 \\ -0.18}\right)\times10^6M_\odot$, with $M_{\rm{BH}}/M_\ast = 0.00024\substack{+0.0021\\ -0.00018}$.
Clearly, our BH--stellar mass ratio of $M_{\rm{BH}}/M_\ast = 0.63\substack{+0.54\\ -0.31}$ is much larger than these values, by order $\sim3$ dex, indicating that \qso\ has a very overmassive BH. There is significant scatter in the \citet{Reines2015} and \citet{Greene2020} relations. Given the conservative uncertainty in our measured stellar and BH mass measurements, we find that \qso\ is 2.7$\sigma$ above the \citet{Greene2020} relation, and 3.8$\sigma$ above the \citet{Reines2015} relation.

Such an overmassive BH relative to its host galaxy is not unprecedented at high-$z$, as seen in Figure \ref{fig:BHstellarMass}.
Observations with JWST have been finding overmassive BHs at high-$z$.
The quasar J0148+0600 at $z=5.98$ has a BH--stellar mass ratio of $\sim0.18$ \citep{Yue2023}. 
\citet{Kokorev2023} discovered an AGN at $z=8.50$ with a BH--stellar mass ratio of $>0.3$.
A $z=6.68$ AGN was measured to have an extreme BH--stellar mass ratio of $\sim0.4$ \citep{Juodzbalis2024}.
\citet{Maiolino2023} have found a range of low BH-mass ($M_{\rm{BH}}<10^{7.5}M_\odot$) $4<z<11$ sources with BH--stellar mass ratios of $>0.1$.
At lower redshifts, \citet{Trakhtenbrot2015} found an AGN in a typical star-forming galaxy at $z=3.3$ with $M_{\rm{BH}}/M_\ast=0.12$.
\citet{Mezcua2023} found that multiple broad-line AGNs in dwarf galaxies at $z<1$ have $M_{\rm{BH}}/M_\ast>0.1$, with the largest two in the sample having $M_{\rm{BH}}/M_\ast=0.32$. 
\citet{VanDenBosch2012} measured the local disc galaxy NGC 1277 to have $M_{\rm{BH}}/M_\ast=0.14$. 
Objects with such large $M_{\rm{BH}}/M_\ast$ are rare, and so while \qso\ may have the largest reported $M_{\rm{BH}}/M_\ast$ ratio to date of $\simeq0.63$, this extreme ratio is not unprecedented.

Given the number of independent observations finding that $M_{\rm{BH}}\simeq(1-2)\times10^9M_\odot$, the most likely way for us to have overestimated the BH--stellar mass ratio is if we have significantly underestimated the stellar mass. The bright quasar makes it significantly difficult to observe the host galaxy, with some non-detections in the NIRCam photometry. If the quasar was hiding a very bright, massive galaxy that was undetectable due to the difficulty in quasar subtraction, this could alleviate the very high BH--stellar mass ratio measured here. 
However, we would not expect a larger stellar mass than the dynamical mass we have measured, $M_{\ast}<M_{\rm{dyn}}=\left(1.3\substack{ +1.9\\-1.1 }\right)\times10^{10}M_\odot$. 
Thus the ratio $M_{\rm{BH}}/M_{\ast}>M_{\rm{BH}}/M_{\rm{dyn}}=0.15\substack{+0.22\\-0.13}$, which further supports the BH being overmassive relative to its host.
If we assume that our dynamical mass estimate is also biased due to the difficulty of the quasar subtraction, and instead use the ALMA dynamical mass estimate, at wavelengths where the quasar contribution is negligible, we have $M_{\ast}<M_{\rm{dyn,ALMA}}<(4.3\pm0.9)\times10^{10}M_\odot$, or $M_{\rm{BH}}/M_{\ast}>0.044$. This would be a very conservative underestimate, as the dynamical mass has a significant contribution from the gas component of the galaxy.
We could also consider the `galaxy stellar mass' to be that of the host and companion galaxy combined: $M_\ast=M_{*,\rm{host}}+M_{*,\rm{companion}}=\left(5.7\substack{+3.0\\-1.9}\right)\times10^9M_\odot$, and so $M_{\rm{BH}}/M_{\ast}=0.33\substack{ +0.19\\-0.12}$.
These various host mass estimates all support this being an overmassive BH.
An alternative way for these ratios to be overestimated is if the BH masses are systematically overestimated. The BH mass scaling relations Equations \ref{eq:MBH_5100} and \ref{eq:MBH_HB} are calibrated using low-$z$ AGNs, and so these may not accurately describe high-$z$ sources; if these scaling relations overestimate BH mass at high-$z$, this could also reduce the BH--stellar mass ratio for this system. A recent dynamical BH mass measurement at $z=2.3$ found that the \ha-based scaling relation overestimated the BH mass by 0.43 dex \citep{Abuter2024}, however, this is consistent with the known scatter in the relation that our uncertainties account for.

\subsection{Comparison with emission seen with ALMA}
\label{sec:ALMA}
We obtained the archival data from the \citet{Venemans2017a} ALMA observations (ALMA project code 2012.1.00882.S) to compare their \cii\ properties to our measurements.
We made a \cii\ integrated intensity (moment 0) map with a natural weighting and resolution of 0\farcs3 as well as a FIR continuum intensity map of the 158$\mu$m rest-frame dust emission. These are overlaid onto our \oiiia\ map in Figure \ref{fig:ALMAoverlay}. This \cii\ emission matches the spatial location of the \oiiia\ emission from the host galaxy well, although shows a more symmetrical, round structure than the elongated \oiiia\ system.
No \cii\ emission is seen from the companion galaxy at this sensitivity.
The FIR continuum emission follows a very similar shape, with no emission from the companion galaxy detected.

Based on the \cii\ line detection of the quasar host, with flux $F_{\rm{{[CII]}}}=1.11\pm0.1$ Jy km/s and $L_{\rm{{[CII]}}}=(1.5\pm0.1)\times10^9L_\odot$, \citet{Venemans2017a} derive a SFR of 80-500 $M_\odot$/yr. This uses the high-redshift $L_{\rm{{[CII]}}}$--SFR scaling relation of \citet{DeLooze2014}.
Given the non-detection of the companion in the ALMA data, the companion galaxy must have a lower SFR than the quasar host. 
Following the \citet{Venemans2017a} methodology, and using their quoted RMS noise level of $0.042$ Jy km/s, for the companion galaxy we estimate $L_{\rm{{[CII]}}}<1.75\times10^8L_\odot$ at the $3\sigma$ level, which corresponds to SFR $<40M_\odot$/yr.
If the companion galaxy is on the star forming main sequence, with stellar mass $M_{*,\rm{companion}}=\left(2.7\substack{+0.5\\-0.5}\right)\times10^9M_\odot$ it would have a SFR of $15\substack{+5\\-4}\,M_\odot$/yr \citep{Popesso2023}, below the ALMA detection limit. Thus the ALMA non-detection only confirms that, unlike the quasar host, the companion galaxy is not an extreme star forming galaxy above the main sequence.

We also estimated the SFRs from the integrated flux of the narrow \hb\ line (Table \ref{tab:HostFlux}) using 
\begin{equation}
\textrm{SFR} = 1.53\times 10^{-41} \frac{L_{\rm{H}\beta}}{(\textrm{erg/s})} \rm{M}_\odot / \rm{yr}
\label{eq:Kenn}
\end{equation}
\citep{Kennicutt2012}.
This conversion assumes no dust extinction, a solar abundance, and a Kroupa IMF, and uses the theoretical conversion from \ha\ to \hb\ flux of $F_{\rm{H}\alpha}/F_{\rm{H}\beta}=2.86$, which assumes a temperature $T = 10^4$ K and an electron density $n_e = 10^2 \rm{cm}^{-3}$ for Case B recombination \citep{Osterbrock1989,Dominguez2013}. If we assume that all of the quasar-subtracted \hb\ narrow-line flux measured in Table \ref{tab:HostFlux} is from star formation, with no contribution from quasar photoionisation, we obtain SFRs of $<20\,M_\odot$/yr for the host galaxy, and $3.6\,M_\odot$/yr for the companion galaxy. 
For the quasar host, the \hb\ SFR is much lower than the \cii-estimated value of 80-500 $M_\odot$/yr. Some of this discrepancy is likely explained by \hb\ and \cii\ tracing star formation on different time scales. In addition, \hb\ will be affected by dust-attenuation, which we have not corrected for---\hb\ would give a lower estimate of the SFR as the \cii\ measure includes dust-obscured star formation. 
With a very low \hb\ SFR of $3.6\,M_\odot$/yr, the companion galaxy is either significantly below the star forming main sequence or has substantial dust-obscured star formation. The ALMA limit of $<40M_\odot$/yr does not differentiate between these two possibilities.

This major merger may be triggering both a starburst and AGN activity in the quasar host galaxy. Theory predicts that galaxy mergers can induce AGN activity, due to torques funnelling gas into the centre of the galaxy \citep{Hernquist1995}, as well as star formation episodes \citep[e.g.][]{Doyon1994,Mihos1996}. In gas-rich mergers, an AGN phase followed by a starburst is a common evolutionary sequence \citep{Hopkins2006,Melnick2015}.

\subsection{Location of the black hole relative to the gas and stellar components}
\label{sec:offset}
When comparing the spatial location of the quasar to the full structure of this system from PSF-subtracted emission maps in Figures \ref{fig:ALMAoverlay} and \ref{fig:RegionMaps}, the quasar is offset to the north-east of the centre of the emission. However, the detail of the IFU observations reveals that this is because there is a separate companion galaxy in the south-west, and that the quasar is indeed at the centre of its host galaxy, as typically expected.
From the NIRCam imaging the quasar similarly appeared offset from the peak of the continuum emission; without the information gained from the IFU, this could be incorrectly interpreted as the quasar being offset from the centre of its host galaxy.

GA-NIFS observations of \DELS\ and \VDES\ at $z=6.8$ also revealed that these quasars are undergoing mergers \citep{Marshall2023}. For \DELS\ there are emission-line structures on both sides of the quasar host, meaning that the quasar still appears in the centre of the structure by chance. The companion galaxies around \VDES\ appear only to its south and west, and so without the detail of the IFU observations this quasar would appear to be offset from the centre of the system. Crucially,  the detailed spectral decomposition provided by the IFU reveals that all of the GA-NIFS quasars analysed so far are located approximately at the centre of their host galaxies.

In general, a physical offset of the BH from the galaxy centre is possible. Observations have found a range of AGNs that are offset from their host's centres on scales of up to tens of kiloparsecs \citep[e.g.][]{Clements2009,Koss2014,Liu2024}.
These offset BHs may be in the process of merging into the central nucleus \citep[e.g.][]{Uebler2024} or may be recoiling, due to the emission of gravitational waves or a slingshot effect from triple BH systems \citep[e.g.][]{Komossa2008,Civano2010,Chiaberge2025}.
However, if a high-$z$ quasar has an apparent spatial offset from the peak of observed galaxy emission, we caution this from being attributed to a potential offset BH from the centre of its host; the presence of close companions and irregular galaxy structures due to interactions may be a more likely option given the quasars observed with the IFU so far.

\section{Conclusions}
\label{sec:Conclusions}
In this paper we have presented the GA-NIFS NIRSpec IFU observations of the $z=7.08$ quasar \qso. This gives spatially resolved imaging spectroscopy covering the $3''\times3''$ FOV around the quasar from 2.9--5.3$\mu$m, which covers the key emission lines \hb\ and \oiii\ for this system.
From the integrated quasar spectrum, we measured the BH mass and Eddington ratio from the broad \hb\ line to be $M_{\rm{BH}}=\left(1.9\substack{+0.4\\-0.3}\right)\times10^9 M_\odot$ and $\lambda_{\rm{Edd}}=0.7\pm0.1$. These values are $M_{\rm{BH}}=\left(1.9\substack{+2.9\\-1.1}\right)\times10^9 M_\odot$ and  $\lambda_{\rm{Edd}}=0.7\substack{ +1.1\\-0.4}$ when considering the scatter in the BH scaling relations, which are consistent with previous JWST-based estimates.

We studied the extended host galaxy emission in \oiii\ and \hb\ by removing the quasar emission from the data cube.
We discovered a companion galaxy interacting with the quasar host, $\sim$0\farcs5 south-west of the quasar, with a velocity offset of $21\pm49$ km/s. 
The small projected distance to the quasar host and the low relative velocity imply that both galaxies form a merging system. 
Assuming a rotating disc geometry for both galaxies, we estimate dynamical masses of $M_{\rm{dyn,host}}=\left(1.3\substack{ +1.9\\-1.1}\right)\times10^{10}M_\odot$ for the quasar host and $M_{\rm{dyn,companion}}=\left(1.8\substack{ +2.1\\-1.3}\right)\times10^{10}M_\odot$ for the companion galaxy. These estimates indicate that both galaxies have similar masses, and therefore they will experience a `major' merger event.

By combining the IFU data with NIRCam imaging from the EIGER programme \citep{Yue2023}, we re-estimated the stellar mass of the host galaxy.
We fixed the modelled continuum morphology to match the separate host and companion galaxy morphologies measured from our \oiiia\ map, and we used the measured \hb\ and \oiii\ line fluxes in the SED fitting. We estimate the host stellar mass to be $M_{*,\rm{host}}=\left(3.0\substack{+2.5\\-1.4}\right)\times10^9M_\odot$, with $M_{*,\rm{companion}}=\left(2.7\substack{+0.5\\-0.5}\right)\times10^9M_\odot$ for the companion galaxy, which is consistent with this being a major merger system.
The SED fitting favours a model where both galaxies have undergone a recent starburst.
The addition of the IFU data provides a more accurate host stellar mass measurement than the original \cite{Yue2023} estimate of $M_\ast=\left(6.5\substack{+4.5 \\ -3.3}\right)\times10^{9}M_\odot$.

With a stellar mass estimate of $M_{*,\rm{host}}=\left(3.0\substack{+2.5\\-1.4}\right)\times10^9M_\odot$ and our \hb\ BH mass estimate of $M_{\rm{BH}}=\left(1.9\substack{+0.4\\-0.3}\right)\times10^9 M_\odot$, we measure an extreme BH--stellar mass ratio of $M_{\rm{BH}}/M_\ast = 0.63\substack{+0.54\\ -0.31}$.
\qso\ has an overmassive BH relative to its host galaxy. Notably, it is the largest BH--stellar mass ratio measured to date.
This ratio is $\sim3$ dex larger than expected from the local \citet{Reines2015} and \citet{Greene2020} scaling relations, but such an overmassive BH is not unprecedented at these redshifts \citep{Yue2023,Kokorev2023,Maiolino2023,Juodzbalis2024}.

Overall, this work highlights the power of using both NIRSpec IFU and NIRCam imaging together to understand high-$z$ quasars, their host galaxies, and local environments. The IFU allows for measurements of BH mass as well as key galaxy emission lines. While the spatial sampling is coarser than that of NIRCam images, the relative ease of IFU quasar subtraction makes it easier to see the structure of the host galaxy and any companion galaxies, and it also provides kinematics. However, the high spectral resolution IFU mode is not sensitive to the faint continuum emission of most high-redshift systems. In such cases, NIRCam imaging can detect the continuum emission, which can be used to determine the host stellar mass. With the NIRSpec IFU providing the spatial structure and emission-line fluxes to improve the SED fitting, combining these observations gives more accurate stellar mass estimates. Used in this way, the two JWST instruments can work in synergy to provide a deeper understanding of high-$z$ quasar hosts.

\begin{acknowledgements}
This work is based on observations made with the NASA/ESA/CSA James Webb Space Telescope. The data were obtained from the Mikulski Archive for Space Telescopes at the Space Telescope Science Institute, which is operated by the Association of Universities for Research in Astronomy, Inc., under NASA contract NAS 5-03127 for JWST. These observations are associated with program \#1263, as part of the Galaxy Assembly with NIRSpec Integral Field Spectroscopy GTO program, and program \#1243, as part of the Emission-line galaxies and Intergalactic Gas in the Epoch of Reionization GTO program.

We thank Ignas Juod\v{z}balis for helping with the compilation of BH--stellar mass measurements from the literature. We thank the referee for their helpful feedback.

MAM acknowledges support by the Laboratory Directed Research and Development program of Los Alamos National Laboratory under project number 20240752PRD1.
The project leading to this publication has received support from ORP, that is funded by the European Union’s Horizon 2020 research and innovation programme under grant agreement No 101004719 [ORP].
MP, SA and BRdP acknowledge grant PID2021-127718NB-I00 funded by the Spanish Ministry of Science and Innovation/State Agency of Research (MICIN/AEI/ 10.13039/501100011033).
JS, RM and FDE acknowledge support by the Science and Technology Facilities Council (STFC), from the ERC Advanced Grant 695671 "QUENCH".
JS and FDE acknowledge the UKRI Frontier Research grant RISEandFALL.
RM acknowledges funding from a research professorship from the Royal Society.
H\"U acknowledges funding by the European Union (ERC APEX, 101164796). Views and opinions expressed are however those of the authors only and do not necessarily reflect those of the European Union or the European Research Council Executive Agency. Neither the European Union nor the granting authority can be held responsible for them.
SC and GV acknowledge support from the European Union (ERC, WINGS,101040227).
AJB and GCJ acknowledge funding from the "FirstGalaxies" Advanced Grant from the European Research Council (ERC) under the European Union’s Horizon 2020 research and innovation programme (Grant agreement No. 789056).
DK acknowledges funding from JSPS KAKENHI Grant Number JP21K13956.

This research has made use of the Astrophysics Data System, funded by NASA under Cooperative Agreement 80NSSC21M00561, QFitsView \citep{Ott2012}, and SAOImageDS9, developed by Smithsonian Astrophysical Observatory.

This paper made use of Python packages and software
AstroPy \citep{Astropy2013},
jwst \citep{Bushouse2022},
Matplotlib \citep{Matplotlib2007},
NumPy \citep{Numpy2011},
Pandas \citep{reback2020pandas}, 
Photutils \citep{photutils},
Prospector \citep{Johnson2021},
psfMC \citep{psfMC},
Regions \citep{Bradley2022},
SciPy \citep{2020SciPy-NMeth},
Seaborn \citep{Waskom2021},
Spectral Cube \citep{Ginsburg2019}, 
QDeblend3D \citep{Husemann2013,Husemann2014}, 
QubeSpec\footnote{\url{https://github.com/honzascholtz/Qubespec}},
and
WebbPSF \citep{Perrin2015}.
 
\end{acknowledgements}

\bibliography{aa52650-24}{}

\begin{thebibliography}{190}
\expandafter\ifx\csname natexlab\endcsname\relax\def\natexlab#1{#1}\fi

\bibitem[{Abazajian {et~al.}(2009)Abazajian, Adelman-McCarthy, Agüeros, Allam, Prieto, An, Anderson, Anderson, Annis, Bahcall, Bailer-Jones, Barentine, Bassett, Becker, Beers, Bell, Belokurov, Berlind, Berman, Bernardi, Bickerton, Bizyaev, Blakeslee, Blanton, Bochanski, Boroski, Brewington, Brinchmann, Brinkmann, Brunner, Budav{\'{a}}ri, Carey, Carliles, Carr, Castander, Cinabro, Connolly, Csabai, Cunha, Czarapata, Davenport, de~Haas, Dilday, Doi, Eisenstein, Evans, Evans, Fan, Friedman, Frieman, Fukugita, Gänsicke, Gates, Gillespie, Gilmore, Gonzalez, Gonzalez, Grebel, Gunn, Györy, Hall, Harding, Harris, Harvanek, Hawley, Hayes, Heckman, Hendry, Hennessy, Hindsley, Hoblitt, Hogan, Hogg, Holtzman, Hyde, ichi Ichikawa, Ichikawa, Im, Ivezi{\'{c}}, Jester, Jiang, Johnson, Jorgensen, Juri{\'{c}}, Kent, Kessler, Kleinman, Knapp, Konishi, Kron, Krzesinski, Kuropatkin, Lampeitl, Lebedeva, Lee, Lee, Leger, L{\'{e}}pine, Li, Lima, Lin, Long, Loomis, Loveday, Lupton, Magnier, Malanushenko, Malanushenko, Mandelbaum,
  Margon, Marriner, Mart{\'{\i}}nez-Delgado, Matsubara, McGehee, McKay, Meiksin, Morrison, Mullally, Munn, Murphy, Nash, Nebot, Neilsen, Newberg, Newman, Nichol, Nicinski, Nieto-Santisteban, Nitta, Okamura, Oravetz, Ostriker, Owen, Padmanabhan, Pan, Park, Pauls, Peoples, Percival, Pier, Pope, Pourbaix, Price, Purger, Quinn, Raddick, Fiorentin, Richards, Richmond, Riess, Rix, Rockosi, Sako, Schlegel, Schneider, Scholz, Schreiber, Schwope, Seljak, Sesar, Sheldon, Shimasaku, Sibley, Simmons, Sivarani, Smith, Smith, Smol{\v{c}}i{\'{c}}, Snedden, Stebbins, Steinmetz, Stoughton, Strauss, SubbaRao, Suto, Szalay, Szapudi, Szkody, Tanaka, Tegmark, Teodoro, Thakar, Tremonti, Tucker, Uomoto, Berk, Vandenberg, Vidrih, Vogeley, Voges, Vogt, Wadadekar, Watters, Weinberg, West, White, Wilhite, Wonders, Yanny, Yocum, York, Zehavi, Zibetti, \& Zucker}]{Abazajian2009}
Abazajian, K.~N., Adelman-McCarthy, J.~K., Agüeros, M.~A., {et~al.} 2009, ApJS, 182, 543

\bibitem[{{Abuter} {et~al.}(2024){Abuter}, {Allouche}, {Amorim}, {Bailet}, {Berdeu}, {Berger}, {Berio}, {Bigioli}, {Boebion}, {Bolzer}, {Bonnet}, {Bourdarot}, {Bourget}, {Brandner}, {Cao}, {Conzelmann}, {Comin}, {Cl{\'e}net}, {Courtney-Barrer}, {Davies}, {Defr{\`e}re}, {Delboulb{\'e}}, {Delplancke-Str{\"o}bele}, {Dembet}, {Dexter}, {de Zeeuw}, {Drescher}, {Eckart}, {{\'E}douard}, {Eisenhauer}, {Fabricius}, {Feuchtgruber}, {Finger}, {F{\"o}rster Schreiber}, {Garcia}, {Garcia Lopez}, {Gao}, {Gendron}, {Genzel}, {Gil}, {Gillessen}, {Gomes}, {Gont{\'e}}, {Gouvret}, {Guajardo}, {Guieu}, {Hackenberg}, {Haddad}, {Hartl}, {Haubois}, {Hau{\ss}mann}, {Hei{\ss}el}, {Henning}, {Hippler}, {H{\"o}nig}, {Horrobin}, {Hubin}, {Jacqmart}, {Jocou}, {Kaufer}, {Kervella}, {Kolb}, {Korhonen}, {Lacour}, {Lagarde}, {Lai}, {Lapeyr{\`e}re}, {Laugier}, {Le Bouquin}, {Leftley}, {L{\'e}na}, {Lewis}, {Liu}, {Lopez}, {Lutz}, {Magnard}, {Mang}, {Marcotto}, {Maurel}, {M{\'e}rand}, {Millour}, {More}, {Netzer}, {Nowacki}, {Nowak}, {Oberti},
  {Ott}, {Pallanca}, {Paumard}, {Perraut}, {Perrin}, {Petrov}, {Pfuhl}, {Pourr{\'e}}, {Rabien}, {Rau}, {Riquelme}, {Robbe-Dubois}, {Rochat}, {Salman}, {Sanchez-Bermudez}, {Santos}, {Scheithauer}, {Sch{\"o}ller}, {Schubert}, {Schuhler}, {Shangguan}, {Shchekaturov}, {Shimizu}, {Sevin}, {Soulez}, {Spang}, {Stadler}, {Sternberg}, {Straubmeier}, {Sturm}, {Sykes}, {Tacconi}, {Tristram}, {Vincent}, {von Fellenberg}, {Uysal}, {Widmann}, {Wieprecht}, {Wiezorrek}, {Woillez}, \& {Zins}}]{Abuter2024}
{Abuter}, R., {Allouche}, F., {Amorim}, A., {et~al.} 2024, \nat, 627, 281

\bibitem[{{Anderson} {et~al.}(2021){Anderson}, {Fall}, \& {Astrometry Working Group}}]{Anderson2021}
{Anderson}, J., {Fall}, S.~M., \& {Astrometry Working Group}. 2021, {The JWST Calibration Field: Absolute Astrometry and Proper Motions with GAIA and a Second HST Epoch}, Technical Report JWST-STScI-007716

\bibitem[{{Astropy Collaboration} {et~al.}(2013){Astropy Collaboration}, Robitaille, Tollerud, Greenfield, Droettboom, Bray, Aldcroft, Davis, Ginsburg, Price-Whelan, Kerzendorf, Conley, Crighton, Barbary, Muna, Ferguson, Grollier, Parikh, Nair, Günther, Deil, Woillez, Conseil, Kramer, Turner, Singer, Fox, Weaver, Zabalza, Edwards, Bostroem, Burke, Casey, Crawford, Dencheva, Ely, Jenness, Labrie, Lim, Pierfederici, Pontzen, Ptak, Refsdal, Servillat, \& Streicher}]{Astropy2013}
{Astropy Collaboration}, Robitaille, T.~P., Tollerud, E.~J., {et~al.} 2013, A{\&}A, 558, A33

\bibitem[{Baldwin {et~al.}(1981)Baldwin, Phillips, \& Terlevich}]{Baldwin1981}
Baldwin, J.~A., Phillips, M.~M., \& Terlevich, R. 1981, PASP, 93, 5

\bibitem[{Ba{\~n}ados {et~al.}(2015)Ba{\~n}ados, {Decarli}, {Walter}, {Venemans}, {Farina}, \& {Fan}}]{Banados2015}
Ba{\~n}ados, E., {Decarli}, R., {Walter}, F., {et~al.} 2015, \apjl, 805, L8

\bibitem[{Ba{\~{n}}ados {et~al.}(2022)Ba{\~{n}}ados, Schindler, Venemans, Connor, Decarli, Farina, Mazzucchelli, Meyer, Stern, Walter, Fan, Hennawi, Khusanova, Morrell, Nanni, Noirot, Pensabene, Rix, Simon, Kleijn, Xie, Yang, \& Connor}]{Banados2022}
Ba{\~{n}}ados, E., Schindler, J.-T., Venemans, B.~P., {et~al.} 2022, ApJS, 265, 29

\bibitem[{Ba{\~{n}}ados {et~al.}(2016)Ba{\~{n}}ados, Venemans, Decarli, Farina, Mazzucchelli, Walter, Fan, Stern, Schlafly, Chambers, Rix, Jiang, McGreer, Simcoe, Wang, Yang, Morganson, Rosa, Greiner, Balokovi{\'{c}}, Burgett, Cooper, Draper, Flewelling, Hodapp, Jun, Kaiser, Kudritzki, Magnier, Metcalfe, Miller, Schindler, Tonry, Wainscoat, Waters, \& Yang}]{Banados2016}
Ba{\~{n}}ados, E., Venemans, B.~P., Decarli, R., {et~al.} 2016, ApJS, 227, 11

\bibitem[{Ba{\~{n}}ados {et~al.}(2018)Ba{\~{n}}ados, Venemans, Mazzucchelli, Farina, Walter, Wang, Decarli, Stern, Fan, Davies, Hennawi, Simcoe, Turner, Rix, Yang, Kelson, Rudie, \& Winters}]{Banados2017}
Ba{\~{n}}ados, E., Venemans, B.~P., Mazzucchelli, C., {et~al.} 2018, Nat, 553, 473

\bibitem[{Bennert {et~al.}(2011)Bennert, Auger, Treu, Woo, \& Malkan}]{Bennert2011}
Bennert, V.~N., Auger, M.~W., Treu, T., Woo, J.-H., \& Malkan, M.~A. 2011, ApJ, 726, 59

\bibitem[{Bertoldi {et~al.}(2003)Bertoldi, Cox, Neri, Carilli, Walter, Omont, Beelen, Henkel, Fan, Strauss, \& Menten}]{Bertoldi2003}
Bertoldi, F., Cox, P., Neri, R., {et~al.} 2003, A\&A, 409, L47

\bibitem[{{Boroson} \& {Green}(1992)}]{BG92}
{Boroson}, T.~A. \& {Green}, R.~F. 1992, \apjs, 80, 109

\bibitem[{{Bosman} {et~al.}(2024){Bosman}, {{\'A}lvarez-M{\'a}rquez}, {Colina}, {Walter}, {Alonso-Herrero}, {Ward}, {{\"O}stlin}, {Greve}, {Wright}, {Bik}, {Boogaard}, {Caputi}, {Costantin}, {Eckart}, {Garc{\'\i}a-Mar{\'\i}n}, {Gillman}, {Hjorth}, {Iani}, {Ilbert}, {Jermann}, {Labiano}, {Langeroodi}, {Pei{\ss}ker}, {Rinaldi}, {Topinka}, {van der Werf}, {G{\"u}del}, {Henning}, {Lagage}, {Ray}, {van Dishoeck}, \& {Vandenbussche}}]{Bosman2024}
{Bosman}, S. E.~I., {{\'A}lvarez-M{\'a}rquez}, J., {Colina}, L., {et~al.} 2024, Nature Astronomy, 8, 1054–1065

\bibitem[{Bradley {et~al.}(2022)Bradley, Deil, Ginsburg, {Sushobhana Patra}, Robitaille, Sipőcz, King, {P. L. Lim}, Homeier, Singer, De~Val-Borro, Jenness, Baumann, {Yash Gondhalekar}, Donath, Tollerud, {Jae-Joon Lee}, Leinweber, \& {Zé Vinícius}}]{Bradley2022}
Bradley, L., Deil, C., Ginsburg, A., {et~al.} 2022, astropy/regions: v0.7

\bibitem[{Bradley {et~al.}(2018)Bradley, Sipocz, Robitaille, Vinícius, Tollerud, Deil, Barbary, Günther, Cara, Busko, Conseil, Droettboom, Bostroem, Bray, Bratholm, Craig, Barentsen, Pascual, Donath, Greco, Perren, Kerzendorf, de~Val-Borro, StuartLittlefair, Ogaz, Lim, Ferreira, D'Eugenio, \& Weaver}]{photutils}
Bradley, L., Sipocz, B., Robitaille, T., {et~al.} 2018, astropy/photutils: v0.5

\bibitem[{Bushouse {et~al.}(2022)Bushouse, Eisenhamer, Dencheva, Davies, Greenfield, Morrison, Hodge, Simon, Grumm, Droettboom, Slavich, Sosey, Pauly, Miller, Jedrzejewski, Hack, Davis, Crawford, Law, Gordon, Regan, Cara, MacDonald, Bradley, Shanahan, Jamieson, Teodoro, \& Williams}]{Bushouse2022}
Bushouse, H., Eisenhamer, J., Dencheva, N., {et~al.} 2022, JWST Calibration Pipeline

\bibitem[{{Byler} {et~al.}(2017){Byler}, {Dalcanton}, {Conroy}, \& {Johnson}}]{Byler2017}
{Byler}, N., {Dalcanton}, J.~J., {Conroy}, C., \& {Johnson}, B.~D. 2017, \apj, 840, 44

\bibitem[{Böker {et~al.}(2022)Böker, Arribas, Lützgendorf, de~Oliveira, Beck, Birkmann, Bunker, Charlot, de~Marchi, Ferruit, Giardino, Jakobsen, Kumari, L{\'{o}}pez-Caniego, Maiolino, Manjavacas, Marston, Moseley, Muzerolle, Ogle, Pirzkal, Rauscher, Rawle, Rix, Sabbi, Sargent, Sirianni, te~Plate, Valenti, Willott, \& Zeidler}]{Boeker2022}
Böker, T., Arribas, S., Lützgendorf, N., {et~al.} 2022, A\&A, 661, A82

\bibitem[{Böker {et~al.}(2023)Böker, Beck, Birkmann, Giardino, Keyes, Kumari, Muzerolle, Rawle, Zeidler, Abul-Huda, de~Oliveira, Arribas, Bechtold, Bhatawdekar, Bonaventura, Bunker, Cameron, Carniani, Charlot, Curti, Espinoza, Ferruit, Franx, Jakobsen, Karakla, L{\'{o}}pez-Caniego, Lützgendorf, Maiolino, Manjavacas, Marston, Moseley, Ogle, Perna, Pe{\~{n}}a-Guerrero, Pirzkal, Plesha, Proffitt, Rauscher, Rix, del Pino, Rustamkulov, Sabbi, Sing, Sirianni, te~Plate, {\'{U}}beda, Wahlgren, Wislowski, Wu, \& Willott}]{Boeker2023}
Böker, T., Beck, T.~L., Birkmann, S.~M., {et~al.} 2023, PASP, 135, 038001

\bibitem[{{Calzetti} {et~al.}(2000){Calzetti}, {Armus}, {Bohlin}, {Kinney}, {Koornneef}, \& {Storchi-Bergmann}}]{Calzetti2000}
{Calzetti}, D., {Armus}, L., {Bohlin}, R.~C., {et~al.} 2000, ApJ, 533, 682

\bibitem[{{Cameron} {et~al.}(2023){Cameron}, {Saxena}, {Bunker}, {D'Eugenio}, {Carniani}, {Maiolino}, {Curtis-Lake}, {Ferruit}, {Jakobsen}, {Arribas}, {Bonaventura}, {Charlot}, {Chevallard}, {Curti}, {Looser}, {Maseda}, {Rawle}, {Rodr{\'\i}guez Del Pino}, {Smit}, {{\"U}bler}, {Willott}, {Witstok}, {Egami}, {Eisenstein}, {Johnson}, {Hainline}, {Rieke}, {Robertson}, {Stark}, {Tacchella}, {Williams}, {Willmer}, {Bhatawdekar}, {Bowler}, {Boyett}, {Circosta}, {Helton}, {Jones}, {Kumari}, {Ji}, {Nelson}, {Parlanti}, {Sandles}, {Scholtz}, \& {Sun}}]{Cameron2023}
{Cameron}, A.~J., {Saxena}, A., {Bunker}, A.~J., {et~al.} 2023, \aap, 677, A115

\bibitem[{Cappellari {et~al.}(2006)Cappellari, Bacon, Bureau, Damen, Davies, Zeeuw, Emsellem, Falcon-Barroso, Krajnovic, Kuntschner, McDermid, Peletier, Sarzi, Bosch, \& Ven}]{Cappellari2006}
Cappellari, M., Bacon, R., Bureau, M., {et~al.} 2006, MNRAS, 366, 1126

\bibitem[{{Chabrier}(2003)}]{Chabrier2003}
{Chabrier}, G. 2003, \pasp, 115, 763

\bibitem[{{Chiaberge} {et~al.}(2025){Chiaberge}, {Morishita}, {Boschini}, {Bianchi}, {Capetti}, {Castignani}, {Gerosa}, {Konishi}, {Koyama}, {Kushibiki}, {Lambrides}, {Meyer}, {Motohara}, {Stiavelli}, {Takahashi}, {Tremblay}, \& {Norman}}]{Chiaberge2025}
{Chiaberge}, M., {Morishita}, T., {Boschini}, M., {et~al.} 2025, arXiv e-prints, arXiv:2501.18730

\bibitem[{{Civano} {et~al.}(2010){Civano}, {Elvis}, {Lanzuisi}, {Jahnke}, {Zamorani}, {Blecha}, {Bongiorno}, {Brusa}, {Comastri}, {Hao}, {Leauthaud}, {Loeb}, {Mainieri}, {Piconcelli}, {Salvato}, {Scoville}, {Trump}, {Vignali}, {Aldcroft}, {Bolzonella}, {Bressert}, {Finoguenov}, {Fruscione}, {Koekemoer}, {Cappelluti}, {Fiore}, {Giodini}, {Gilli}, {Impey}, {Lilly}, {Lusso}, {Puccetti}, {Silverman}, {Aussel}, {Capak}, {Frayer}, {Le Floch}, {McCracken}, {Sanders}, {Schiminovich}, \& {Taniguchi}}]{Civano2010}
{Civano}, F., {Elvis}, M., {Lanzuisi}, G., {et~al.} 2010, \apj, 717, 209

\bibitem[{{Clements} {et~al.}(2009){Clements}, {Petitpas}, {Farrah}, {Hatziminaoglou}, {Babbedge}, {Rowan-Robinson}, {P{\'e}rez-Fournon}, {Hern{\'a}n-Caballero}, {Castro-Rodr{\'\i}guez}, {Lonsdale}, {Surace}, {Franceschini}, {Wilkes}, \& {Smith}}]{Clements2009}
{Clements}, D.~L., {Petitpas}, G., {Farrah}, D., {et~al.} 2009, \apjl, 698, L188

\bibitem[{Coatman {et~al.}(2016)Coatman, Hewett, Banerji, \& Richards}]{Coatman2016}
Coatman, L., Hewett, P.~C., Banerji, M., \& Richards, G.~T. 2016, MNRAS, 461, 647

\bibitem[{Conselice {et~al.}(2009)Conselice, Yang, \& Bluck}]{Conselice2009}
Conselice, C.~J., Yang, C., \& Bluck, A. F.~L. 2009, MNRAS, 394, 1956

\bibitem[{Cresci {et~al.}(2015)Cresci, Mainieri, Brusa, Marconi, Perna, Mannucci, Piconcelli, Maiolino, Feruglio, Fiore, Bongiorno, Lanzuisi, Merloni, Schramm, Silverman, \& Civano}]{Cresci2015}
Cresci, G., Mainieri, V., Brusa, M., {et~al.} 2015, ApJ, 799, 82

\bibitem[{Croton(2006)}]{Croton2006b}
Croton, D.~J. 2006, MNRAS, 369, 1808

\bibitem[{{da Silva} {et~al.}(2011){da Silva}, {Prochaska}, {Rosario}, {Tumlinson}, \& {Tripp}}]{daSilva2011}
{da Silva}, R.~L., {Prochaska}, J.~X., {Rosario}, D., {Tumlinson}, J., \& {Tripp}, T.~M. 2011, \apj, 735, 54

\bibitem[{{De Looze} {et~al.}(2014){De Looze}, {Cormier}, {Lebouteiller}, {Madden}, {Baes}, {Bendo}, {Boquien}, {Boselli}, {Clements}, {Cortese}, {Cooray}, {Galametz}, {Galliano}, {Graci{\'a}-Carpio}, {Isaak}, {Karczewski}, {Parkin}, {Pellegrini}, {R{\'e}my-Ruyer}, {Spinoglio}, {Smith}, \& {Sturm}}]{DeLooze2014}
{De Looze}, I., {Cormier}, D., {Lebouteiller}, V., {et~al.} 2014, \aap, 568, A62

\bibitem[{{Decarli} {et~al.}(2019){Decarli}, {Dotti}, {Ba{\~n}ados}, {Farina}, {Walter}, {Carilli}, {Fan}, {Mazzucchelli}, {Neeleman}, {Novak}, {Riechers}, {Strauss}, {Venemans}, {Yang}, \& {Wang}}]{Decarli2019}
{Decarli}, R., {Dotti}, M., {Ba{\~n}ados}, E., {et~al.} 2019, \apj, 880, 157

\bibitem[{{Decarli} {et~al.}(2024){Decarli}, {Loiacono}, {Farina}, {Dotti}, {Lupi}, {Meyer}, {Mignoli}, {Pensabene}, {Strauss}, {Venemans}, {Yang}, {Walter}, {Wolf}, {Ba{\~n}ados}, {Blecha}, {Bosman}, {Carilli}, {Comastri}, {Connor}, {Costa}, {Eilers}, {Fan}, {Gilli}, {Jun}, {Liu}, {Marshall}, {Mazzucchelli}, {Neeleman}, {Onoue}, {Overzier}, {Pudoka}, {Riechers}, {Rix}, {Schindler}, {Trakhtenbrot}, {Trebitsch}, {Vestergaard}, {Volonteri}, {Wang}, {Zhang}, \& {Zou}}]{Decarli2024}
{Decarli}, R., {Loiacono}, F., {Farina}, E.~P., {et~al.} 2024, \aap, 689, A219

\bibitem[{{Decarli} {et~al.}(2017){Decarli}, {Walter}, {Venemans}, {Ba{\~n}ados}, {Bertoldi}, {Carilli}, {Fan}, {Farina}, {Mazzucchelli}, {Riechers}, {Rix}, {Strauss}, {Wang}, \& {Yang}}]{Decarli2017}
{Decarli}, R., {Walter}, F., {Venemans}, B.~P., {et~al.} 2017, Nature, 545, 457

\bibitem[{Decarli {et~al.}(2018)Decarli, Walter, Venemans, Ba{\~{n}}ados, Bertoldi, Carilli, Fan, Farina, Mazzucchelli, Riechers, Rix, Strauss, Wang, \& Yang}]{Decarli2018}
Decarli, R., Walter, F., Venemans, B.~P., {et~al.} 2018, ApJ, 854, 97

\bibitem[{Decarli {et~al.}(2012)Decarli, Walter, Yang, Carilli, Fan, Hennawi, Kurk, Riechers, Rix, Strauss, \& Venemans}]{Decarli2012}
Decarli, R., Walter, F., Yang, Y., {et~al.} 2012, ApJ, 756, 150

\bibitem[{{D'Eugenio} {et~al.}(2024){D'Eugenio}, {P{\'e}rez-Gonz{\'a}lez}, {Maiolino}, {Scholtz}, {Perna}, {Circosta}, {{\"U}bler}, {Arribas}, {B{\"o}ker}, {Bunker}, {Carniani}, {Charlot}, {Chevallard}, {Cresci}, {Curtis-Lake}, {Jones}, {Kumari}, {Lamperti}, {Looser}, {Parlanti}, {Rix}, {Robertson}, {Rodr{\'\i}guez Del Pino}, {Tacchella}, {Venturi}, \& {Willott}}]{D'Eugenio2024}
{D'Eugenio}, F., {P{\'e}rez-Gonz{\'a}lez}, P.~G., {Maiolino}, R., {et~al.} 2024, Nature Astronomy, 8, 1443

\bibitem[{{Di Matteo} {et~al.}(2005){Di Matteo}, Springel, \& Hernquist}]{Matteo2005}
{Di Matteo}, T., Springel, V., \& Hernquist, L. 2005, Nature, 433, 604

\bibitem[{Ding {et~al.}(2023)Ding, Onoue, Silverman, Matsuoka, Izumi, Strauss, Jahnke, Andika, Aoki, Baba, Bieri, Bosman, Eilers, Fujimoto, Habouzit, Haiman, Imanishi, Inayoshi, Iwasawa, Kashikawa, Kawaguchi, Kohno, Lee, Li, Lupi, Lyu, Nagao, Overzier, Phillips, Schindler, Schramm, Shimasaku, Toba, Trakhtenbrot, Trebitsch, Treu, Umehata, Venemans, Vestergaard, Volonteri, Walter, Wang, \& Yang}]{Ding2023}
Ding, X., Onoue, M., Silverman, J.~D., {et~al.} 2023, Nature, 621, 51

\bibitem[{Dom{\'{\i}}nguez {et~al.}(2013)Dom{\'{\i}}nguez, Siana, Henry, Scarlata, Bedregal, Malkan, Atek, Ross, Colbert, Teplitz, Rafelski, McCarthy, Bunker, Hathi, Dressler, Martin, \& Masters}]{Dominguez2013}
Dom{\'{\i}}nguez, A., Siana, B., Henry, A.~L., {et~al.} 2013, ApJ, 763, 145

\bibitem[{{Doyon} {et~al.}(1994){Doyon}, {Joseph}, \& {Wright}}]{Doyon1994}
{Doyon}, R., {Joseph}, R.~D., \& {Wright}, G.~S. 1994, \apj, 421, 101

\bibitem[{Dunlop {et~al.}(2003)Dunlop, McLure, Kukula, Baum, O'Dea, \& Hughes}]{dunlop_2003}
Dunlop, J.~S., McLure, R.~J., Kukula, M.~J., {et~al.} 2003, MNRAS, 340, 1095

\bibitem[{Eilers {et~al.}(2023)Eilers, Simcoe, Yue, Mackenzie, Matthee, {\v{D}}urov{\v{c}}{\'{\i}}kov{\'{a}}, Kashino, Bordoloi, \& Lilly}]{Eilers2023}
Eilers, A.-C., Simcoe, R.~A., Yue, M., {et~al.} 2023, ApJ, 950, 68

\bibitem[{Fan {et~al.}(2001)Fan, Narayanan, Lupton, Strauss, Knapp, Becker, White, Pentericci, Leggett, Haiman, Gunn, Ivezi{\'{c}}, Schneider, Anderson, Brinkmann, Bahcall, Connolly, Csabai, Doi, Fukugita, Geballe, Grebel, Harbeck, Hennessy, Lamb, Miknaitis, Munn, Nichol, Okamura, Pier, Prada, Richards, Szalay, \& York}]{fan_2001}
Fan, X., Narayanan, V.~K., Lupton, R.~H., {et~al.} 2001, AJ, 122, 2833

\bibitem[{Fan {et~al.}(2003)Fan, Strauss, Schneider, Becker, White, Haiman, Gregg, Pentericci, Grebel, Narayanan, Loh, Richards, Gunn, Lupton, Knapp, Ivezi{\'{c}}, Brandt, Collinge, Hao, Harbeck, Prada, Schaye, Strateva, Zakamska, Anderson, Brinkmann, Bahcall, Lamb, Okamura, Szalay, \& York}]{Fan2003}
Fan, X., Strauss, M.~A., Schneider, D.~P., {et~al.} 2003, AJ, 125, 1649

\bibitem[{Fan {et~al.}(2000)Fan, White, Davis, Becker, Strauss, Haiman, Schneider, Gregg, Gunn, Knapp, Lupton, Anderson, Anderson, Annis, Bahcall, Boroski, Brunner, Chen, Connolly, Csabai, Doi, Fukugita, Hennessy, Hindsley, Ichikawa, Ivezić, Loveday, Meiksin, McKay, Munn, Newberg, Nichol, Okamura, Pier, Sekiguchi, Shimasaku, Stoughton, Szalay, Szokoly, Thakar, Vogeley, \& York}]{fan_2000}
Fan, X., White, R.~L., Davis, M., {et~al.} 2000, AJ, 120, 1167

\bibitem[{Farina {et~al.}(2022)Farina, Schindler, Walter, Ba{\~{n}}ados, Davies, Decarli, Eilers, Fan, Hennawi, Mazzucchelli, Meyer, Trakhtenbrot, Volonteri, Wang, Worseck, Yang, Gutcke, Venemans, Bosman, Costa, Rosa, Drake, \& Onoue}]{Farina2022}
Farina, E.~P., Schindler, J.-T., Walter, F., {et~al.} 2022, ApJ, 941, 106

\bibitem[{Floyd {et~al.}(2013)Floyd, Dunlop, Kukula, Brown, McLure, Baum, \& O'Dea}]{floyd_2013}
Floyd, D. J.~E., Dunlop, J.~S., Kukula, M.~J., {et~al.} 2013, MNRAS, 429, 2

\bibitem[{{Furtak} {et~al.}(2024){Furtak}, {Labb{\'e}}, {Zitrin}, {Greene}, {Dayal}, {Chemerynska}, {Kokorev}, {Miller}, {Goulding}, {de Graaff}, {Bezanson}, {Brammer}, {Cutler}, {Leja}, {Pan}, {Price}, {Wang}, {Weaver}, {Whitaker}, {Atek}, {Bogd{\'a}n}, {Charlot}, {Curtis-Lake}, {van Dokkum}, {Endsley}, {Feldmann}, {Fudamoto}, {Fujimoto}, {Glazebrook}, {Juneau}, {Marchesini}, {Maseda}, {Nelson}, {Oesch}, {Plat}, {Setton}, {Stark}, \& {Williams}}]{Furtak2024}
{Furtak}, L.~J., {Labb{\'e}}, I., {Zitrin}, A., {et~al.} 2024, \nat, 628, 57

\bibitem[{{Gaia Collaboration} {et~al.}(2018){Gaia Collaboration}, {Brown}, {Vallenari}, {Prusti}, {de Bruijne}, {Babusiaux}, {Bailer-Jones}, {Biermann}, {Evans}, {Eyer}, {Jansen}, {Jordi}, {Klioner}, {Lammers}, {Lindegren}, {Luri}, {Mignard}, {Panem}, {Pourbaix}, {Randich}, {Sartoretti}, {Siddiqui}, {Soubiran}, {van Leeuwen}, {Walton}, {Arenou}, {Bastian}, {Cropper}, {Drimmel}, {Katz}, {Lattanzi}, {Bakker}, {Cacciari}, {Casta{\~n}eda}, {Chaoul}, {Cheek}, {De Angeli}, {Fabricius}, {Guerra}, {Holl}, {Masana}, {Messineo}, {Mowlavi}, {Nienartowicz}, {Panuzzo}, {Portell}, {Riello}, {Seabroke}, {Tanga}, {Th{\'e}venin}, {Gracia-Abril}, {Comoretto}, {Garcia-Reinaldos}, {Teyssier}, {Altmann}, {Andrae}, {Audard}, {Bellas-Velidis}, {Benson}, {Berthier}, {Blomme}, {Burgess}, {Busso}, {Carry}, {Cellino}, {Clementini}, {Clotet}, {Creevey}, {Davidson}, {De Ridder}, {Delchambre}, {Dell'Oro}, {Ducourant}, {Fern{\'a}ndez-Hern{\'a}ndez}, {Fouesneau}, {Fr{\'e}mat}, {Galluccio}, {Garc{\'\i}a-Torres},
  {Gonz{\'a}lez-N{\'u}{\~n}ez}, {Gonz{\'a}lez-Vidal}, {Gosset}, {Guy}, {Halbwachs}, {Hambly}, {Harrison}, {Hern{\'a}ndez}, {Hestroffer}, {Hodgkin}, {Hutton}, {Jasniewicz}, {Jean-Antoine-Piccolo}, {Jordan}, {Korn}, {Krone-Martins}, {Lanzafame}, {Lebzelter}, {L{\"o}ffler}, {Manteiga}, {Marrese}, {Mart{\'\i}n-Fleitas}, {Moitinho}, {Mora}, {Muinonen}, {Osinde}, {Pancino}, {Pauwels}, {Petit}, {Recio-Blanco}, {Richards}, {Rimoldini}, {Robin}, {Sarro}, {Siopis}, {Smith}, {Sozzetti}, {S{\"u}veges}, {Torra}, {van Reeven}, {Abbas}, {Abreu Aramburu}, {Accart}, {Aerts}, {Altavilla}, {{\'A}lvarez}, {Alvarez}, {Alves}, {Anderson}, {Andrei}, {Anglada Varela}, {Antiche}, {Antoja}, {Arcay}, {Astraatmadja}, {Bach}, {Baker}, {Balaguer-N{\'u}{\~n}ez}, {Balm}, {Barache}, {Barata}, {Barbato}, {Barblan}, {Barklem}, {Barrado}, {Barros}, {Barstow}, {Bartholom{\'e} Mu{\~n}oz}, {Bassilana}, {Becciani}, {Bellazzini}, {Berihuete}, {Bertone}, {Bianchi}, {Bienaym{\'e}}, {Blanco-Cuaresma}, {Boch}, {Boeche}, {Bombrun}, {Borrachero},
  {Bossini}, {Bouquillon}, {Bourda}, {Bragaglia}, {Bramante}, {Breddels}, {Bressan}, {Brouillet}, {Br{\"u}semeister}, {Brugaletta}, {Bucciarelli}, {Burlacu}, {Busonero}, {Butkevich}, {Buzzi}, {Caffau}, {Cancelliere}, {Cannizzaro}, {Cantat-Gaudin}, {Carballo}, {Carlucci}, {Carrasco}, {Casamiquela}, {Castellani}, {Castro-Ginard}, {Charlot}, {Chemin}, {Chiavassa}, {Cocozza}, {Costigan}, {Cowell}, {Crifo}, {Crosta}, {Crowley}, {Cuypers}, {Dafonte}, {Damerdji}, {Dapergolas}, {David}, {David}, {de Laverny}, {De Luise}, {De March}, {de Martino}, {de Souza}, {de Torres}, {Debosscher}, {del Pozo}, {Delbo}, {Delgado}, {Delgado}, {Di Matteo}, {Diakite}, {Diener}, {Distefano}, {Dolding}, {Drazinos}, {Dur{\'a}n}, {Edvardsson}, {Enke}, {Eriksson}, {Esquej}, {Eynard Bontemps}, {Fabre}, {Fabrizio}, {Faigler}, {Falc{\~a}o}, {Farr{\`a}s Casas}, {Federici}, {Fedorets}, {Fernique}, {Figueras}, {Filippi}, {Findeisen}, {Fonti}, {Fraile}, {Fraser}, {Fr{\'e}zouls}, {Gai}, {Galleti}, {Garabato}, {Garc{\'\i}a-Sedano}, {Garofalo},
  {Garralda}, {Gavel}, {Gavras}, {Gerssen}, {Geyer}, {Giacobbe}, {Gilmore}, {Girona}, {Giuffrida}, {Glass}, {Gomes}, {Granvik}, {Gueguen}, {Guerrier}, {Guiraud}, {Guti{\'e}rrez-S{\'a}nchez}, {Haigron}, {Hatzidimitriou}, {Hauser}, {Haywood}, {Heiter}, {Helmi}, {Heu}, {Hilger}, {Hobbs}, {Hofmann}, {Holland}, {Huckle}, {Hypki}, {Icardi}, {Jan{\ss}en}, {Jevardat de Fombelle}, {Jonker}, {Juh{\'a}sz}, {Julbe}, {Karampelas}, {Kewley}, {Klar}, {Kochoska}, {Kohley}, {Kolenberg}, {Kontizas}, {Kontizas}, {Koposov}, {Kordopatis}, {Kostrzewa-Rutkowska}, {Koubsky}, {Lambert}, {Lanza}, {Lasne}, {Lavigne}, {Le Fustec}, {Le Poncin-Lafitte}, {Lebreton}, {Leccia}, {Leclerc}, {Lecoeur-Taibi}, {Lenhardt}, {Leroux}, {Liao}, {Licata}, {Lindstr{\o}m}, {Lister}, {Livanou}, {Lobel}, {L{\'o}pez}, {Managau}, {Mann}, {Mantelet}, {Marchal}, {Marchant}, {Marconi}, {Marinoni}, {Marschalk{\'o}}, {Marshall}, {Martino}, {Marton}, {Mary}, {Massari}, {Matijevi{\v{c}}}, {Mazeh}, {McMillan}, {Messina}, {Michalik}, {Millar}, {Molina}, {Molinaro},
  {Moln{\'a}r}, {Montegriffo}, {Mor}, {Morbidelli}, {Morel}, {Morris}, {Mulone}, {Muraveva}, {Musella}, {Nelemans}, {Nicastro}, {Noval}, {O'Mullane}, {Ord{\'e}novic}, {Ord{\'o}{\~n}ez-Blanco}, {Osborne}, {Pagani}, {Pagano}, {Pailler}, {Palacin}, {Palaversa}, {Panahi}, {Pawlak}, {Piersimoni}, {Pineau}, {Plachy}, {Plum}, {Poggio}, {Poujoulet}, {Pr{\v{s}}a}, {Pulone}, {Racero}, {Ragaini}, {Rambaux}, {Ramos-Lerate}, {Regibo}, {Reyl{\'e}}, {Riclet}, {Ripepi}, {Riva}, {Rivard}, {Rixon}, {Roegiers}, {Roelens}, {Romero-G{\'o}mez}, {Rowell}, {Royer}, {Ruiz-Dern}, {Sadowski}, {Sagrist{\`a} Sell{\'e}s}, {Sahlmann}, {Salgado}, {Salguero}, {Sanna}, {Santana-Ros}, {Sarasso}, {Savietto}, {Schultheis}, {Sciacca}, {Segol}, {Segovia}, {S{\'e}gransan}, {Shih}, {Siltala}, {Silva}, {Smart}, {Smith}, {Solano}, {Solitro}, {Sordo}, {Soria Nieto}, {Souchay}, {Spagna}, {Spoto}, {Stampa}, {Steele}, {Steidelm{\"u}ller}, {Stephenson}, {Stoev}, {Suess}, {Surdej}, {Szabados}, {Szegedi-Elek}, {Tapiador}, {Taris}, {Tauran}, {Taylor},
  {Teixeira}, {Terrett}, {Teyssandier}, {Thuillot}, {Titarenko}, {Torra Clotet}, {Turon}, {Ulla}, {Utrilla}, {Uzzi}, {Vaillant}, {Valentini}, {Valette}, {van Elteren}, {Van Hemelryck}, {van Leeuwen}, {Vaschetto}, {Vecchiato}, {Veljanoski}, {Viala}, {Vicente}, {Vogt}, {von Essen}, {Voss}, {Votruba}, {Voutsinas}, {Walmsley}, {Weiler}, {Wertz}, {Wevers}, {Wyrzykowski}, {Yoldas}, {{\v{Z}}erjal}, {Ziaeepour}, {Zorec}, {Zschocke}, {Zucker}, {Zurbach}, \& {Zwitter}}]{Gaia2018}
{Gaia Collaboration}, {Brown}, A.~G.~A., {Vallenari}, A., {et~al.} 2018, \aap, 616, A1

\bibitem[{Gardner {et~al.}(2023)Gardner, Mather, Abbott, Abell, Abernathy, Abney, Abraham, Abraham, Abul-Huda, Acton, Adams, Adams, Adler, Adriaensen, Aguilar, Ahmed, Ahmed, Ahmed, Albat, Albert, Alberts, Aldridge, Allen, Allen, Altenburg, Altunc, Alvarez, Álvarez Márquez, de~Oliveira, Ambrose, Anandakrishnan, Andersen, Anderson, Anderson, Anderson, Anderson, Aprea, Archer, Arenberg, Argyriou, Arribas, Artigau, Arvai, Atcheson, Atkinson, Averbukh, Aymergen, Bacinski, Baggett, Bagnasco, Baker, Balzano, Banks, Baran, Barker, Barrett, Barringer, Barto, Bast, Baudoz, Baum, Beatty, Beaulieu, Bechtold, Beck, Beddard, Beichman, Bellagama, Bely, Berger, Bergeron, Darveau-Bernier, Bertch, Beskow, Betz, Biagetti, Birkmann, Bjorklund, Blackwood, Blazek, Blossfeld, Bluth, Boccaletti, Boegner, Bohlin, Boia, Böker, Bonaventura, Bond, Bosley, Boucarut, Bouchet, Bouwman, Bower, Bowers, Bowers, Boyce, Boyer, Boyer, Boyer, Boyer, Bradley, Brady, Brandl, Brannen, Breda, Bremmer, Brennan, Bresnahan, Bright, Broiles,
  Bromenschenkel, Brooks, Brooks, Brown, Brown, Brown, Bruce, Bryson, Bujanda, Bullock, Bunker, Bureo, Burt, Bush, Bushouse, Bussman, Cabaud, Cale, Calhoon, Calvani, Canipe, Caputo, Cara, Carey, Case, Cesari, Cetorelli, Chance, Chandler, Chaney, Chapman, Charlot, Chayer, Cheezum, Chen, Chen, Cherinka, Chichester, Chilton, Chittiraibalan, Clampin, Clark, Clark, Clark, Claybrooks, Cleveland, Cohen, Cohen, Colón, Coleman, Colina, Comber, Comeau, Comer, Reis, Connolly, Conroy, Contos, Contreras, Cook, Cooper, Cooper, Correia, Correnti, Cossou, Costanza, Coulais, Cox, Coyle, Cracraft, Noriega-Crespo, Crew, Curtis, Cusveller, Maciel, Dailey, Daugeron, Davidson, Davies, Davis, Davis, Day, de~Chambure, de~Jong, De~Marchi, Dean, Decker, Delisa, Dell, Dellagatta, Dembinska, Demosthenes, Dencheva, Deneu, DePriest, Deschenes, Dethienne, Detre, Diaz, Dicken, DiFelice, Dillman, Disharoon, van Dishoeck, Dixon, Doggett, Dominguez, Donaldson, Doria-Warner, Santos, Doty, Douglas, Doyon, Dressler, Driggers, Driggers, Dunn,
  DuPrie, Dupuis, Durning, Dutta, Earl, Eccleston, Ecobichon, Egami, Ehrenwinkler, Eisenhamer, Eisenhower, Eisenstein, Hamel, Elie, Elliott, Elliott, Engesser, Espinoza, Etienne, Etxaluze, Evans, Fabreguettes, Falcolini, Falini, Fatig, Feeney, Feinberg, Fels, Ferdous, Ferguson, Ferrarese, Ferreira, Ferruit, Ferry, Filippazzo, Firre, Fix, Flagey, Flanagan, Fleming, Florian, Flynn, Foiadelli, Fontaine, Fontanella, Forshay, Fortner, Fox, Framarini, Francisco, Franck, Franx, Franz, Friedman, Friend, Frost, Fu, Fullerton, Gaillard, Galkin, Gallagher, Galyer, Marín, Gardner, Garland, Garrett, Gasman, Gáspár, Gastaud, Gaudreau, Gauthier, Geers, Geithner, Gennaro, Gerber, Gereau, Giampaoli, Giardino, Gibbons, Gilbert, Gilman, Girard, Giuliano, Gkountis, Glasse, Glassmire, Glauser, Glazer, Goldberg, Golimowski, Gonzaga, Gordon, Gordon, Goudfrooij, Gough, Graham, Grau, Green, Greene, Greene, Greenfield, Greenhouse, Greve, Greville, Grimaldi, Groe, Groebner, Grumm, Grundy, Güdel, Guillard, Guldalian, Gunn, Gurule,
  Gutman, Guy, Guyot, Hack, Haderlein, Hagan, Hagedorn, Hainline, Haley, Hami, Hamilton, Hammann, Hammel, Hanley, Hansen, Hardy, Harnisch, Harr, Harris, Hart, Hartig, Hasan, Hashim, Hashimoto, Haskins, Hawkins, Hayden, Hayden, Healy, Hecht, Heeg, Hejal, Helm, Hengemihle, Henning, Henry, Henry, Henshaw, Hernandez, Herrington, Heske, Hesman, Hickey, Hilbert, Hines, Hinz, Hirsch, Hitcho, Hodapp, Hodge, Hoffman, Holfeltz, Holler, Hoppa, Horner, Howard, Howard, Huber, Hunkeler, Hunter, Hunter, Hurd, Hurst, Hutchings, Hylan, Ignat, Illingworth, Irish, Isaacs, Jackson, Jaffe, Jahic, Jahromi, Jakobsen, James, James, James, Jamieson, Jandra, Jayawardhana, Jedrzejewski, Jeffers, Jensen, Joanne, Johns, Johnson, Johnson, Johnson, Johnson, Johnson, Johnson, Johnstone, Jollet, Jones, Jones, Jones, Jones, Jones, Jordan, Jordan, Jue, Jurkowski, Justis, Justtanont, Kaleida, Kalirai, Kalmanson, Kaltenegger, Kammerer, Kan, Kanarek, Kao, Karakla, Karl, Kassin, Kauffman, Kavanagh, Kelley, Kelly, Kendrew, Kennedy, Kenny,
  Keski-Kuha, Keyes, Khan, Kidwell, Kimble, King, King, Kinzel, Kirk, Kirkpatrick, Klaassen, Klingemann, Klintworth, Knapp, Knight, Knollenberg, Knutsen, Koehler, Koekemoer, Kofler, Kontson, Kovacs, Kozhurina-Platais, Krause, Kriss, Krist, Kristoffersen, Krogel, Krueger, Kulp, Kumari, Kwan, Kyprianou, Labador, Labiano, Lafrenière, Lagage, Laidler, Laine, Laird, Lajoie, Lallo, Lam, LaMassa, Lambros, Lampenfield, Lander, Langston, Larson, Larson, LaVerghetta, Law, Lawrence, Lee, Lee, Lee, Leisenring, Leveille, Levenson, Levi, Levine, Lewis, Lewis, Lewis, Libralato, Lidon, Liebrecht, Lightsey, Lilly, Lim, Lim, Ling, Link, Link, Lipinski, Liu, Lo, Lobmeyer, Logue, Long, Long, Long, Long, López-Caniego, Lotz, Love-Pruitt, Lubskiy, Luers, Luetgens, Luevano, Lui, Lund, Lundquist, Lunine, Lützgendorf, Lynch, MacDonald, MacDonald, Macias, Macklis, Maghami, Maharaja, Maiolino, Makrygiannis, Malla, Malumuth, Manjavacas, Marini, Marrione, Marston, Martel, Martin, Martin, Martinez, Maschmann, Masci, Masetti,
  Maszkiewicz, Matthews, Matuskey, McBrayer, McCarthy, McCaughrean, McClare, McClare, McCloskey, McClurg, McCoy, McElwain, McGregor, McGuffey, McKay, McKenzie, McLean, McMaster, McNeil, De~Meester, Mehalick, Meixner, Meléndez, Menzel, Menzel, Merz, Mesterharm, Meyer, Meyett, Meza, Midwinter, Milam, Miller, Miller, Miskey, Misselt, Mitchell, Mohan, Montoya, Moran, Morishita, Moro-Martín, Morrison, Morrison, Morse, Moschos, Moseley, Mosier, Mosner, Mountain, Muckenthaler, Mueller, Mueller, Muhiem, Mühlmann, Mullally, Mullen, Munger, Murphy, Murray, Muzerolle, Mycroft, Myers, Myers, Myers, Myers, Myrick, Nagle, Nayak, Naylor, Neff, Nelan, Nella, Nguyen, Nguyen, Nickson, Nidhiry, Niedner, Nieto-Santisteban, Nikolov, Nishisaka, Noriega-Crespo, Nota, O'Mara, Oboryshko, O'Brien, Ochs, Offenberg, Ogle, Ohl, Olmsted, Osborne, O'Shaughnessy, Östlin, O'Sullivan, Otor, Ottens, Ouellette, Outlaw, Owens, Pacifici, Page, Paranilam, Park, Parrish, Paschal, Patapis, Patel, Patrick, Pattishall, Paul, Paul, Pauly,
  Pavlovsky, Peña-Guerrero, Pedder, Peek, Pelham, Penanen, Perriello, Perrin, Perrine, Perrygo, Peslier, Petach, Peterson, Pfarr, Pierson, Pietraszkiewicz, Pilchen, Pipher, Pirzkal, Pitman, Player, Plesha, Plitzke, Pohner, Poletis, Pollizzi, Polster, Pontius, Pontoppidan, Porges, Potter, Prescott, Proffitt, Pueyo, Neira, Radich, Rager, Rameau, Ramey, Alarcon, Rampini, Rapp, Rashford, Rauscher, Ravindranath, Rawle, Rawlings, Ray, Regan, Rehm, Rehm, Reid, Reis, Renk, Reoch, Ressler, Rest, Reynolds, Richon, Richon, Ridgaway, Riedel, Rieke, Rieke, Rifelli, Rigby, Riggs, Ringel, Ritchie, Rix, Robberto, Robinson, Robinson, Rock, Rodriguez, del Pino, Roellig, Rohrbach, Roman, Romelfanger, Romo, Rosales, Rose, Roteliuk, Roth, Rothwell, Rouzaud, Rowe, Rowlands, Roy, Royer, Rui, Rumler, Rumpl, Russ, Ryan, Ryan, Saad, Sabata, Sabatino, Sabbi, Sabelhaus, Sabia, Sahu, Saif, Salvignol, Samara-Ratna, Samuelson, Sanders, Sappington, Sargent, Sauer, Savadkin, Sawicki, Schappell, Scheffer, Scheithauer, Scherer, Schiff,
  Schlawin, Schmeitzky, Schmitz, Schmude, Schneider, Schreiber, Schroeven-Deceuninck, Schultz, Schwab, Schwartz, Scoccimarro, Scott, Scott, Seaton, Seely, Seery, Seidleck, Sembach, Shanahan, Shaughnessy, Shaw, Shay, Sheehan, Sheth, Shih, Shivaei, Siegel, Sienkiewicz, Simmons, Simon, Sirianni, Sivaramakrishnan, Slade, Sloan, Slocum, Slowinski, Smith, Smith, Smith, Smith, Smith, Smith, Smolik, Soderblom, Sohn, Sokol, Sonneborn, Sontag, Sooy, Soummer, Southwood, Spain, Sparmo, Speer, Spencer, Sprofera, Stallcup, Stanley, Stansberry, Stark, Starr, Stassi, Steck, Steeley, Stephens, Stephenson, Stewart, Stiavelli, Stockman, Strada, Straughn, Streetman, Strickland, Strobele, Stuhlinger, Stys, Such, Sukhatme, Sullivan, Sullivan, Sumner, Sun, Sunnquist, Swade, Swam, Swenton, Swoish, Litten, Tamas, Tao, Taylor, Taylor, Plate, Van~Tea, Teague, Telfer, Temim, Texter, Thatte, Thompson, Thompson, Thomson, Thronson, Tierney, Tikkanen, Tinnin, Tippet, Todd, Tran, Trauger, Trejo, Truong, Tsukamoto, Tufail, Tumlinson, Tustain,
  Tyra, Ubeda, Underwood, Uzzo, Vaclavik, Valenduc, Valenti, Van~Campen, van~de Wetering, Van Der~Marel, van Haarlem, Vandenbussche, Vanterpool, Vernoy, Costas, Volk, Voorzaat, Voyton, Vydra, Waddy, Waelkens, Wahlgren, Walker, Wander, Warfield, Warner, Wasiak, Wasiak, Wehner, Weiler, Weilert, Weiss, Wells, Welty, Wheate, Wheeler, White, Whitehouse, Whiteleather, Whitman, Williams, Willmer, Willott, Willoughby, Wilson, Wilson, Wilson, Windhorst, Wislowski, Wolfe, Wolfe, Wolff, Wondel, Woo, Woods, Worden, Workman, Wright, Wu, Wu, Wun, Wymer, Yadetie, Yan, Yang, Yates, Yeager, Yerger, Young, Young, Yu, Yu, Zak, Zeidler, Zepp, Zhou, Zincke, Zonak, \& Zondag}]{Gardner2023}
Gardner, J.~P., Mather, J.~C., Abbott, R., {et~al.} 2023, PASP, 135, 068001

\bibitem[{Gardner {et~al.}(2006)Gardner, Mather, Clampin, Doyon, Greenhouse, Hammel, Hutchings, Jakobsen, Lilly, Long, Lunine, Mccaughrean, Mountain, Nella, Rieke, Rieke, Rix, Smith, Sonneborn, Stiavelli, Stockman, Windhorst, \& Wright}]{Gardner2006}
Gardner, J.~P., Mather, J.~C., Clampin, M., {et~al.} 2006, SSRv, 123, 485

\bibitem[{Ginsburg {et~al.}(2019)Ginsburg, Koch, Robitaille, Beaumont, {Adamginsburg}, Sipőcz, ZuHone, {Sushobhana Patra}, Jones, {P. L. Lim}, Stern, Rosolowsky, Earl, Val-Borro, {Jrobbfed}, {Shuokong}, Kepley, {Vlas Sokolov}, Badger, Maret, Garrido, Booker, \& Tollerud}]{Ginsburg2019}
Ginsburg, A., Koch, E., Robitaille, T., {et~al.} 2019, radio-astro-tools/spectral-cube: Release v0.4.5

\bibitem[{Greene \& Ho(2005)}]{Greene2005}
Greene, J.~E. \& Ho, L.~C. 2005, ApJ, 630, 122

\bibitem[{{Greene} {et~al.}(2020){Greene}, {Strader}, \& {Ho}}]{Greene2020}
{Greene}, J.~E., {Strader}, J., \& {Ho}, L.~C. 2020, \araa, 58, 257

\bibitem[{{Harikane} {et~al.}(2023){Harikane}, {Zhang}, {Nakajima}, {Ouchi}, {Isobe}, {Ono}, {Hatano}, {Xu}, \& {Umeda}}]{Harikane2023}
{Harikane}, Y., {Zhang}, Y., {Nakajima}, K., {et~al.} 2023, \apj, 959, 39

\bibitem[{H{\"a}ring \& Rix(2004)}]{Haring2004}
H{\"a}ring, N. \& Rix, H.-W. 2004, ApJL, 604, L89

\bibitem[{{Hernquist} \& {Mihos}(1995)}]{Hernquist1995}
{Hernquist}, L. \& {Mihos}, J.~C. 1995, \apj, 448, 41

\bibitem[{{Hinshaw} {et~al.}(2013){Hinshaw}, {Larson}, {Komatsu}, {Spergel}, {Bennett}, {Dunkley}, {Nolta}, {Halpern}, {Hill}, {Odegard}, {Page}, {Smith}, {Weiland}, {Gold}, {Jarosik}, {Kogut}, {Limon}, {Meyer}, {Tucker}, {Wollack}, \& {Wright}}]{Hinshaw2013}
{Hinshaw}, G., {Larson}, D., {Komatsu}, E., {et~al.} 2013, ApJS, 208, 19

\bibitem[{{Hopkins} {et~al.}(2006){Hopkins}, {Hernquist}, {Cox}, {Di Matteo}, {Robertson}, \& {Springel}}]{Hopkins2006}
{Hopkins}, P.~F., {Hernquist}, L., {Cox}, T.~J., {et~al.} 2006, \apjs, 163, 1

\bibitem[{Hunter(2007)}]{Matplotlib2007}
Hunter, J.~D. 2007, CiSE, 9, 90

\bibitem[{Husemann {et~al.}(2014)Husemann, Jahnke, S{\'{a}}nchez, Wisotzki, Nugroho, Kupko, \& Schramm}]{Husemann2014}
Husemann, B., Jahnke, K., S{\'{a}}nchez, S.~F., {et~al.} 2014, MNRAS, 443, 755

\bibitem[{Husemann {et~al.}(2013)Husemann, Wisotzki, S{\'{a}}nchez, \& Jahnke}]{Husemann2013}
Husemann, B., Wisotzki, L., S{\'{a}}nchez, S.~F., \& Jahnke, K. 2013, A\&A, 549, A43

\bibitem[{Hutchings(2003)}]{hutchings_2003}
Hutchings, J.~B. 2003, AJ, 125, 1053

\bibitem[{Izumi {et~al.}(2019)Izumi, Onoue, Matsuoka, Nagao, Strauss, Imanishi, Kashikawa, Fujimoto, Kohno, Toba, Umehata, Goto, Ueda, Shirakata, Silverman, Greene, Harikane, Hashimoto, Ikarashi, Iono, Iwasawa, Lee, Minezaki, Nakanishi, Tamura, Tang, \& Taniguchi}]{Izumi2019}
Izumi, T., Onoue, M., Matsuoka, Y., {et~al.} 2019, PASJ, 71, 6, 111

\bibitem[{Izumi {et~al.}(2018)Izumi, Onoue, Shirakata, Nagao, Kohno, Matsuoka, Imanishi, Strauss, Kashikawa, Schulze, Silverman, Fujimoto, Harikane, Toba, Umehata, Nakanishi, Greene, Tamura, Taniguchi, Yamaguchi, Goto, Hashimoto, Ikarashi, Iono, Iwasawa, Lee, Makiya, Minezaki, \& Tang}]{Izumi2018}
Izumi, T., Onoue, M., Shirakata, H., {et~al.} 2018, PASJ, 70, 3, 36

\bibitem[{Jakobsen {et~al.}(2022)Jakobsen, Ferruit, de~Oliveira, Arribas, Bagnasco, Barho, Beck, Birkmann, Böker, Bunker, Charlot, de~Jong, de~Marchi, Ehrenwinkler, Falcolini, Fels, Franx, Franz, Funke, Giardino, Gnata, Holota, Honnen, Jensen, Jentsch, Johnson, Jollet, Karl, Kling, Köhler, Kolm, Kumari, Lander, Lemke, L{\'{o}}pez-Caniego, Lützgendorf, Maiolino, Manjavacas, Marston, Maschmann, Maurer, Messerschmidt, Moseley, Mosner, Mott, Muzerolle, Pirzkal, Pittet, Plitzke, Posselt, Rapp, Rauscher, Rawle, Rix, Rödel, Rumler, Sabbi, Salvignol, Schmid, Sirianni, Smith, Strada, te~Plate, Valenti, Wettemann, Wiehe, Wiesmayer, Willott, Wright, Zeidler, \& Zincke}]{Jakobsen2022}
Jakobsen, P., Ferruit, P., de~Oliveira, C.~A., {et~al.} 2022, A\&A, 661, A80

\bibitem[{{Ji} {et~al.}(2021){Ji}, {Li}, {Yan}, {Mo}, {Lin}, {Zou}, {Lian}, {Stark}, {Riffel}, {Pan}, {Bizyaev}, \& {Bundy}}]{Ji2021}
{Ji}, X., {Li}, C., {Yan}, R., {et~al.} 2021, \mnras, 508, 3943

\bibitem[{{Ji} {et~al.}(2024){Ji}, {{\"U}bler}, {Maiolino}, {D'Eugenio}, {Arribas}, {Bunker}, {Charlot}, {Perna}, {Rodr{\'\i}guez Del Pino}, {B{\"o}ker}, {Cresci}, {Curti}, {Kumari}, \& {Lamperti}}]{Ji2024}
{Ji}, X., {{\"U}bler}, H., {Maiolino}, R., {et~al.} 2024, \mnras, 535, 881

\bibitem[{{Johnson} {et~al.}(2021){Johnson}, {Leja}, {Conroy}, \& {Speagle}}]{Johnson2021}
{Johnson}, B.~D., {Leja}, J., {Conroy}, C., \& {Speagle}, J.~S. 2021, \apjs, 254, 22

\bibitem[{{Jones} {et~al.}(2024){Jones}, {{\"U}bler}, {Perna}, {Arribas}, {Bunker}, {Carniani}, {Charlot}, {Maiolino}, {Del Pino}, {Willott}, {Bowler}, {B{\"o}ker}, {Cameron}, {Chevallard}, {Cresci}, {Curti}, {D'Eugenio}, {Kumari}, {Saxena}, {Scholtz}, {Venturi}, \& {Witstok}}]{Jones2024}
{Jones}, G.~C., {{\"U}bler}, H., {Perna}, M., {et~al.} 2024, \aap, 682, A122

\bibitem[{{Juod{\v{z}}balis} {et~al.}(2024){Juod{\v{z}}balis}, {Maiolino}, {Baker}, {Tacchella}, {Scholtz}, {D'Eugenio}, {Witstok}, {Schneider}, {Trinca}, {Valiante}, {DeCoursey}, {Curti}, {Carniani}, {Chevallard}, {de Graaff}, {Arribas}, {Bennett}, {Bourne}, {Bunker}, {Charlot}, {Jiang}, {Koudmani}, {Perna}, {Robertson}, {Sijacki}, {{\"U}bler}, {Williams}, \& {Willott}}]{Juodzbalis2024}
{Juod{\v{z}}balis}, I., {Maiolino}, R., {Baker}, W.~M., {et~al.} 2024, \nat, 636, 594

\bibitem[{Kashikawa {et~al.}(2015)Kashikawa, Ishizaki, Willott, Onoue, Im, Furusawa, Toshikawa, Ishikawa, Niino, Shimasaku, Ouchi, \& Hibon}]{Kashikawa2015}
Kashikawa, N., Ishizaki, Y., Willott, C.~J., {et~al.} 2015, ApJ, 798, 28

\bibitem[{Kashino {et~al.}(2023)Kashino, Lilly, Matthee, Eilers, Mackenzie, Bordoloi, \& Simcoe}]{Kashino2022}
Kashino, D., Lilly, S.~J., Matthee, J., {et~al.} 2023, ApJ, 950, 66

\bibitem[{Kaspi {et~al.}(2021)Kaspi, Brandt, Maoz, Netzer, Schneider, Shemmer, \& Grier}]{Kaspi2021}
Kaspi, S., Brandt, W.~N., Maoz, D., {et~al.} 2021, ApJ, 915, 129

\bibitem[{Kauffmann {et~al.}(2003)Kauffmann, Heckman, Tremonti, Brinchmann, Charlot, White, Ridgway, Brinkmann, Fukugita, Hall, IveziÄ, Richards, \& Schneider}]{Kauffmann2003}
Kauffmann, G., Heckman, T.~M., Tremonti, C., {et~al.} 2003, MNRAS, 346, 1055

\bibitem[{{Keel} {et~al.}(2019){Keel}, {Bennert}, {Pancoast}, {Harris}, {Nierenberg}, {Chojnowski}, {Moiseev}, {Oparin}, {Lintott}, {Schawinski}, {Mitchell}, \& {Cornen}}]{Keel2019}
{Keel}, W.~C., {Bennert}, V.~N., {Pancoast}, A., {et~al.} 2019, \mnras, 483, 4847

\bibitem[{Kennicutt \& Evans(2012)}]{Kennicutt2012}
Kennicutt, R.~C. \& Evans, N.~J. 2012, ARA\&A, 50, 531

\bibitem[{Kewley {et~al.}(2001)Kewley, Dopita, Sutherland, Heisler, \& Trevena}]{Kewley2001}
Kewley, L.~J., Dopita, M.~A., Sutherland, R.~S., Heisler, C.~A., \& Trevena, J. 2001, ApJ, 556, 121

\bibitem[{{Kokorev} {et~al.}(2023){Kokorev}, {Fujimoto}, {Labbe}, {Greene}, {Bezanson}, {Dayal}, {Nelson}, {Atek}, {Brammer}, {Caputi}, {Chemerynska}, {Cutler}, {Feldmann}, {Fudamoto}, {Furtak}, {Goulding}, {de Graaff}, {Leja}, {Marchesini}, {Miller}, {Nanayakkara}, {Oesch}, {Pan}, {Price}, {Setton}, {Smit}, {Stefanon}, {Wang}, {Weaver}, {Whitaker}, {Williams}, \& {Zitrin}}]{Kokorev2023}
{Kokorev}, V., {Fujimoto}, S., {Labbe}, I., {et~al.} 2023, \apjl, 957, L7

\bibitem[{{Komossa} {et~al.}(2008){Komossa}, {Zhou}, \& {Lu}}]{Komossa2008}
{Komossa}, S., {Zhou}, H., \& {Lu}, H. 2008, \apjl, 678, L81

\bibitem[{Kormendy \& Ho(2013)}]{Kormendy2013}
Kormendy, J. \& Ho, L.~C. 2013, ARA\&A, 51, 511

\bibitem[{{Koss} {et~al.}(2014){Koss}, {Blecha}, {Mushotzky}, {Hung}, {Veilleux}, {Trakhtenbrot}, {Schawinski}, {Stern}, {Smith}, {Li}, {Man}, {Filippenko}, {Mauerhan}, {Stanek}, \& {Sanders}}]{Koss2014}
{Koss}, M., {Blecha}, L., {Mushotzky}, R., {et~al.} 2014, \mnras, 445, 515

\bibitem[{Kovacevic {et~al.}(2010)Kovacevic, Popovic, \& Dimitrijevic}]{Kovacevic2010}
Kovacevic, J., Popovic, L.~C., \& Dimitrijevic, M.~S. 2010, ApJS, 189, 15

\bibitem[{Kuraszkiewicz {et~al.}(2002)Kuraszkiewicz, Green, Forster, Aldcroft, Evans, \& Koratkar}]{Kuraszkiewicz2002}
Kuraszkiewicz, J.~K., Green, P.~J., Forster, K., {et~al.} 2002, ApJS, 143, 257

\bibitem[{{Lamperti} {et~al.}(2024){Lamperti}, {Arribas}, {Perna}, {Rodr{\'\i}guez Del Pino}, {Circosta}, {P{\'e}rez-Gonz{\'a}lez}, {Bunker}, {Carniani}, {Charlot}, {D'Eugenio}, {Maiolino}, {{\"U}bler}, {Willott}, {Bertola}, {B{\"o}ker}, {Cresci}, {Curti}, {Jones}, {Kumari}, {Parlanti}, {Scholtz}, \& {Venturi}}]{Lamperti2024}
{Lamperti}, I., {Arribas}, S., {Perna}, M., {et~al.} 2024, \aap, 691, A153

\bibitem[{Larson {et~al.}(2023)Larson, Finkelstein, Kocevski, Hutchison, Trump, Haro, Bromm, Cleri, Dickinson, Fujimoto, Kartaltepe, Koekemoer, Papovich, Pirzkal, Tacchella, Zavala, Bagley, Behroozi, Champagne, Cole, Jung, Morales, Yang, Zhang, Zitrin, Amor{\'{\i}}n, Burgarella, Casey, Ortiz, Cox, Chworowsky, Fontana, Gawiser, Grazian, Grogin, Harish, Hathi, Hirschmann, Holwerda, Juneau, Leung, Lucas, McGrath, P{\'{e}}rez-Gonz{\'{a}}lez, Rigby, Seill{\'{e}}, Simons, de~la Vega, Weiner, Wilkins, \& Yung}]{Larson2023}
Larson, R.~L., Finkelstein, S.~L., Kocevski, D.~D., {et~al.} 2023, ApJ, 953, L29

\bibitem[{Law {et~al.}(2023)Law, Morrison, Argyriou, Patapis, {\'{A}}lvarez-M{\'{a}}rquez, Labiano, \& Vandenbussche}]{Law2023}
Law, D.~D., Morrison, J.~E., Argyriou, I., {et~al.} 2023, AJ, 166, 45

\bibitem[{{Lawrence} {et~al.}(2007){Lawrence}, {Warren}, {Almaini}, {Edge}, {Hambly}, {Jameson}, {Lucas}, {Casali}, {Adamson}, {Dye}, {Emerson}, {Foucaud}, {Hewett}, {Hirst}, {Hodgkin}, {Irwin}, {Lodieu}, {McMahon}, {Simpson}, {Smail}, {Mortlock}, \& {Folger}}]{Lawrence2007}
{Lawrence}, A., {Warren}, S.~J., {Almaini}, O., {et~al.} 2007, \mnras, 379, 1599

\bibitem[{{Li} {et~al.}(2025){Li}, {Silverman}, {Shen}, {Volonteri}, {Jahnke}, {Zhuang}, {Scoggins}, {Ding}, {Harikane}, {Onoue}, \& {Tanaka}}]{Li2024}
{Li}, J., {Silverman}, J.~D., {Shen}, Y., {et~al.} 2025, \apj, 981, 19

\bibitem[{{Liu} {et~al.}(2024){Liu}, {Dai}, {Omont}, {Liu}, {Cox}, {Neri}, {Krips}, {Yang}, {Wu}, \& {Huang}}]{Liu2024}
{Liu}, F.-Y., {Dai}, Y.~S., {Omont}, A., {et~al.} 2024, \apj, 964, 136

\bibitem[{{Loiacono} {et~al.}(2024){Loiacono}, {Decarli}, {Mignoli}, {Farina}, {Ba{\~n}ados}, {Bosman}, {Eilers}, {Schindler}, {Strauss}, {Vestergaard}, {Wang}, {Blecha}, {Carilli}, {Comastri}, {Connor}, {Costa}, {Dotti}, {Fan}, {Gilli}, {Jun}, {Liu}, {Lupi}, {Marshall}, {Mazzucchelli}, {Meyer}, {Neeleman}, {Overzier}, {Pensabene}, {Riechers}, {Trakhtenbrot}, {Trebitsch}, {Venemans}, {Walter}, \& {Yang}}]{Loiacono2024}
{Loiacono}, F., {Decarli}, R., {Mignoli}, M., {et~al.} 2024, \aap, 685, A121

\bibitem[{Magorrian {et~al.}(1998)Magorrian, Tremaine, Richstone, Bender, Bower, Dressler, Faber, Gebhardt, Green, Grillmair, Kormendy, \& Lauer}]{Magorrian1998}
Magorrian, J., Tremaine, S., Richstone, D., {et~al.} 1998, AJ, 115, 2285

\bibitem[{{Maiolino} {et~al.}(2024){Maiolino}, {Scholtz}, {Curtis-Lake}, {Carniani}, {Baker}, {de Graaff}, {Tacchella}, {{\"U}bler}, {D'Eugenio}, {Witstok}, {Curti}, {Arribas}, {Bunker}, {Charlot}, {Chevallard}, {Eisenstein}, {Egami}, {Ji}, {Jones}, {Lyu}, {Rawle}, {Robertson}, {Rujopakarn}, {Perna}, {Sun}, {Venturi}, {Williams}, \& {Willott}}]{Maiolino2023}
{Maiolino}, R., {Scholtz}, J., {Curtis-Lake}, E., {et~al.} 2024, \aap, 691, A145

\bibitem[{Marconi \& Hunt(2003)}]{Marconi2003}
Marconi, A. \& Hunt, L.~K. 2003, ApJ, 589, L21

\bibitem[{Marshall {et~al.}(2020)Marshall, Mechtley, Windhorst, Cohen, Jansen, Jiang, Jones, Wyithe, Fan, Hathi, Jahnke, Keel, Koekemoer, Marian, Ren, Robinson, Röttgering, Ryan, Scannapieco, Schneider, Schneider, Smith, \& Yan}]{Marshall2019c}
Marshall, M.~A., Mechtley, M., Windhorst, R.~A., {et~al.} 2020, ApJ, 900, 21

\bibitem[{{Marshall} {et~al.}(2023){Marshall}, {Perna}, {Willott}, {Maiolino}, {Scholtz}, {{\"U}bler}, {Carniani}, {Arribas}, {L{\"u}tzgendorf}, {Bunker}, {Charlot}, {Ferruit}, {Jakobsen}, {Rix}, {Rodr{\'\i}guez Del Pino}, {B{\"o}ker}, {Cameron}, {Cresci}, {Curtis-Lake}, {Jones}, {Kumari}, {P{\'e}rez-Gonz{\'a}lez}, \& {Reed}}]{Marshall2023}
{Marshall}, M.~A., {Perna}, M., {Willott}, C.~J., {et~al.} 2023, \aap, 678, A191

\bibitem[{Matsuoka {et~al.}(2018)Matsuoka, Onoue, Kashikawa, Iwasawa, Strauss, Nagao, Imanishi, Lee, Akiyama, Asami, Bosch, Foucaud, Furusawa, Goto, Gunn, Harikane, Ikeda, Izumi, Kawaguchi, Kikuta, Kohno, Komiyama, Lupton, Minezaki, Miyazaki, Morokuma, Murayama, Niida, Nishizawa, Oguri, Ono, Ouchi, Price, Sameshima, Schulze, Shirakata, Silverman, Sugiyama, Tait, Takada, Takata, Tanaka, Tang, Toba, Utsumi, \& Wang}]{Matsuoka2018}
Matsuoka, Y., Onoue, M., Kashikawa, N., {et~al.} 2018, PASJ, 70

\bibitem[{{Matthee} {et~al.}(2023){Matthee}, {Mackenzie}, {Simcoe}, {Kashino}, {Lilly}, {Bordoloi}, \& {Eilers}}]{Matthee2023}
{Matthee}, J., {Mackenzie}, R., {Simcoe}, R.~A., {et~al.} 2023, \apj, 950, 67

\bibitem[{{Matthee} {et~al.}(2024){Matthee}, {Naidu}, {Brammer}, {Chisholm}, {Eilers}, {Goulding}, {Greene}, {Kashino}, {Labbe}, {Lilly}, {Mackenzie}, {Oesch}, {Weibel}, {Wuyts}, {Xiao}, {Bordoloi}, {Bouwens}, {van Dokkum}, {Illingworth}, {Kramarenko}, {Maseda}, {Mason}, {Meyer}, {Nelson}, {Reddy}, {Shivaei}, {Simcoe}, \& {Yue}}]{Matthee2024}
{Matthee}, J., {Naidu}, R.~P., {Brammer}, G., {et~al.} 2024, \apj, 963, 129

\bibitem[{Mazzucchelli {et~al.}(2017)Mazzucchelli, Ba{\~{n}}ados, Venemans, Decarli, Farina, Walter, Eilers, Rix, Simcoe, Stern, Fan, Schlafly, Rosa, Hennawi, Chambers, Greiner, Burgett, Draper, Kaiser, Kudritzki, Magnier, Metcalfe, Waters, \& Wainscoat}]{Mazzucchelli2017}
Mazzucchelli, C., Ba{\~{n}}ados, E., Venemans, B.~P., {et~al.} 2017, ApJ, 849, 91

\bibitem[{{Mazzucchelli} {et~al.}(2025){Mazzucchelli}, {Decarli}, {Belladitta}, {Ba{\~n}ados}, {Meyer}, {Connor}, {Momjian}, {Rojas-Ruiz}, {Eilers}, {Khusanova}, {Farina}, {Drake}, {Walter}, {Wang}, {Onoue}, \& {Venemans}}]{Mazzucchelli2025}
{Mazzucchelli}, C., {Decarli}, R., {Belladitta}, S., {et~al.} 2025, \aap, 694, A171

\bibitem[{McElwain {et~al.}(2023)McElwain, Feinberg, Perrin, Clampin, Mountain, Lallo, Lajoie, Kimble, Bowers, Stark, Acton, Atkinson, Barinek, Barto, Basinger, Beck, Bergkoetter, Bluth, Boucarut, Brady, Brooks, Brown, Byard, Carey, Carrasquilla, Chae, Chaney, Chayer, Chonis, Cohen, Cole, Comeau, Coon, Coppock, Coyle, Dean, Dziak, Eisenhower, Flagey, Franck, Gallagher, Gilman, Glassman, Green, Grieco, Haase, Hadjimichael, Hagopian, Hahn, Hartig, Havey, Hayden, Hellekson, Hicks, Holfeltz, Howard, Huguet, Jahne, Johnson, Johnston, Jurling, Kegley, Kennard, Keski-Kuha, Knight, Kulp, Levi, Levine, Lightsey, Luetgens, Mather, Matthews, McKay, Mehalick, Mel{\'{e}}ndez, Mosier, Murphy, Nelan, Niedner, Nol, Ohara, Ohl, Olczak, Osborne, Park, Perrygo, Pueyo, Redding, Regan, Reynolds, Rifelli, Rigby, Sabatke, Saif, Scorse, Seo, Shi, Sigrist, Smith, Smith, Smith, Sohn, Stahl, Telfer, Terlecki, Texter, Buren, Campen, Vila, Voyton, Waldman, Walker, Weiser, Wells, West, Whitman, Wolf, \& Zielinski}]{McElwain2023}
McElwain, M.~W., Feinberg, L.~D., Perrin, M.~D., {et~al.} 2023, PASP, 135, 058001

\bibitem[{McGreer {et~al.}(2014)McGreer, Fan, Strauss, Haiman, Richards, Jiang, Bian, \& Schneider}]{McGreer2014}
McGreer, I.~D., Fan, X., Strauss, M.~A., {et~al.} 2014, AJ, 148, 73

\bibitem[{McLeod \& Rieke(1994)}]{mcleod_1994}
McLeod, K.~K. \& Rieke, G.~H. 1994, ApJ, 420, 58

\bibitem[{Mechtley(2019)}]{psfMC}
Mechtley, M. 2019, psfMC

\bibitem[{Mechtley {et~al.}(2016)Mechtley, Jahnke, Windhorst, Andrae, Cisternas, Cohen, Hewlett, Koekemoer, Schramm, Schulze, Silverman, Villforth, van~der Wel, \& Wisotzki}]{Mechtley2016}
Mechtley, M., Jahnke, K., Windhorst, R.~A., {et~al.} 2016, ApJ, 830, 156

\bibitem[{Mechtley {et~al.}(2012)Mechtley, Windhorst, Ryan, Schneider, Cohen, Jansen, Fan, Hathi, Keel, Koekemoer, Röttgering, Scannapieco, Schneider, Strauss, \& Yan}]{Mechtley2012}
Mechtley, M., Windhorst, R.~A., Ryan, R.~E., {et~al.} 2012, ApJ, 756, L38

\bibitem[{{Melnick} {et~al.}(2015){Melnick}, {Telles}, {De Propris}, \& {Chu}}]{Melnick2015}
{Melnick}, J., {Telles}, E., {De Propris}, R., \& {Chu}, Z.-H. 2015, \aap, 582, A37

\bibitem[{{Merluzzi} {et~al.}(2018){Merluzzi}, {Busarello}, {Dopita}, {Thomas}, {Haines}, {Grado}, {Limatola}, \& {Mercurio}}]{Merluzzi2018}
{Merluzzi}, P., {Busarello}, G., {Dopita}, M.~A., {et~al.} 2018, \apj, 852, 113

\bibitem[{{Mezcua} {et~al.}(2023){Mezcua}, {Siudek}, {Suh}, {Valiante}, {Spinoso}, \& {Bonoli}}]{Mezcua2023}
{Mezcua}, M., {Siudek}, M., {Suh}, H., {et~al.} 2023, \apjl, 943, L5

\bibitem[{{Mihos} \& {Hernquist}(1996)}]{Mihos1996}
{Mihos}, J.~C. \& {Hernquist}, L. 1996, \apj, 464, 641

\bibitem[{{Moiseev} {et~al.}(2023){Moiseev}, {Smirnova}, \& {Movsessian}}]{Moiseev2023}
{Moiseev}, A.~V., {Smirnova}, A.~A., \& {Movsessian}, T.~A. 2023, Universe, 9, 493

\bibitem[{{Moran} {et~al.}(1992){Moran}, {Halpern}, {Bothun}, \& {Becker}}]{Moran1992}
{Moran}, E.~C., {Halpern}, J.~P., {Bothun}, G.~D., \& {Becker}, R.~H. 1992, \aj, 104, 990

\bibitem[{Mortlock {et~al.}(2011)Mortlock, Warren, Venemans, Patel, Hewett, McMahon, Simpson, Theuns, GonzÃ¡les-Solares, Adamson, Dye, Hambly, Hirst, Irwin, Kuiper, Lawrence, \& RÃ¶ttgering}]{Mortlock2011}
Mortlock, D.~J., Warren, S.~J., Venemans, B.~P., {et~al.} 2011, Nature, 474, 616

\bibitem[{Moseley {et~al.}(2010)Moseley, Arendt, Fixsen, Lindler, Loose, \& Rauscher}]{Moseley2010}
Moseley, S.~H., Arendt, R.~G., Fixsen, D.~J., {et~al.} 2010, Proceedings of SPIE, 7742, 77421B

\bibitem[{Nagao {et~al.}(2006)Nagao, Marconi, \& Maiolino}]{Nagao2006}
Nagao, T., Marconi, A., \& Maiolino, R. 2006, A\&A, 447, 157

\bibitem[{{Napolitano} {et~al.}(2024){Napolitano}, {Castellano}, {Pentericci}, {Vignali}, {Gilli}, {Fontana}, {Santini}, {Treu}, {Calabr{\`o}}, {Llerena}, {Piconcelli}, {Zappacosta}, {Mascia}, {Tripodi}, {Arrabal Haro}, {Bergamini}, {Bakx}, {Dickinson}, {Glazebrook}, {Henry}, {Leethochawalit}, {Mazzolari}, {Merlin}, {Morishita}, {Nanayakkara}, {Paris}, {Puccetti}, {Roberts-Borsani}, {Rojas Ruiz}, {Rosati}, {Vanzella}, {Vito}, {Vulcani}, {Wang}, {Yoon}, \& {Zavala}}]{Napolitano2024}
{Napolitano}, L., {Castellano}, M., {Pentericci}, L., {et~al.} 2024, accepted for publication in ApJ, arXiv:2410.18763

\bibitem[{Neeleman {et~al.}(2021)Neeleman, Novak, Venemans, Walter, Decarli, Kaasinen, Schindler, Ba{\~{n}}ados, Carilli, Drake, Fan, \& Rix}]{Neeleman2021}
Neeleman, M., Novak, M., Venemans, B.~P., {et~al.} 2021, ApJ, 911, 141

\bibitem[{Osterbrock(1989)}]{Osterbrock1989}
Osterbrock, D.~E. 1989, Astrophysics of gaseous nebulae and active galactic nuclei (University Science Books)

\bibitem[{{Ott}(2012)}]{Ott2012}
{Ott}, T. 2012, {QFitsView: FITS file viewer}, Astrophysics Source Code Library, record ascl:1210.019

\bibitem[{{Pacucci} {et~al.}(2023){Pacucci}, {Nguyen}, {Carniani}, {Maiolino}, \& {Fan}}]{Pacucci2023}
{Pacucci}, F., {Nguyen}, B., {Carniani}, S., {Maiolino}, R., \& {Fan}, X. 2023, \apjl, 957, L3

\bibitem[{{Pan} {et~al.}(2020){Pan}, {Lin}, {Hsieh}, {Micha{\l}owski}, {Bothwell}, {Huang}, {Moiseev}, {Oparin}, {O'Sullivan}, {Worrall}, {S{\'a}nchez}, {Gwyn}, {Law}, {Stark}, {Bizyaev}, {Li}, {Lee}, {Fu}, {Belfiore}, {Bundy}, {Fern{\'a}ndez-Trincado}, {Gelfand}, \& {Peirani}}]{Pan2020}
{Pan}, H.-A., {Lin}, L., {Hsieh}, B.-C., {et~al.} 2020, \apj, 903, 16

\bibitem[{{Pandas Development Team}(2020)}]{reback2020pandas}
{Pandas Development Team}. 2020, pandas-dev/pandas: Pandas

\bibitem[{{Park} {et~al.}(2022){Park}, {Barth}, {Ho}, \& {Laor}}]{Park2022}
{Park}, D., {Barth}, A.~J., {Ho}, L.~C., \& {Laor}, A. 2022, \apjs, 258, 38

\bibitem[{Patton {et~al.}(2000)Patton, Carlberg, Marzke, Pritchet, da~Costa, \& Pellegrini}]{Patton2000}
Patton, D.~R., Carlberg, R.~G., Marzke, R.~O., {et~al.} 2000, ApJ, 536, 153

\bibitem[{Pensabene {et~al.}(2020)Pensabene, Carniani, Perna, Cresci, Decarli, Maiolino, \& Marconi}]{Pensabene2020}
Pensabene, A., Carniani, S., Perna, M., {et~al.} 2020, A\&A, 637, A84

\bibitem[{{Perna} {et~al.}(2025){Perna}, {Arribas}, {Lamperti}, {Circosta}, {Bertola}, {P{\'e}rez-Gonz{\'a}lez}, {D'Eugenio}, {{\"U}bler}, {Cresci}, {Volonteri}, {Mannucci}, {Maiolino}, {Rodr{\'\i}guez Del Pino}, {B{\"o}ker}, {Bunker}, {Charlot}, {Willott}, {Carniani}, {Curti}, {Jones}, {Kumari}, {Marshall}, {Venturi}, {Saxena}, {Scholtz}, \& {Witstok}}]{Perna2023b}
{Perna}, M., {Arribas}, S., {Lamperti}, I., {et~al.} 2025, \aap, 696, A59

\bibitem[{{Perna} {et~al.}(2023){Perna}, {Arribas}, {Marshall}, {D'Eugenio}, {{\"U}bler}, {Bunker}, {Charlot}, {Carniani}, {Jakobsen}, {Maiolino}, {Rodr{\'\i}guez Del Pino}, {Willott}, {B{\"o}ker}, {Circosta}, {Cresci}, {Curti}, {Husemann}, {Kumari}, {Lamperti}, {P{\'e}rez-Gonz{\'a}lez}, \& {Scholtz}}]{Perna2023}
{Perna}, M., {Arribas}, S., {Marshall}, M., {et~al.} 2023, \aap, 679, A89

\bibitem[{Perrin {et~al.}(2015)Perrin, Long, Sivaramakrishnan, Lajoie, Elliot, Pueyo, \& Albert}]{Perrin2015}
Perrin, M.~D., Long, J., Sivaramakrishnan, A., {et~al.} 2015, WebbPSF: James Webb Space Telescope PSF Simulation Tool

\bibitem[{Peterson(2009)}]{Peterson2009}
Peterson, B.~M. 2009, Proc. Int. Astron. Union, 5, 151

\bibitem[{{Popesso} {et~al.}(2023){Popesso}, {Concas}, {Cresci}, {Belli}, {Rodighiero}, {Inami}, {Dickinson}, {Ilbert}, {Pannella}, \& {Elbaz}}]{Popesso2023}
{Popesso}, P., {Concas}, A., {Cresci}, G., {et~al.} 2023, \mnras, 519, 1526

\bibitem[{{Protu{\v{s}}ov{\'a}} {et~al.}(2024){Protu{\v{s}}ov{\'a}}, {Bosman}, {Wang}, {Meyer}, {Champagne}, {Davies}, {Eilers}, {Fan}, {Hennawi}, {Jin}, {Jun}, {Kakiichi}, {Li}, {Liu}, \& {Yang}}]{Protusova2024}
{Protu{\v{s}}ov{\'a}}, K., {Bosman}, S. E.~I., {Wang}, F., {et~al.} 2024, submitted to A\&A, arXiv:2412.12256

\bibitem[{Rauscher {et~al.}(2017)Rauscher, Arendt, Fixsen, Greenhouse, Lander, Lindler, Loose, Moseley, Mott, Wen, Wilson, \& Xenophontos}]{Rauscher2017}
Rauscher, B.~J., Arendt, R.~G., Fixsen, D.~J., {et~al.} 2017, PASP, 129, 105003

\bibitem[{Reines \& Volonteri(2015)}]{Reines2015}
Reines, A.~E. \& Volonteri, M. 2015, ApJ, 813, 82

\bibitem[{{Richardson} {et~al.}(2014){Richardson}, {Allen}, {Baldwin}, {Hewett}, \& {Ferland}}]{Richardson2014}
{Richardson}, C.~T., {Allen}, J.~T., {Baldwin}, J.~A., {Hewett}, P.~C., \& {Ferland}, G.~J. 2014, \mnras, 437, 2376

\bibitem[{{Rieke} {et~al.}(2015){Rieke}, {Wright}, {B{\"o}ker}, {Bouwman}, {Colina}, {Glasse}, {Gordon}, {Greene}, {G{\"u}del}, {Henning}, {Justtanont}, {Lagage}, {Meixner}, {N{\o}rgaard-Nielsen}, {Ray}, {Ressler}, {van Dishoeck}, \& {Waelkens}}]{Rieke2015}
{Rieke}, G.~H., {Wright}, G.~S., {B{\"o}ker}, T., {et~al.} 2015, \pasp, 127, 584

\bibitem[{{Rieke} {et~al.}(2023){Rieke}, {Kelly}, {Misselt}, {Stansberry}, {Boyer}, {Beatty}, {Egami}, {Florian}, {Greene}, {Hainline}, {Leisenring}, {Roellig}, {Schlawin}, {Sun}, {Tinnin}, {Williams}, {Willmer}, {Wilson}, {Clark}, {Rohrbach}, {Brooks}, {Canipe}, {Correnti}, {DiFelice}, {Gennaro}, {Girard}, {Hartig}, {Hilbert}, {Koekemoer}, {Nikolov}, {Pirzkal}, {Rest}, {Robberto}, {Sunnquist}, {Telfer}, {Wu}, {Ferry}, {Lewis}, {Baum}, {Beichman}, {Doyon}, {Dressler}, {Eisenstein}, {Ferrarese}, {Hodapp}, {Horner}, {Jaffe}, {Johnstone}, {Krist}, {Martin}, {McCarthy}, {Meyer}, {Rieke}, {Trauger}, \& {Young}}]{Rieke2023}
{Rieke}, M.~J., {Kelly}, D.~M., {Misselt}, K., {et~al.} 2023, \pasp, 135, 028001

\bibitem[{{Rodr{\'\i}guez Del Pino} {et~al.}(2024){Rodr{\'\i}guez Del Pino}, {Perna}, {Arribas}, {D'Eugenio}, {Lamperti}, {P{\'e}rez-Gonz{\'a}lez}, {{\"U}bler}, {Bunker}, {Carniani}, {Charlot}, {Maiolino}, {Willott}, {B{\"o}ker}, {Chevallard}, {Cresci}, {Curti}, {Jones}, {Parlanti}, {Scholtz}, \& {Venturi}}]{DelPino2024}
{Rodr{\'\i}guez Del Pino}, B., {Perna}, M., {Arribas}, S., {et~al.} 2024, \aap, 684, A187

\bibitem[{{Salvestrini} {et~al.}(2025){Salvestrini}, {Feruglio}, {Tripodi}, {Fontanot}, {Bischetti}, {De Lucia}, {Fiore}, {Hirschmann}, {Maio}, {Piconcelli}, {Saccheo}, {Tortosa}, {Valiante}, {Xie}, \& {Zappacosta}}]{Salvestrini2025}
{Salvestrini}, F., {Feruglio}, C., {Tripodi}, R., {et~al.} 2025, \aap, 695, A23

\bibitem[{Sanders {et~al.}(1988)Sanders, Soifer, Elias, Madore, Matthews, Neugebauer, \& Scoville}]{Sanders1988}
Sanders, D.~B., Soifer, B.~T., Elias, J.~H., {et~al.} 1988, ApJ, 325, 74

\bibitem[{Schmidt(1963)}]{schmidt_1963}
Schmidt, M. 1963, Nature, 197, 1040

\bibitem[{{Scholtz} {et~al.}(2025){Scholtz}, {Maiolino}, {D'Eugenio}, {Curtis-Lake}, {Carniani}, {Charlot}, {Curti}, {Silcock}, {Arribas}, {Baker}, {Bhatawdekar}, {Boyett}, {Bunker}, {Chevallard}, {Circosta}, {Eisenstein}, {Hainline}, {Hausen}, {Ji}, {Ji}, {Johnson}, {Kumari}, {Looser}, {Lyu}, {Maseda}, {Parlanti}, {Perna}, {Rieke}, {Robertson}, {Del Pino}, {Sun}, {Tacchella}, {{\"U}bler}, {Venturi}, {Williams}, {Willmer}, {Willott}, \& {Witstok}}]{Scholtz2023}
{Scholtz}, J., {Maiolino}, R., {D'Eugenio}, F., {et~al.} 2025, \aap, 697, A175

\bibitem[{Shen {et~al.}(2008)Shen, Greene, Strauss, Richards, \& Schneider}]{shen2008}
Shen, Y., Greene, J.~E., Strauss, M.~A., Richards, G.~T., \& Schneider, D.~P. 2008, ApJ, 680, 169

\bibitem[{Shen \& Kelly(2012)}]{Shen2012}
Shen, Y. \& Kelly, B.~C. 2012, ApJ, 746, 169

\bibitem[{{Shen} \& {Liu}(2012)}]{Shen2012b}
{Shen}, Y. \& {Liu}, X. 2012, \apj, 753, 125

\bibitem[{Shen {et~al.}(2011)Shen, Richards, Strauss, Hall, Schneider, Snedden, Bizyaev, Brewington, Malanushenko, Malanushenko, Oravetz, Pan, \& Simmons}]{Shen2011}
Shen, Y., Richards, G.~T., Strauss, M.~A., {et~al.} 2011, ApJS, 194, 45

\bibitem[{Stone {et~al.}(2023)Stone, Lyu, Rieke, \& Alberts}]{Stone2023}
Stone, M.~A., Lyu, J., Rieke, G.~H., \& Alberts, S. 2023, ApJ, 953, 180

\bibitem[{{Stone} {et~al.}(2024){Stone}, {Lyu}, {Rieke}, {Alberts}, \& {Hainline}}]{Stone2024}
{Stone}, M.~A., {Lyu}, J., {Rieke}, G.~H., {Alberts}, S., \& {Hainline}, K.~N. 2024, \apj, 964, 90

\bibitem[{{Storey} \& {Zeippen}(2000)}]{Storey2000}
{Storey}, P.~J. \& {Zeippen}, C.~J. 2000, \mnras, 312, 813

\bibitem[{Sun(2020)}]{SunGithub}
Sun, J. 2020, Sun\_Astro\_Tools, GitHub

\bibitem[{Sun {et~al.}(2018)Sun, Leroy, Schruba, Rosolowsky, Hughes, Kruijssen, Meidt, Schinnerer, Blanc, Bigiel, Bolatto, Chevance, Groves, Herrera, Hygate, Pety, Querejeta, Usero, \& Utomo}]{Sun2018}
Sun, J., Leroy, A.~K., Schruba, A., {et~al.} 2018, ApJ, 860, 172

\bibitem[{{Sun} {et~al.}(2025){Sun}, {Lyu}, {Rieke}, {Ji}, {Sun}, {Zhu}, {Bunker}, {Cargile}, {Circosta}, {D'Eugenio}, {Egami}, {Hainline}, {Helton}, {Rinaldi}, {Robertson}, {Scholtz}, {Shivaei}, {Stone}, {Tacchella}, {Williams}, {Willmer}, \& {Willott}}]{Sun2024}
{Sun}, Y., {Lyu}, J., {Rieke}, G.~H., {et~al.} 2025, \apj, 978, 98

\bibitem[{{Trakhtenbrot} {et~al.}(2017){Trakhtenbrot}, {Lira}, {Netzer}, {Cicone}, {Maiolino}, \& {Shemmer}}]{Trakhtenbrot2017}
{Trakhtenbrot}, B., {Lira}, P., {Netzer}, H., {et~al.} 2017, ApJ, 836, 8

\bibitem[{{Trakhtenbrot} {et~al.}(2015){Trakhtenbrot}, {Urry}, {Civano}, {Rosario}, {Elvis}, {Schawinski}, {Suh}, {Bongiorno}, \& {Simmons}}]{Trakhtenbrot2015}
{Trakhtenbrot}, B., {Urry}, C.~M., {Civano}, F., {et~al.} 2015, Science, 349, 168

\bibitem[{Trevese {et~al.}(2014)Trevese, Perna, Vagnetti, Saturni, \& Dadina}]{Trevese2014}
Trevese, D., Perna, M., Vagnetti, F., Saturni, F.~G., \& Dadina, M. 2014, ApJ, 795, 164

\bibitem[{{{\"U}bler} {et~al.}(2023){{\"U}bler}, {Maiolino}, {Curtis-Lake}, {P{\'e}rez-Gonz{\'a}lez}, {Curti}, {Perna}, {Arribas}, {Charlot}, {Marshall}, {D'Eugenio}, {Scholtz}, {Bunker}, {Carniani}, {Ferruit}, {Jakobsen}, {Rix}, {Rodr{\'\i}guez Del Pino}, {Willott}, {Boeker}, {Cresci}, {Jones}, {Kumari}, \& {Rawle}}]{Uebler2023}
{{\"U}bler}, H., {Maiolino}, R., {Curtis-Lake}, E., {et~al.} 2023, \aap, 677, A145

\bibitem[{{{\"U}bler} {et~al.}(2024){{\"U}bler}, {Maiolino}, {P{\'e}rez-Gonz{\'a}lez}, {D'Eugenio}, {Perna}, {Curti}, {Arribas}, {Bunker}, {Carniani}, {Charlot}, {Rodr{\'\i}guez Del Pino}, {Baker}, {B{\"o}ker}, {Cresci}, {Dunlop}, {Grogin}, {Jones}, {Kumari}, {Lamperti}, {Laporte}, {Marshall}, {Mazzolari}, {Parlanti}, {Rawle}, {Scholtz}, {Venturi}, \& {Witstok}}]{Uebler2024}
{{\"U}bler}, H., {Maiolino}, R., {P{\'e}rez-Gonz{\'a}lez}, P.~G., {et~al.} 2024, \mnras, 531, 355

\bibitem[{{van den Bosch} {et~al.}(2012){van den Bosch}, {Gebhardt}, {G{\"u}ltekin}, {van de Ven}, {van der Wel}, \& {Walsh}}]{VanDenBosch2012}
{van den Bosch}, R. C.~E., {Gebhardt}, K., {G{\"u}ltekin}, K., {et~al.} 2012, \nat, 491, 729

\bibitem[{van~der Walt {et~al.}(2011)van~der Walt, Colbert, \& Varoquaux}]{Numpy2011}
van~der Walt, S., Colbert, S.~C., \& Varoquaux, G. 2011, CiSE, 13, 22

\bibitem[{van~der Wel {et~al.}(2022)van~der Wel, van Houdt, Bezanson, Franx, D'Eugenio, Straatman, Bell, Muzzin, Sobral, Maseda, de~Graaff, \& Holden}]{VanDerWel2022}
van~der Wel, A., van Houdt, J., Bezanson, R., {et~al.} 2022, ApJ, 936, 9

\bibitem[{van Dokkum(2001)}]{Dokkum2001}
van Dokkum, P.~G. 2001, PASP, 113, 1420

\bibitem[{Venemans {et~al.}(2012)Venemans, McMahon, Walter, Decarli, Cox, Neri, Hewett, Mortlock, Simpson, \& Warren}]{Venemans2012}
Venemans, B.~P., McMahon, R.~G., Walter, F., {et~al.} 2012, ApJ, 751, L25

\bibitem[{Venemans {et~al.}(2017)Venemans, Walter, Decarli, Ba{\~{n}}ados, Hodge, Hewett, McMahon, Mortlock, \& Simpson}]{Venemans2017a}
Venemans, B.~P., Walter, F., Decarli, R., {et~al.} 2017, ApJ, 837, 146

\bibitem[{{Venemans} {et~al.}(2020){Venemans}, {Walter}, {Neeleman}, {Novak}, {Otter}, {Decarli}, {Ba{\~n}ados}, {Drake}, {Farina}, {Kaasinen}, {Mazzucchelli}, {Carilli}, {Fan}, {Rix}, \& {Wang}}]{Venemans2020}
{Venemans}, B.~P., {Walter}, F., {Neeleman}, M., {et~al.} 2020, \apj, 904, 130

\bibitem[{Venemans {et~al.}(2015)Venemans, Walter, Zschaechner, Decarli, Rosa, Findlay, McMahon, \& Sutherland}]{Venemans2015}
Venemans, B.~P., Walter, F., Zschaechner, L., {et~al.} 2015, ApJ, 816, 37

\bibitem[{{V{\'e}ron-Cetty} {et~al.}(2004){V{\'e}ron-Cetty}, {Joly}, \& {V{\'e}ron}}]{VeronCetty2004}
{V{\'e}ron-Cetty}, M.~P., {Joly}, M., \& {V{\'e}ron}, P. 2004, \aap, 417, 515

\bibitem[{Vestergaard \& Osmer(2009)}]{Vestergaard2009}
Vestergaard, M. \& Osmer, P.~S. 2009, ApJ, 699, 800

\bibitem[{Vestergaard \& Peterson(2006)}]{Vestergaard2006}
Vestergaard, M. \& Peterson, B.~M. 2006, ApJ, 641, 689

\bibitem[{{Virtanen} {et~al.}(2020){Virtanen}, {Gommers}, {Oliphant}, {Haberland}, {Reddy}, {Cournapeau}, {Burovski}, {Peterson}, {Weckesser}, {Bright}, {van der Walt}, {Brett}, {Wilson}, {Jarrod Millman}, {Mayorov}, {Nelson}, {Jones}, {Kern}, {Larson}, {Carey}, {Polat}, {Feng}, {Moore}, {Vand erPlas}, {Laxalde}, {Perktold}, {Cimrman}, {Henriksen}, {Quintero}, {Harris}, {Archibald}, {Ribeiro}, {Pedregosa}, {van Mulbregt}, \& {Contributors}}]{2020SciPy-NMeth}
{Virtanen}, P., {Gommers}, R., {Oliphant}, T.~E., {et~al.} 2020, Nat. Methods, 17, 261

\bibitem[{Wagg {et~al.}(2012)Wagg, Wiklind, Carilli, Espada, Peck, Riechers, Walter, Wootten, Aravena, Barkats, Cortes, Hills, Hodge, Impellizzeri, Iono, Leroy, Mart{\'{\i}}n, Rawlings, Maiolino, McMahon, Scott, Villard, \& Vlahakis}]{Wagg2012}
Wagg, J., Wiklind, T., Carilli, C.~L., {et~al.} 2012, ApJ, 752, L30

\bibitem[{Walter {et~al.}(2003)Walter, Bertoldi, Carilli, Cox, Lo, Neri, Fan, Omont, Strauss, \& Menten}]{Walter2003}
Walter, F., Bertoldi, F., Carilli, C., {et~al.} 2003, Nature, 424, 406

\bibitem[{Walter {et~al.}(2009)Walter, Riechers, Cox, Neri, Carilli, Bertoldi, Weiss, \& Maiolino}]{Walter2009}
Walter, F., Riechers, D., Cox, P., {et~al.} 2009, Nature, 457, 699

\bibitem[{{Wang} {et~al.}(2021){Wang}, {Yang}, {Fan}, {Hennawi}, {Barth}, {Banados}, {Bian}, {Boutsia}, {Connor}, {Davies}, {Decarli}, {Eilers}, {Farina}, {Green}, {Jiang}, {Li}, {Mazzucchelli}, {Nanni}, {Schindler}, {Venemans}, {Walter}, {Wu}, \& {Yue}}]{Wang2021b}
{Wang}, F., {Yang}, J., {Fan}, X., {et~al.} 2021, \apjl, 907, L1

\bibitem[{Wang {et~al.}(2019)Wang, Yang, Fan, Wu, Yue, Li, Bian, Jiang, Ba{\~{n}}ados, Schindler, Findlay, Davies, Decarli, Farina, Green, Hennawi, Huang, Mazzuccheli, McGreer, Venemans, Walter, Dye, Lyke, Myers, \& Nunez}]{Wang2019}
Wang, F., Yang, J., Fan, X., {et~al.} 2019, ApJ, 884, 30

\bibitem[{Wang {et~al.}(2013)Wang, Wagg, Carilli, Walter, Lentati, Fan, Riechers, Bertoldi, Narayanan, Strauss, Cox, Omont, Menten, Knudsen, Neri, \& Jiang}]{Wang2013}
Wang, R., Wagg, J., Carilli, C.~L., {et~al.} 2013, ApJ, 773, 44

\bibitem[{Waskom(2021)}]{Waskom2021}
Waskom, M. 2021, JOSS, 6, 3021

\bibitem[{Willott {et~al.}(2010)Willott, Albert, Arzoumanian, Bergeron, Crampton, Delorme, Hutchings, Omont, Reyl{\'{e}}, \& Schade}]{Willott2010}
Willott, C.~J., Albert, L., Arzoumanian, D., {et~al.} 2010, AJ, 140, 546

\bibitem[{Willott {et~al.}(2015)Willott, Bergeron, \& Omont}]{Willott2015}
Willott, C.~J., Bergeron, J., \& Omont, A. 2015, ApJ, 801, 123

\bibitem[{Willott {et~al.}(2017)Willott, Bergeron, \& Omont}]{Willott2017}
Willott, C.~J., Bergeron, J., \& Omont, A. 2017, ApJ, 850, 108

\bibitem[{Willott {et~al.}(2009)Willott, Delorme, Reyl\'e, Albert, Bergeron, Crampton, Delfosse, Forveille, Hutchings, McLure, Omont, \& Schade}]{willott_2009}
Willott, C.~J., Delorme, P., Reyl\'e, C., {et~al.} 2009, AJ, 137, 3541

\bibitem[{Willott {et~al.}(2013)Willott, Omont, \& Bergeron}]{willott_2013b}
Willott, C.~J., Omont, A., \& Bergeron, J. 2013, ApJ, 770, 13

\bibitem[{Willott {et~al.}(2005)Willott, Percival, McLure, Crampton, Hutchings, Jarvis, Sawicki, \& Simard}]{willott_2005}
Willott, C.~J., Percival, W.~J., McLure, R.~J., {et~al.} 2005, ApJ, 626, 657

\bibitem[{{Yang} {et~al.}(2023){Yang}, {Fan}, {Gupta}, {Myers}, {Palanque-Delabrouille}, {Wang}, {Y{\`e}che}, {Aguilar}, {Ahlen}, {Alexander}, {Brooks}, {Dawson}, {de la Macorra}, {Dey}, {Dhungana}, {Fanning}, {Font-Ribera}, {Gontcho}, {Guy}, {Honscheid}, {Juneau}, {Kisner}, {Kremin}, {Le Guillou}, {Levi}, {Magneville}, {Martini}, {Meisner}, {Miquel}, {Moustakas}, {Nie}, {Percival}, {Poppett}, {Prada}, {Schlafly}, {Tarl{\'e}}, {Vargas Magana}, {Weaver}, {Wechsler}, {Zhou}, {Zhou}, \& {Zou}}]{Yang2023}
{Yang}, J., {Fan}, X., {Gupta}, A., {et~al.} 2023, \apjs, 269, 27

\bibitem[{Yang {et~al.}(2021)Yang, Wang, Fan, Barth, Hennawi, Nanni, Bian, Davies, Farina, Schindler, Ba{\~{n}}ados, Decarli, Eilers, Green, Guo, Jiang, Li, Venemans, Walter, Wu, \& Yue}]{Yang2021a}
Yang, J., Wang, F., Fan, X., {et~al.} 2021, ApJ, 923, 262

\bibitem[{Yang {et~al.}(2023)Yang, Wang, Fan, Hennawi, Barth, Bañados, Sun, Liu, Cai, Jiang, Li, Onoue, Schindler, Shen, Wu, Bhowmick, Bieri, Blecha, Bosman, Champagne, Colina, Connor, Costa, Davies, Decarli, De~Rosa, Drake, Egami, Eilers, Evans, Farina, Habouzit, Haiman, Jin, Jun, Kakiichi, Khusanova, Kulkarni, Loiacono, Lupi, Mazzucchelli, Pan, Rojas-Ruiz, Strauss, Tee, Trakhtenbrot, Trebitsch, Venemans, Vestergaard, Volonteri, Walter, Xie, Yue, Zhang, Zhang, \& Zou}]{Yang2023a}
Yang, J., Wang, F., Fan, X., {et~al.} 2023, ApJL, 951, L5

\bibitem[{Yang {et~al.}(2020)Yang, Wang, Fan, Hennawi, Davies, Yue, Banados, Wu, Venemans, Barth, Bian, Boutsia, Decarli, Farina, Green, Jiang, Li, Mazzucchelli, \& Walter}]{Yang2020}
Yang, J., Wang, F., Fan, X., {et~al.} 2020, ApJ, 897, L14

\bibitem[{{Yue} {et~al.}(2024){Yue}, {Eilers}, {Simcoe}, {Mackenzie}, {Matthee}, {Kashino}, {Bordoloi}, {Lilly}, \& {Naidu}}]{Yue2023}
{Yue}, M., {Eilers}, A.-C., {Simcoe}, R.~A., {et~al.} 2024, \apj, 966, 176

\bibitem[{{Zhuang} \& {Shen}(2024)}]{zhuangshen2023}
{Zhuang}, M.-Y. \& {Shen}, Y. 2024, \apj, 962, 139

\end{thebibliography}
\bibliographystyle{aa}
\nopagebreak
\appendix
\counterwithin{figure}{section}

\section{PSF estimate}

\begin{figure}[h!]
\begin{center}
\includegraphics[scale=0.8]{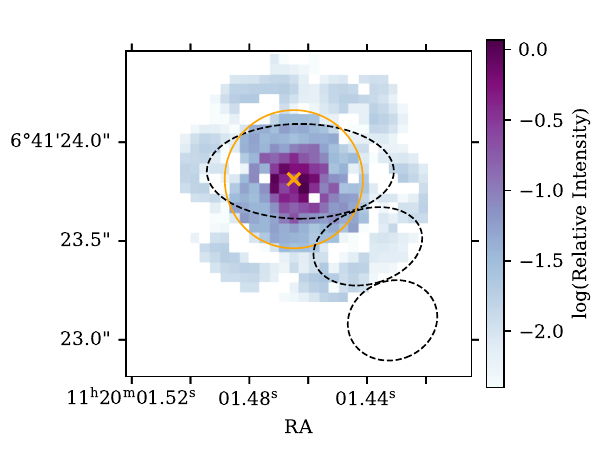}
\caption{NIRSpec IFU PSF at 4$\mu$m as traced by the quasar \hb\ broad-line flux. The location of the peak of the quasar is marked as an orange cross, with the 0\farcs35 radius aperture used to integrate the quasar spectrum shown as an orange circle. The black ellipses are the same as in Figure \ref{fig:RegionMaps} to aid in a spatial comparison to the extended galaxy emission.
}
\label{fig:PSF}
\end{center}
\end{figure}

Using \textrm{QDeblend3D} \citep{Husemann2013,Husemann2014}, we estimated the NIRSpec IFU PSF at 4$\mu$m by tracing the relative strength of the quasar \hb\ broad line across each spaxel, as discussed in Section \ref{sec:IFUsubtraction}. We plot this spatial PSF in Figure \ref{fig:PSF}, for reference.
Figure \ref{fig:PSF} also shows our integration aperture of radius $0\farcs35$, centred on the peak of the quasar emission, selected to maximise the $\rm{S/N}$ while containing the majority of the quasar flux. This contains the main core of the PSF, yet not the ring of reduced flux that occurs prior to the second radial PSF peak.

\section{BPT diagnostics}
\label{sec:BPT}

\begin{figure}[h!]
\begin{center}
\includegraphics[scale=0.8]{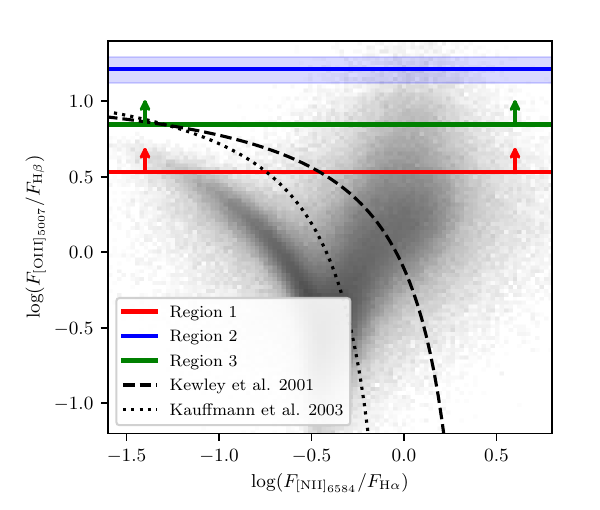}
\caption{BPT diagnostic diagram showing the \oiiia/\hb\ flux ratio measured for the companion galaxy (Region 2), and the lower limits for the host galaxy (Region 1) and Region 3, as listed in Table \ref{tab:HostFlux}.
For reference, we show the \citet{Kauffmann2003} (black dotted) and \citet{Kewley2001} (black dashed) curves which attempt to differentiate between galaxies primarily ionised by star formation (lower left) and AGN (upper right) at low-$z$.
 We also show local sources from the SDSS \citep[][2D histogram]{Abazajian2009} for comparison.
}
\label{fig:BPT}
\end{center}
\end{figure}

Using our measured \hb\ and \oiiia\ flux values and limits from Table \ref{tab:HostFlux}, we place the emission-line regions on a BPT diagram \citep{Baldwin1981}, which we show in Figure \ref{fig:BPT}.
This diagnostic aims to characterise the main photoionisation source based on the narrow-line flux; we use the fluxes measured from the quasar-subtracted cube, where the narrow-line flux is emitted by the extended galaxy component and the point source quasar contribution has been removed.
As there are no existing measurements of \niia\, we can only constrain the location of these galaxies on the vertical \oiiia/\hb\  axis.

\label{LastPage}
\end{document}